\documentclass[12pt,letterpaper]{article}
\usepackage[margin=1in]{geometry}
\usepackage[T1]{fontenc}
\usepackage[utf8]{inputenc}
\usepackage{lmodern}
\usepackage{microtype}
\usepackage{amsmath, amssymb}
\usepackage{booktabs}
\usepackage{threeparttable}
\usepackage{makecell}
\usepackage{graphicx}
\usepackage{float}
\usepackage{caption}
\usepackage{subcaption}
\usepackage{setspace}
\usepackage{natbib}
\usepackage{array}
\usepackage{tabularx}
\usepackage{adjustbox}
\usepackage{multirow}
\usepackage{longtable}
\usepackage{pdflscape}
\usepackage[section]{placeins}
\usepackage{xurl}
\usepackage{rotating}
\usepackage{hyperref}
\hypersetup{
  hidelinks,
  pdftitle={Do Sanctions Backfire? New Evidence on the Macroeconomic Effects of Supporting Ukraine},
  pdfauthor={Vicente Rios, Izaskun Barba, Lisa Gianmoena},
  pdfsubject={Sender-side macroeconomic effects of sanctions on Russia},
  pdfkeywords={economic sanctions; Russia; Ukraine; consumer prices; common correlated effects; difference-in-differences},
  bookmarksnumbered=true
}
\newcommand{\sym}[1]{\ifmmode^{#1}\else\(^{#1}\)\fi}

\title{Do Sanctions Backfire? New Evidence on the Macroeconomic Effects of Supporting Ukraine}
\author{%
	Vicente Rios$^{1,*}$ \quad
	Izaskun Barba$^{2}$ \quad
	Lisa Gianmoena$^{1}$
}
\date{September 11, 2026}

\begin{document}
	
	\pagenumbering{gobble}
	\maketitle
	
	\begin{center}
		\footnotesize
		\noindent $^{1}$ Department of Economics and Management, University of Pisa, Italy.\\
		\noindent $^{2}$ Department of Economics, Universidad P\'ublica de Navarra, Spain. Email: \href{mailto:izaskun.barba@unavarra.es}{izaskun.barba@unavarra.es}\\
		\noindent $^{*}$ Corresponding author: \href{mailto:vicente.rios@unipi.it}{vicente.rios@unipi.it}.\\
		\noindent Lisa Gianmoena: \href{mailto:lisa.gianmoena@unipi.it}{lisa.gianmoena@unipi.it}.
	\end{center}
	
	\begin{abstract}
		 Using a balanced panel of 129 countries from 2000 to 2024, we estimate the macroeconomic costs of sanctions on Russia. To that end, we employ a novel two-wave common-correlated-effects  imputation design that separates the 2014 sanctions from the 2022 escalation and accounts for heterogeneous exposure to global shocks. We find the main boomerang effect of these sanctions operates through prices rather than output. Specifically, relative to 2019-2021, we find a significant post-2022 increase in prices among sanctioning countries that remains robust under alternative counterfactuals, reference windows, and inference procedures. Conversely, we find no robust evidence of aggregate GDP-per-capita losses.
		\end{abstract}
	\medskip
	\begin{quote}\small
	\end{quote}
	\clearpage
	\pagenumbering{arabic} 
	
	\section{Introduction \label{sec:introduction}}
	
	Economic sanctions are intended to raise the target country's cost of noncompliance. However, they usually come with consequences for the states that impose them. Sanctions that produce trade restrictions eliminate mutually beneficial exchanges, financial sanctions interrupt cross-border capital flows, and energy-related measures can raise input and consumer prices in economies that previously depended on the target as a supplier. 
	
	These reciprocal costs are of major importance for the credibility and durability of sanctions. Nevertheless, they have received much less attention than the damage inflicted on target economies \citep{Wagner1988,EatonEngers1992,Felbermayr2021}. For policy design, the relevant question is not only whether sanctions damage the target but also how the burden is transmitted back to the countries implementing them.
	
	Sender-side costs need not show up as a recession. Substitution, inventories, and fiscal support can keep real GDP per capita up while import costs and consumer prices rise. Monitoring GDP alone would then understate the domestic incidence of sanctions.
	
	The measures adopted against Russia after the annexation of Crimea in 2014 and the full-scale invasion of Ukraine in 2022 provide a unique setting in which to examine such sender-side effects.
	
	Existing work on the 2014 and 2022 Russia episodes has not established whether the countries that sanctioned Russia paid a measurable national macroeconomic price, in the consumer-price level or in real GDP per capita. Research on the 2014 episode documents losses in bilateral trade and finance, especially in Europe, but comparatively limited average output effects \citep{Besedes2017,KholodilinNetsunajev2019,CrozetHinz2020}. Research after 2022 shows large and heterogeneous energy-price, welfare, and sectoral effects \citep{Ari2022,YagiManagi2023,Bachmann2024,DiBella2024}. Two limits remain. Many 2022 aggregate output estimates are model-based rather than realized, and most realized studies cover firms, sectors, households, or single countries rather than a coalition-wide macroeconomic panel. Moreover, studies generally examine either the 2014 episode or the 2022 war escalation, even though the countries involved in the latter had often been exposed to the former for almost a decade.
	
Estimating those sender-side effects raises three empirical problems. First, sanctioning countries are not a random subset of the world economy: they are disproportionately high-income, democratic, trade-integrated, and exposed to European energy markets. Second, the post-2022 period also coincided with pandemic recovery, supply-chain disruption, commodity-price shocks, and global monetary tightening, thus involving multiple sources of confounding. In this context, relying on a standard two-way fixed-effects panel estimator would absorb these global shocks and trends into common year effects, forcing an unrealistic assumption of homogeneous sensitivity across countries. Finally, treating the 2022 escalation as a singular intervention would rely on a pre-treatment baseline already contaminated by the post-2014 sanctions regime.

To address these modeling challenges, we assembled a balanced annual panel of
129 countries for 2000--2024 and estimated a parsimonious two-wave Common Correlated Effects
Difference-in-Differences (CCEDID) imputation model combining the approaches of
\cite{BrownButtsWesterlund2026} and \cite{Pesaran2006}. The model constructs a
country-specific counterfactual simulation of the macroeconomic path that each
coalition member would have followed under continued non-participation in the sanctions regime. Using only pre-2014 data, it learns how each country's outcomes and macroeconomic
controls historically moved with international common factors observed among
never-treated donors, while allowing each country to respond differently to
those common movements. It then applies these pre-treatment relationships to
the post-2013 donor path to simulate untreated controls and outcomes. Comparing
observed outcomes with these simulated untreated counterfactual paths yields separate gaps for the 2014--2021 sanctions regime and for the incremental shift after the 2022
escalation. 

Throughout the paper, we estimate the macroeconomic effect of
participation in the sanctions coalition, rather than the effect of a single
legal measure. Our results reveal a clear contrast between prices and output. Relative to each country's 2019--2021 gap, log CPI in sanctioning countries lies 0.072 log points (about 7.5 percent) above its imputed untreated path in 2022--2024. That incremental gap is not the same object as the Wave-2 level ATT of 0.022. This finding remains stable across a wide range of robustness checks. By contrast, we do not find robust evidence of an aggregate GDP-per-capita loss. Furthermore, we find that cross-country differences in the estimated post-2022 CPI and GDP-per-capita growth differentials are related to pre-war energy exposure. Within the sanctioning country group, a one-standard-deviation increase in Russian fossil-energy imports relative to GDP over 2018--2021 is associated with a larger post-2022 CPI differential of approximately 2.2 percentage points and 0.96 percentage points lower annual GDP-per-capita growth. These  estimates reveal a source of vulnerability: the post-2022 price and growth burden was larger among the sanctioning coalition members that entered the escalation with greater measured exposure to Russian energy.
	
The paper contributes to the existing literature in three ways. First, it measures realized sender-side incidence at the national macroeconomic level and distinguishes price costs from output costs. Second, it provides a replicable two-wave policy-evaluation framework for interventions that overlap with large common shocks and follow an earlier treatment regime. Third, it connects aggregate incidence to a pre-existing and policy-relevant source of vulnerability: exposure to Russian energy.
	
The remainder of the paper proceeds as follows. Section~\ref{sec:literature_review} reviews the related literature. Sections~\ref{sec:data} and~\ref{sec:econometric_strategy} describe the data and empirical strategy. Section~\ref{sec:results} presents the main results, and Section~\ref{sec:robustness} reports the robustness checks. Section~\ref{sec:russia_energy} examines the role of pre-war Russian-energy exposure, and Section~\ref{sec:conclusion} concludes.

	\section{Literature Review \label{sec:literature_review}} 
	
	\subsection{Economic sanctions and the incidence of sender costs}

	Economic sanctions are instruments of coercive diplomacy that restrict commercial, financial, or political relations in order to raise the target's cost of noncompliance. They range from comprehensive measures that curtail broad economic relations to more targeted or ``smart'' sanctions directed at particular sectors, firms, officials, or financial transactions \citep{CortrightLopez2002,Drezner2011,Felbermayr2021}. The sanctions literature has traditionally focused on whether these measures impose sufficient costs on the target and on the conditions under which economic pressure produces political concessions \citep{Drezner1999,Hufbauer2007,Felbermayr2021}. The same framework implies reciprocal costs for senders. Disrupted trade and finance eliminate mutual gains from exchange and can therefore impose costs on both sender and target.
    In bargaining models, the sender’s willingness to bear those costs helps determine what it can extract from the target \citep{EatonEngers1992}. At the same time, asymmetric interdependence alone (i.e, having the upper hand in trade) does not guarantee political leverage \citep{Wagner1988} and in multilateral alliances, unequal domestic burdens may also complicate coalition participation, unity and enforcement \citep{Felbermayr2021,ItskhokiRibakova2024}. The expected distribution of sender-side losses therefore affects bargaining power, the credibility of the threat, and whether the coalition holds.

	Sender costs are a policy-design problem because aggregate economic resilience can coexist with concentrated losses. Firms, industries, workers, households, and regions can bear substantial adjustment costs even when national output changes little. In turn, concentrated losses can generate lobbying, evasion, demands for exemptions, and pressure to relax sanctions \citep{Kirshner1997,McLeanWhang2014,Early2015}. Thus, a coalition's capacity to sustain economic coercion depends on both the size and the incidence of domestic costs.

	Most empirical work nevertheless concentrates on the target. Sanctions are associated with lower output, investment, consumption, and trade in sanctioned economies, although effects vary with coalition breadth, prior economic ties, policy instruments, and adaptation capacity of the target \citep{Neuenkirch2015,Gutmann2023,Felbermayr2021}. Sender-side evidence is scant and is observed most clearly in bilateral trade, cross-border finance, firms, and exposed sectors \citep{Besedes2017,CrozetHinz2020}. This leaves open the question of whether the burden appears in national prices, aggregate production, or both.
	
	\subsection{The Russia episodes: trade, energy, and macroeconomic adjustment}
	
	The 2014 measures against Russia after the  annexation of Crimea in 2014 were targeted, focusing on finance, arms, selected dual-use and energy technologies, firms, and individuals. The clearest sender-side effects were correspondingly bilateral. \citet{CrozetHinz2020} estimate sizable export losses for sanctioning economies, concentrated in the European Union, while \citet{Besedes2017} document sharp reductions and diversion in German cross-border financial flows. \citet{Sedrakyan2022} documents trade and foreign-investment spillovers across transition economies, while \citet{Dai2021} show that sanctions can have persistent effects on bilateral trade relationships even after formal restrictions are removed. Evidence on aggregate euro-area output is much weaker \citep{KholodilinNetsunajev2019}. General-equilibrium analyses likewise imply that average sender losses were smaller than losses in Russia, although highly heterogeneous across trading partners \citep{Simola2023,Flach2024}.
	
	The 2022 escalation changed both the policy scope and the relevant transmission mechanisms. Restrictions on Russian finance, technology, trade, and central-bank reserves coincided with Russian supply reductions, the reorientation of fossil-fuel imports, war-related uncertainty, and global macroeconomic adjustment. The 2022 episode is a bundle: legal sanctions, Russian supply cuts, energy reallocation, and domestic compensation. Our estimates capture the realized path of coalition members under that bundle, not the ceteris-paribus effect of a single legal instrument.
	
	Energy was the most visible channel in Europe. Retail energy and electricity prices rose sharply, with the incidence depending on Russian dependence, alternative suppliers, interconnection, the energy mix, and fiscal mitigation \citep{Ari2022,DiBella2024}. Structural models generally predict manageable average output losses once substitution and reallocation are incorporated, but much larger costs for highly exposed countries and sectors \citep{Hosoe2023,Bachmann2024}. Input--output and microsimulation studies show that the same shock can generate substantial price, welfare, and distributional effects even when aggregate output remains resilient \citep{YagiManagi2023,BonfattiGiarda2025,Bonfiglio2026}.

    Therefore, this strand of literature cautions against interpreting resilient output as evidence that sender costs were small and suggests the distinction between prices and quantities is relevant. Substitution can preserve production while increasing the cost of replacement inputs, and fiscal support can protect activity while moving part of the burden to public budgets. If only aggregate GDP is monitored, purchasing-power losses and the case for compensation are understated. The same studies also imply that pre-crisis energy exposure is an observable marker of who will need that compensation.
	
	\subsection{The policy-evaluation gap}
	
	Three gaps remain. First, existing ex-post studies do not establish whether the broader coalition experienced systematic deviations in national price levels and real GDP per capita. Second, most analyses study either the post-2014 regime or the post-2022 escalation, although the later intervention occurred after almost a decade of prior sanctions and adjustment. Third, the literature offers limited guidance on which macroeconomic indicators should be monitored and which coalition members are most likely to require mitigation or burden sharing.
	
    This study addresses those gaps by estimating coalition-wide counterfactual
paths over both waves, separating persistent outcome levels from annual rates,
and relating post-2022 heterogeneity to pre-war Russian-energy exposure.

	Table~\ref{tab:literature_gap_design} summarizes the sender-side studies most closely related to the present study.
	
 \begin{sidewaystable}[p]
    \centering
    \caption{Selected sender-side evidence by evaluation design and the policy-evaluation gap
        \label{tab:literature_gap_design}}
\footnotesize
\setlength{\tabcolsep}{6pt}
    \renewcommand{\arraystretch}{1.20}
    \begin{threeparttable}
        \begin{tabularx}{\textwidth}{
                >{\raggedright\arraybackslash}p{3.5cm}
                >{\raggedright\arraybackslash}p{4.0cm}
                X
                >{\raggedright\arraybackslash}p{3.0cm}
                >{\raggedright\arraybackslash}p{4.5cm}
                >{\centering\arraybackslash}p{1.2cm}
            }
            \toprule
            Study & Scope & Main outcome and unit & Evidence type & Method & Wave \\
            \midrule
            
            \multicolumn{6}{@{}l}{\textit{Trade, finance, and first-wave effects}} \\
            
            \citet{Besedes2017}
            & Germany and sanctioned partners
            & Bilateral cross-border capital flows
            & Ex post causal
            & DiD
            & Mult. \\
            
            \citet{KholodilinNetsunajev2019}
            & Russia and aggregate euro area
            & Aggregate GDP and real effective exchange rate
            & Ex post structural
            & SVAR with narrative sign restrictions
            & 2014 \\
            
            \citet{CrozetHinz2020}
            & 37 sanctioning countries; French firms
            & Bilateral exports and firm-level trade adjustment
            & Ex post structural
            & Structural gravity + GE; firm-level PPML
            & 2014 \\
            
            \citet{Flach2024}
            & Russia and trading partners
            & Bilateral trade and country-level real income
            & Ex post structural
            & Structural gravity + multisector GE
            & 2014 \\
            
            \addlinespace[4pt]
            \multicolumn{6}{@{}l}{\textit{Macroeconomic costs under post-2022 disruption scenarios}} \\
            
            \citet{Hosoe2023}
            & World economy and alternative sender coalitions
            & National GDP and welfare
            & Ex ante simulation
            & World CGE model
            & 2022 \\
            
            \citet{DiBella2024}
            & European economies
            & Gas shortages and national output
            & Ex ante simulation
            & Gas-network and macroeconomic scenarios
            & 2022 \\
            
            \citet{Bachmann2024}
            & Germany
            & GDP, welfare, and distribution
            & Ex ante simulation
            & Multisector model and production-function bounds
            & 2022 \\
            
            \addlinespace[4pt]
            \multicolumn{6}{@{}l}{\textit{Prices, household welfare, financial markets, and sectoral incidence}} \\
            
            \citet{Ari2022}
            & European economies and households
            & Household cost of living and policy support
            & Policy/descriptive
            & Incidence simulation and policy assessment
            & 2022 \\
            
            \citet{YagiManagi2023}
            & 44 countries and 56 sectors
            & Sectoral price transmission and welfare spillovers
            & Ex ante simulation
            & Monthly Leontief input--output model
            & 2022 \\
            
            \citet{BonfattiGiarda2025}
            & Italian households
            & Household expenditure, income, and distribution
            & Ex post structural
            & Retrospective microsimulation
            & 2022 \\
            
            \citet{Bonfiglio2026}
            & Italian regions and agri-food sectors
            & Production, trade, and input--output linkages
            & Ex post structural
            & Nonlinear programming + MRIO counterfactuals
            & 2022 \\
            
            \citet{Klose2024}
            & 23 sanctioning and supporting countries
            & Stock prices and exchange rates
            & Ex post structural
            & Recursively identified panel VAR
            & 2022 \\
            
            \bottomrule
        \end{tabularx}
        
        \begin{tablenotes}[flushleft]
            \footnotesize
            \item \textit{Notes:}
            The evidence type refers to the principal design used to estimate
            the sender-side outcome summarized in the table.
            ``Ex post causal'' denotes analysis of realized outcomes using an
            explicit comparison or counterfactual identification strategy.
            ``Ex post structural'' denotes retrospective evaluation of a realized
            shock or policy through an estimated or calibrated structural,
            dynamic, or microsimulation model.
            ``Ex ante simulation'' denotes evaluation of a hypothetical or
            not-yet-observed disruption scenario.
            ``Policy/descriptive'' denotes incidence accounting and policy
            assessment without an identified causal counterfactual.
            Hybrid studies are classified according to the component producing
            the main sender-side estimate reported here.
            ``Mult.'' denotes multiple sanctions episodes rather than a
            Russia-specific wave.
            CGE denotes computable general equilibrium; GE, general equilibrium;
            MRIO, multiregional input--output; PPML, Poisson
            pseudo-maximum likelihood; SVAR, structural vector autoregression;
            VAR, vector autoregression; and DiD, difference-in-differences.
        \end{tablenotes}
    \end{threeparttable}
\end{sidewaystable}

	\section{Data and Policy Definitions \label{sec:data}}
	
	We construct a balanced annual panel of 129 countries for 2000--2024,
	yielding 3,225 country-year observations. The panel contains four outcomes,
	eight macroeconomic controls, and two policy indicators. The main estimations
	use the natural logarithms of CPI and real GDP per capita as key outcomes. 
	Most control variables come from the World Development Indicators (WDI), while rule of law
	comes from the Worldwide Governance Indicators. Appendix
	Table~\ref{tab:A1_variables} reports the definitions, transformations, and
	source codes for every variable.

     \subsection{Outcomes, controls, and transformations}

 Our primary outcomes of interest are the logarithm of the Consumer Price Index (log CPI) and the logarithm of real GDP per capita. We use log CPI and log real GDP per capita as the main outcomes so that a window average is a persistent gap in price and output levels. Annual inflation and growth are complements: they describe year-to-year momentum, not the accumulated wedge. First, geopolitical ruptures and energy supply shocks may induce structural step-level adjustments. While annual inflation and growth rates measure year-to-year changes, log-level gaps allow us to quantify the cumulative and persistent deviation from the estimated counterfactual path over multi-year policy windows. For instance, a temporary inflation surge in 2022 generates a permanent price-level wedge that persists long after annual inflation rates normalize. Thus, the use of log CPI outcome directly captures this lasting loss of domestic purchasing power. Second, log-level gaps yield an intuitive, direct percentage interpretation of structural divergence from counterfactual trajectories. Nevertheless, to distinguish persistent level shifts from changes in annual economic momentum, we also consider annual inflation and GDP-per-capita growth rates as complementary outcomes in our sensitivity analyses.
 
The counterfactual model includes eight controls: oil rents, trade openness,
	terms of trade, the investment rate, government consumption, rule of law,
	population growth, and the fuel-share balance. All controls enter with a
	one-year lag, so the effective estimation period is 2001--2024.\footnote{Appendix Table~\ref{tab:A1_variables} reports their definitions, transformations, and sources. The fuel-share balance is a general energy-related control.} 
	
	\subsection{Sample construction and missing-data completion}
	
	Country eligibility is determined from raw data coverage before any missing
	value is repaired and without using either policy indicator. Starting from the
	available World Bank country database universe, a country is retained in the final sample only if its raw
	outcome and control coverage satisfies prespecified thresholds aiming at maximizing both quality and coverage.\footnote{A
		country is retained if: (i) none of the four outcomes has more than 35\% raw
		missingness over 2000--2024; (ii) at most one of the eight controls has more
		than 25\% raw missingness; (iii) no control has more than 65\% raw
		missingness; and (iv) CPI and real GDP per capita each have no more than 25\%
		raw missingness during the outcome-critical 2019--2024 period. These criteria
		retain countries with isolated or locally missing data gaps while excluding
		countries whose macroeconomic histories are too incomplete to support credible
		counterfactual estimation.} All data completion procedures to generate the balanced panel are applied only after
	the eligible sample has been fixed. Treatment status, estimated effects, and
	imputed values do not determine whether a country enters the
	sample which limits concerns of endogenous sample selection and collider bias.
	
	Raw coverage of the four outcomes is high in the retained sample. Real GDP per
	capita and GDP-per-capita growth contain no raw missing observations. CPI and
	inflation contain 57 missing cells in total, corresponding to 0.44\% of the
	12,900 potential outcome cells. \footnote{Fifty-five are leading observations completed
		using bounded local-trend backcasts. The remaining two are Bosnia and
		Herzegovina's 2024 CPI and inflation observations, which are carried forward
		from their respective 2023 values.}
	
	The eight controls contain 1,471 raw missing cells, or 5.70\% of the potential
	control observations. These cells are imputed using a deterministic
	hierarchy of interpolation and bounded extrapolation rules.\footnote{Internal
		gaps bounded by valid observations are filled by linear interpolation.
		Trailing gaps use the last available observation, and leading gaps of no more than three years use the first available observation. Longer leading gaps are completed using a local linear trend estimated from the first five valid
		observations. Trend extrapolations are bounded by the country-specific
		historical range, extended by one standard deviation on either side. Trade
		openness, investment, and government consumption are additionally constrained to remain non-negative. These control-completion rules are distinct from the cell-level bounded-trend backcasts used for CPI and inflation.} Appendix
	Table~\ref{tab:A1_sample} reports the eligibility and missing data completion
	rules implemented, while Appendix Table~\ref{tab:A6_missing_variables} provides the variable-level missing-data audit. Country-level raw-missingness statistics are reported in Appendix Table ~\ref{tab:A6_countries}.
    
	\subsection{Policy variables}
\label{subsec:policy_variables}

    We code two sender-side membership rules: a strict-sanctions coalition and a narrower direct-arming coalition. Direct arming is a more demanding definition of participation, not a separate estimate of the effect of military aid.
    
	The sanctions indicator equals one from the first year in which a country
	is legally covered by, or formally aligns with, an official sanctions regime
	against Russia. The sample contains 43 sanctions-coded countries.
	However, Ukraine is kept in the descriptive panel but dropped from the sender-side treated group, because its outcomes reflect invasion and wartime destruction. This leaves 42 countries in the sample to compute the sanctions average treatment effect on the treated (ATT).
	
	The direct-arming indicator equals one from the first year in which publicly
	verifiable evidence establishes that a country delivered or committed lethal
	weapons, ammunition, or complete military systems to Ukraine. Broad non-lethal
	assistance, purely financial support, indirect transfers, leaked claims, and
	insufficiently documented commitments are not coded as direct arming. Appendix
	Table~\ref{tab:A4_treatment} reports country-level policy classifications and
	entry years. The direct-arming group contains 29 countries, 28 of which also impose strict
	sanctions.
	
	Russia, Belarus, and Ukraine are excluded from every donor pool. Russia is the
	target of the sanctions and a principal belligerent whereas Belarus is too closely
	implicated in the same geopolitical rupture to provide a credible untreated
	counterfactual. In turn, Ukraine's outcomes are directly affected by the war. These
	are geopolitical design restrictions rather than consequences of missing data.

\begin{figure}[!htbp]
		\centering
		\caption{Strict sanctions and direct-arming categories in the final sample}
		\label{fig:treatment_map}
		\includegraphics[width=\textwidth]{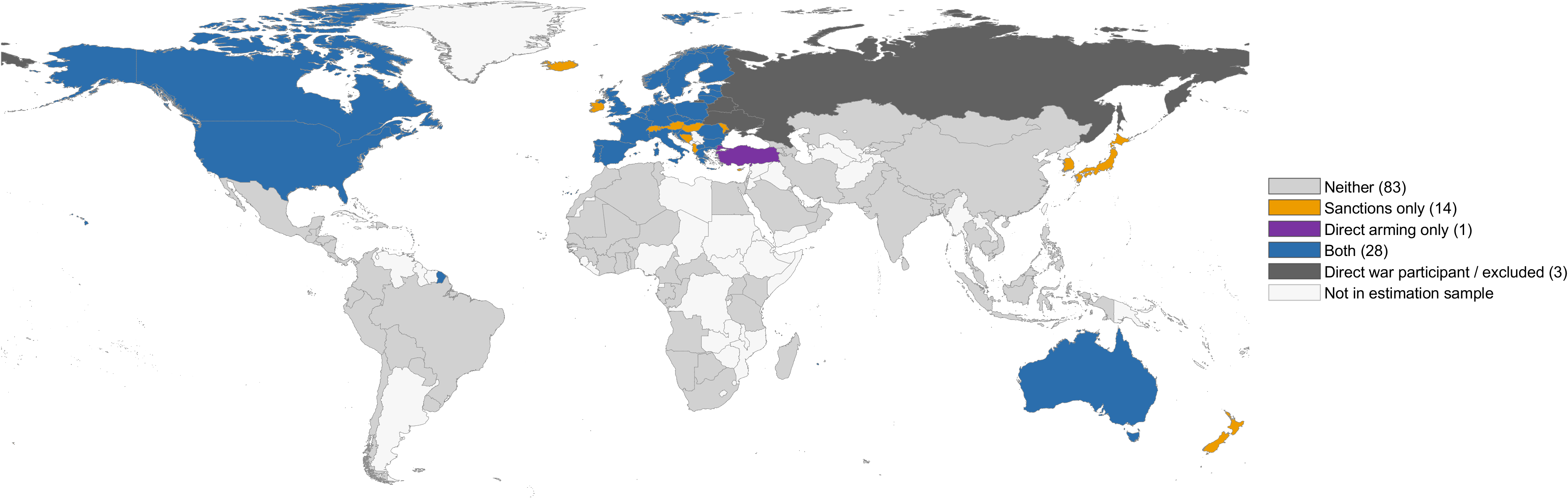}
		
		\begin{minipage}{0.96\textwidth}
			\footnotesize
			\textit{Notes:} The map displays five mutually exclusive categories: neither
			policy, sanctions only, direct arming only, both policies, and the three
			geopolitical exclusions. Russia, Belarus, and Ukraine are displayed separately
			and are excluded from all donor pools. Ukraine remains in the descriptive
			$N=129$ panel but is excluded from the main sanctions ATT.
		\end{minipage}
	\end{figure}

	The resulting descriptive categories are: 
	sanctions only (14), direct arming only (1), neither policy (83 countries), both policies (28), and the three
	geopolitical exclusions shown separately.  Table~\ref{tab:sample_design} summarizes the final design, and
	Figure~\ref{fig:treatment_map} shows its geographic composition.
    
    For the sanctions analysis, the 42 target countries consist of the 14 sanctions-only countries and the 28 countries that implement both policies. Its 84 donors are the 83 countries that implement neither policy and the one country that provides direct arms without imposing sanctions. On the other hand, the direct-arming estimations use 29 targets and 96 or 97 donors, depending on the outcome. For the direct-arming analysis, the 29 targets consist of the 28 both-policy countries and the one direct-arming-only country. The candidate donor pool then contains the 83 neither-policy countries and the 14 sanctions-only countries. All 97 candidates satisfy the log-GDP-per-capita requirements; one lacks the required log-CPI data, leaving 96 CPI donors.

	Figure~\ref{fig:treatment_map} shows that participation is concentrated in Europe, North America, and a few allied economies. Those countries are, on average, richer and institutionally stronger than the donor pool (Appendix Table~\ref{tab:A7_characteristics}). The target groups therefore differ substantially from the full donor
	pools in their pre-treatment income levels and institutional characteristics.\footnote{Appendix Table~\ref{tab:A7_characteristics} compares the pre-treatment macroeconomic and institutional characteristics (averaged over 2009--2013) of the treated countries and their respective donor pools. The descriptive statistics reveal that the sanctioning and direct-arming targets are structurally distinct from the non-sanctioning donors. Specifically, the treated groups are  wealthier (e.g., main sanctions targets average a log real GDP per capita of 10.06 versus 8.18 for their donors), possess stronger institutional frameworks (rule of law index of 1.03 versus -0.36), and have negligible reliance on oil rents (0.38\% versus 5.46\%). These structural differences weaken the case for unadjusted parallel trends. Section~\ref{sec:econometric_strategy} describes the counterfactual design used
to accommodate them.}

	\begin{table}[!htbp]
		\centering
		\caption{Final sample, treatment groups, and donor eligibility}
		\label{tab:sample_design}
		\begin{adjustbox}{max width=\textwidth}
			\begin{threeparttable}
				\begin{tabular}{lr}
					\toprule
					Item & Countries / observations \\
					\midrule
					\multicolumn{2}{l}{\textit{Panel A. Final sample}} \\
					Final balanced panel & 129 countries \\
					Years & 2000--2024 \\
					Country-years & 3,225 \\
					Effective estimation period & 2001--2024 \\
					\addlinespace
					\multicolumn{2}{l}{\textit{Panel B. Treatment groups}} \\
					Sanctions-coded countries (descriptive) & 43 \\
					Main sanctions estimation targets & 42 \\
					Direct-arming countries & 29 \\
					Geopolitical exclusions shown separately & 3 \\
					\addlinespace
					\multicolumn{2}{l}{\textit{Panel C. Donor eligibility}} \\
					Sanctions donors: Log CPI / Log GDPpc & 84 / 84 \\
					Direct-arming donors: Log CPI / Log GDPpc & 96 / 97 \\
					\bottomrule
				\end{tabular}
				\begin{tablenotes}[flushleft]
					\footnotesize
					\item \textit{Notes:} Ukraine remains sanctions-coded in the descriptive data
					but is excluded from the 42-country main sanctions target group because its
					outcomes reflect direct exposure to the war. Russia, Belarus, and Ukraine are
					excluded from all donor pools. Treatment and donor roles are defined separately
					for each policy and outcome.
				\end{tablenotes}
			\end{threeparttable}
		\end{adjustbox}
	\end{table}
	
	\section{Econometric Strategy \label{sec:econometric_strategy}}
	
	Our empirical strategy measures how the macroeconomic paths of countries sanctioning Russia differed from the counterfactual paths predicted for them in the absence of sanctions. Two features of the setting shape the econometric design. First, the 2022 escalation was not the first relevant episode: many countries had already imposed sanctions after the 2014 annexation of Crimea. Second, the years after 2022 also contained major global disturbances, including pandemic recovery, supply-chain disruption, monetary tightening, and energy-market turmoil. These shocks affected many countries, but not necessarily with the same intensity.
	
	We address these features using a two-wave Common Correlated Effects Difference-in-Differences (CCEDID) imputation estimator. Methodologically, the estimator belongs to the broader family of imputation and interactive-factor approaches to panel counterfactuals \citep{BorusyakJaravelSpiess2024,Xu2017}. Specifically, the approach adapts the framework of \citet{BrownButtsWesterlund2026} and uses a donor-based common-factor proxy inspired by \citet{Pesaran2006}. In simple terms, the estimator uses the pre-2014 data to learn how each treated country's outcome normally moves with its macroeconomic controls and with international movements observed among untreated countries. It then uses those relationships to predict the path the country would have followed after 2013 without sanctioning policy participation.\footnote{The procedure is related to generalized
difference-in-differences in the potential-outcomes and imputation sense: it
uses pre-treatment relationships and post-treatment information from
never-treated donors to impute untreated paths for policy-participating
countries and then forms observed-minus-simulated contrasts. However, it is not a
conventional $2\times2$, two-way fixed-effects, or group-time DiD estimator.}
	
	 The complete procedure is run separately for the strict-sanctions and direct-arming definitions described in Section~\ref{subsec:policy_variables}. Each policy--outcome combination uses its own target group, donor pool, and fitted parameters. Thus, to avoid carrying policy and outcome subscripts through every equation, the notation below describes one generic estimation. We restore those subscripts whenever the distinction matters.

	\subsection{Groups and Reporting Windows}
	
	Let $D_{it}$ be the annual policy indicator for country $i$ in year $t$. For a given policy definition, the ever-treated set $\mathcal T$ and the never-treated candidate set $\mathcal C$ are given by:
	\begin{equation}
		\mathcal T=\{i:\max_t D_{it}=1\},
		\qquad
		\mathcal C=\{i:\max_t D_{it}=0\}.
		\label{eq:groups}
	\end{equation}
	The final target group used for aggregation is denoted by $\mathcal T^*\subseteq\mathcal T$. The eligible donor set may differ across outcomes and policy variables considered, so $\mathcal C_Y\subseteq\mathcal C$ denotes the never-treated countries that satisfy the geopolitical and data-quality requirements for outcome $Y$.\footnote{The composition and counts of these sets are reported in Table~\ref{tab:sample_design} and Appendix Table~\ref{tab:A5_donors}.} The baseline estimand is the equally weighted average post-2013 gap among countries that ever join the coalition.
	
	We summarize the estimated gaps over two common geopolitical windows:
	\begin{equation}
		\mathcal W_1=\{2014,\ldots,2021\},
		\qquad
		\mathcal W_2=\{2022,\ldots,2024\}.
		\label{eq:wave_windows}
	\end{equation}
 Wave~1 is the 2014--2021 average gap among countries that ever join the coalition, including years before a late joiner's own entry. Wave~2 is the 2022--2024 average gap for the same ever-treated set.
	
	\subsection{Untreated-Outcome Model and Common-Factor Proxy}
	
	Let $Y_{it}(\infty)$ denote the outcome path that country $i$ would have experienced under the counterfactual scenario of non-participation in the sanctions. We model this potential outcome as:
	\begin{equation}
		Y_{it}(\infty)
		=X_{i,t-1}'\beta+\alpha_i'f_t+\varepsilon_{it},
		\label{eq:untreated_model}
	\end{equation}
\noindent where $X_{i,t-1}$ and $\beta$ are $K$-vectors of lagged macroeconomic controls and common control coefficients, respectively, $\alpha_i$ is a two-element vector of country-specific loadings on the unobserved common factors $f_t$, and $\varepsilon_{it}$ is an idiosyncratic disturbance. Note that this specification is more flexible than common year effects alone. A common year effect absorbs an international shock but imposes the same response on every country. The interactive-factor term in Equation ~\ref{eq:untreated_model} is more flexible: the common movement is shared, while the loading $\alpha_i$ allows its effect to differ across countries \citep{Pesaran2006}.

The vector $X_{i,t-1}$ contains eight one-year-lagged controls: oil rents,
trade openness, terms of trade, the investment rate, government consumption,
the rule of law, population growth, and the fuel-share balance. All eight
enter both the pre-2014 model and the post-2013 counterfactual, but their
source differs. We use observed controls to estimate the model through 2013.
After 2013, we do not condition the target's counterfactual on its realized
controls, because trade, investment, fiscal policy, or energy use may
themselves respond to the geopolitical episodes. To generate the counterfactual in Step 3 we simulate the
path those controls would have followed assuming non-participation in the sanctions policy.
	
For each outcome under analysis, we define the donor mean and construct the associated common-factor proxy vector $\widehat f_t$ as:
	\begin{equation}
		\overline Y_{t,\mathcal C_Y}
		=\frac{1}{|\mathcal C_Y|}
		\sum_{j\in\mathcal C_Y}Y_{jt},
		\qquad
		\widehat f_t
		=
		\begin{pmatrix}
			1\\[2pt]
			\overline Y_{t,\mathcal C_Y}
		\end{pmatrix}.
		\label{eq:factor_proxy}
	\end{equation}
Instead of estimating latent common factors directly, we use $\widehat f_t$ as a donor-based proxy for the common international component of the outcome. The proxy $\widehat f_t=(1,\overline{Y}_{t,C_Y})'$  contains a constant and the cross-sectional average (CSA) among eligible donors at time $t$ . The donor CSA $\overline Y_{t,\mathcal C_Y}$ summarizes the international component of the outcome in each year. For CPI, it captures common inflationary movements whereas for GDP per capita, it captures common output movements. Writing $\alpha_i=(\mu_i,\lambda_i)'$ makes the role of the two factor loadings explicit: $\alpha_i'\widehat f_t=\mu_i+\lambda_i \overline Y_{t,\mathcal C_Y}$. Here $\mu_i$ is a country-specific intercept and plays the role of a country fixed effect (i.e, it absorbs time-invariant unobserved heterogeneity across countries). In turn, the coefficient $\lambda_i$ allows each country to respond differently to common shocks summarized by the donor mean. Specifically, the loading $\lambda_i$ is estimated from country $i$'s own pre-2014
history. The donor pool supplies the common time path whereas the pre-treatment data
determine how strongly each country loads on that path.
 
Our implementation is deliberately parsimonious when compared to \citet{BrownButtsWesterlund2026} . General CCE specifications can include the CSA of several outcomes and regressors, whereas the preferred specification here uses only the CSA of the outcome being modeled. This choice excludes the target coalition's realized post-2013 outcomes from its own factor proxy to reduce endogeneity concerns, and avoids adding several potentially collinear proxies to a pre-treatment period containing only 13 effective years.\footnote{Unlike the fuller CCE proxy used by \citet{BrownButtsWesterlund2026}, which relies on cross-sectional averages of outcomes and covariates, our preferred specification retains only the never-treated donor CSA of the outcome. A richer CCE proxy would fit the 13 pre-treatment years more closely and would be less stable when extrapolated after 2013.}

 	\subsection{Estimation Procedure}
	
	The estimator can be understood in five steps.\footnote{Further details are provided in Appendix~B, Section~S.B.1.}
	
	\paragraph{Step 1: Construct the donor factor.}
	For each policy and outcome, we calculate the donor-based factor proxy $\widehat f_t$. To isolate common unobserved global shocks without contaminating them with treatment effects, the CSA is calculated using only the never-treated donor pool ($\mathcal{C}_Y$) for every year $t$.
	
	\paragraph{Step 2: Estimate the pre-treatment relationship.}
	Using strictly the effective pre-treatment period ($t \leq 2013$), an Ordinary Least Squares (OLS) regression models the relationship between observed outcomes, controls, and the estimated factors:
	\begin{equation}
		Y_{it}
		=X_{i,t-1}'\beta+\widehat\alpha_i'\widehat f_t+u_{it},
		\qquad t \leq 2013.
		\label{eq:pretreatment_ols}
	\end{equation}
To estimate the common control coefficient $\beta$, let $Y_i$ and $X_i$ stack the pre-treatment observations for country $i$, and let $\widehat f$ stack the pre-treatment factor proxies. We project out the donor factor using the residual-maker matrix $M_{\widehat f} = I - \widehat f(\widehat f'\widehat f)^{-1}\widehat f'$.\footnote{The matrix $M_{\widehat f}$ residualizes both $Y_i$ and $X_i$ with respect to
the two columns of the factor matrix: the constant and the donor-outcome CSA.} Let $\mathcal S_Y$ denote the set of target and donor countries. The pooled estimator for the covariate coefficients ($\widehat\beta$) is:\footnote{Equation \ref{eq:beta_hat} estimates the common control coefficient $\beta$ from the
pooled pre-2014 association between these factor-adjusted outcomes and
controls. This isolates variation not linearly explained by the chosen donor
proxy.}

	\begin{equation}
		\widehat\beta
		=
		\left(
		\sum_{i\in\mathcal S_Y}
		X_i'M_{\widehat f}X_i
		\right)^{-1}
		\sum_{i\in\mathcal S_Y}
		X_i'M_{\widehat f}Y_i.
		\label{eq:beta_hat}
	\end{equation}
	The unit-specific factor loadings ($\widehat\alpha_i$) are then estimated from the pre-treatment data for every unit $i$:
	\begin{equation}
		\widehat\alpha_i
		=(\widehat f'\widehat f)^{-1}\widehat f'(Y_i-X_i\widehat\beta).
		\label{eq:alpha_hat}
	\end{equation}
	
\paragraph{Step 3: Construct the post-treatment counterfactual.}
As explained above, target countries' observed post-treatment covariates cannot be used directly because they may themselves have been affected by treatment. We therefore construct, for each target country, the covariate path that would be expected under no treatment. To do so, we first estimate how its eight covariates co-moved with the donor factor during the pre-2014 period:
\begin{equation}
	\widehat\Lambda_i
	=
	(\widehat f'\widehat f)^{-1}\widehat f'X_i.
	\label{eq:lambda_hat}
\end{equation}
Because $\widehat f$ contains a constant and the donor-outcome mean,
$\widehat\Lambda_i$ contains two loadings for each covariate. These loadings
summarize the relationship between country $i$'s covariates and the common
movements observed among never-treated donors before treatment.

For each year after 2013, we then apply these pre-treatment relationships to the contemporaneous donor factor:
\begin{equation}
	\widehat X_{i,t-1}(\infty)
	=
	\widehat\Lambda_i'\widehat f_t,
	\qquad t\geq 2014.
	\label{eq:imputed_x}
\end{equation}
The resulting vector $\widehat X_{i,t-1}(\infty)$ represents the covariate values predicted for target country $i$ under the counterfactual no-treatment scenario.

Finally, we combine this counterfactual covariate path with the outcome relationships estimated in Step~2:
\begin{equation}
	\widehat Y_{it}(\infty)
	=
	\widehat X_{i,t-1}(\infty)'\widehat\beta
	+
	\widehat\alpha_i'\widehat f_t,
	\qquad t\geq 2014.
	\label{eq:imputed_y}
\end{equation}
The first term captures the outcome predicted by the country's counterfactual covariates, while the second captures its pre-treatment sensitivity to common movements in the donor pool. Thus, the country-specific relationships are learned before 2014, and their post-2013 evolution is driven by information from never-treated donors. This ensures no post-treatment outcome or covariate observed for a target country enters the construction of its own counterfactual path.

	\paragraph{Step 4: Calculate country-year gaps.}
The estimated gap is the actual observed outcome minus the imputed counterfactual:
	\begin{equation}
		\widehat g_{it}
		=Y_{it}-\widehat Y_{it}(\infty).
		\label{eq:gap}
	\end{equation}
A positive value means that the observed outcome lies above its imputed untreated path; a negative value means that it lies below it.
	
	\paragraph{Step 5: Aggregate the gaps.}
For a reporting window $\mathcal W\in\{\mathcal W_1,\mathcal W_2\}$, the coalition average is:
	\begin{equation}
		\widehat{\mathrm{ATT}}_{\mathcal W}
		=\frac{1}{N_T}
		\sum_{i\in\mathcal T^*}
		\left(
		\frac{1}{|\mathcal W|}
		\sum_{t\in\mathcal W}\widehat g_{it}
		\right),
		\qquad
		N_T=|\mathcal T^*|.
		\label{eq:wave_att}
	\end{equation}
    
In this application, the average-treatment-effect-on-the-treated (ATT) denotes the equally weighted average counterfactual gap among the countries in the relevant estimation target group, as formally defined in equation~\eqref{eq:wave_att} (i.e, countries receive equal weight, regardless of population or economic size.)

The Wave~2 average counterfactual gap is a total post-2022 gap relative
to non-participation. It may therefore include deviations that had already
emerged during the 2014--2021 period. To characterize the additional
post-2022 departure from the coalition's pre-escalation position, we construct
a stable-gap continuation counterfactual.

For each target country, define its late-Wave-1 baseline gap as:
\begin{equation}
    \widehat b_i
    =
    \frac{1}{3}
    \sum_{t=2019}^{2021}
    \widehat g_{it}.
    \label{eq:baseline_gap}
\end{equation}
This quantity measures country $i$'s average position relative to its
imputed non-participation path immediately before the 2022 escalation.
We then define the post-2022 continuation counterfactual as:
\begin{equation}
    \widehat Y^{\mathrm{cont}}_{it}
    =
    \widehat Y_{it}(\infty)
    +
    \widehat b_i,
    \qquad
    t\in\mathcal W_2.
    \label{eq:continuation_counterfactual}
\end{equation}
This counterfactual preserves the country's average 2019--2021 gap while
allowing its underlying non-participation counterfactual path to continue evolving with the
post-2021 donor factor. It therefore permits the earlier sanctions regime and
any persistent country-specific prediction error to have left a non-zero
pre-2022 gap.

The preferred incremental post-2022 contrast can equivalently be written as
\begin{equation}
\begin{aligned}
    \widehat{\mathrm{ATT}}_{\mathrm{Inc},2}
    &=
    \frac{1}{N_T}
    \sum_{i\in\mathcal T^*}
    \left[
        \frac{1}{3}
        \sum_{t=2022}^{2024}
        \left(
            Y_{it}
            -
            \widehat Y^{\mathrm{cont}}_{it}
        \right)
    \right]                                                   \\[3pt]
    &=
    \frac{1}{N_T}
    \sum_{i\in\mathcal T^*}
    \left[
        \frac{1}{3}
        \sum_{t=2022}^{2024}
        \widehat g_{it}
        -
        \frac{1}{3}
        \sum_{t=2019}^{2021}
        \widehat g_{it}
    \right].
\end{aligned}
\label{eq:incremental}
\end{equation}

The second equality shows that this continuation-counterfactual representation
is identical to subtracting each country's 2019--2021 mean gap
from its 2022--2024 mean gap.  Interpreting this result hinges on a "stable-gap" assumption: without the post-2022 escalation, the coalition’s gap relative to the non-participation baseline would have held steady at its 2019–2021 level. The assumption allows the earlier regime to have generated
a persistent gap, but rules out further post-2021 drift in that inherited
baseline or in the average counterfactual prediction error.

We therefore interpret
$\widehat{\mathrm{ATT}}_{\mathrm{Inc},2}$ as an incremental post-2022 gap
contrast under a stable-gap continuation assumption. For any
estimate $\widehat\theta$ reported in log points, the exact percentage
interpretation is $100[\exp(\widehat\theta)-1]$, while
$100\widehat\theta$ is the usual small-effect approximation.

	\subsection{Interpretation, Identification, and Inference}

The design does not require target and donor countries to follow parallel
trends in their unadjusted outcomes. A causal interpretation instead requires
(i) the donor factor to capture the common shocks relevant to coalition members,
(ii) the relationships estimated before 2014 to remain informative after 2013, and (iii)
the timing, anticipation, and spillover conditions detailed in Appendix~B to
hold.

These conditions are plausible but not guaranteed in this context. The estimates may also pick up coalition-specific shocks that the donor factor misses: war-proximity uncertainty, targeted supply cuts, or a post-2022 change in how coalition members load on global inflation and output. When these maintained conditions fail, the estimates are
more safely interpreted as realized macroeconomic divergences associated with
coalition membership rather than as independent causal effects of the legal
sanctions.

Statistical inference begins with one complete contrast per target country. If $d_i$ denotes either the country-specific window average $\overline g_{i,\mathcal W}$ or the country-specific incremental contrast $d_{i,\mathrm{Inc},2}$, the country-dispersion standard error is:
	\begin{equation}
		\widehat{SE}_{CD}=\frac{sd(d_i)}{\sqrt{N_T}}.
		\label{eq:country_dispersion_se}
	\end{equation}
The country-dispersion standard error is the sample standard deviation of the
$N_T$ complete country contrasts divided by $\sqrt{N_T}$. It produces the
significance stars in the main tables and measures uncertainty about the
equally weighted coalition mean under an independent-country approximation.
Because each country's full time path first becomes one contrast, annual
observations are not incorrectly treated as independent. The calculation
nevertheless treats country contrasts as independent and does not propagate
all first-stage uncertainty from estimating $\widehat\beta$, the loadings,
and the donor factor.
	
For these reasons, we report three complementary inferential checks on the uncertainty of the estimated ATT and the incremental ATT.  First, we implement a country-history bootstrap that resamples whole country paths and re-estimates the complete design, thereby preserving within-country time dependence and propagating model-estimation uncertainty.     
 Secondly, we apply a region-block bootstrap that resamples World Bank regions and keeps countries from the same region together. Both procedures follow the logic of cluster-level resampling \citep{CameronMiller2015}. Third, we adapt the placebo comparisons used in comparative case studies (\citealp{AbadieDiamondHainmueller2010,Abadie2021}). To that end, we randomly create size-matched false coalitions to investigate how often the same method produces an equally extreme contrast for a false coalition. Because coalition membership was not randomly assigned, this placebo statistic is only a diagnostic tail probability. It is not a confidence interval, and it is not the source of the significance stars.\footnote{Further details on the algorithms used to implement these alternative inferential procedures are provided in Appendix B, section B.3.}

	\section{Results \label{sec:results}}
	
	\subsection{Baseline results}
	
	\subsubsection{Dynamic Evidence for Sanctioning Countries}
	
\begin{figure}[!htbp]
		\centering
		\caption{Dynamic CCEDID gaps for sanctioning countries}
		\includegraphics[width=0.86\textwidth]{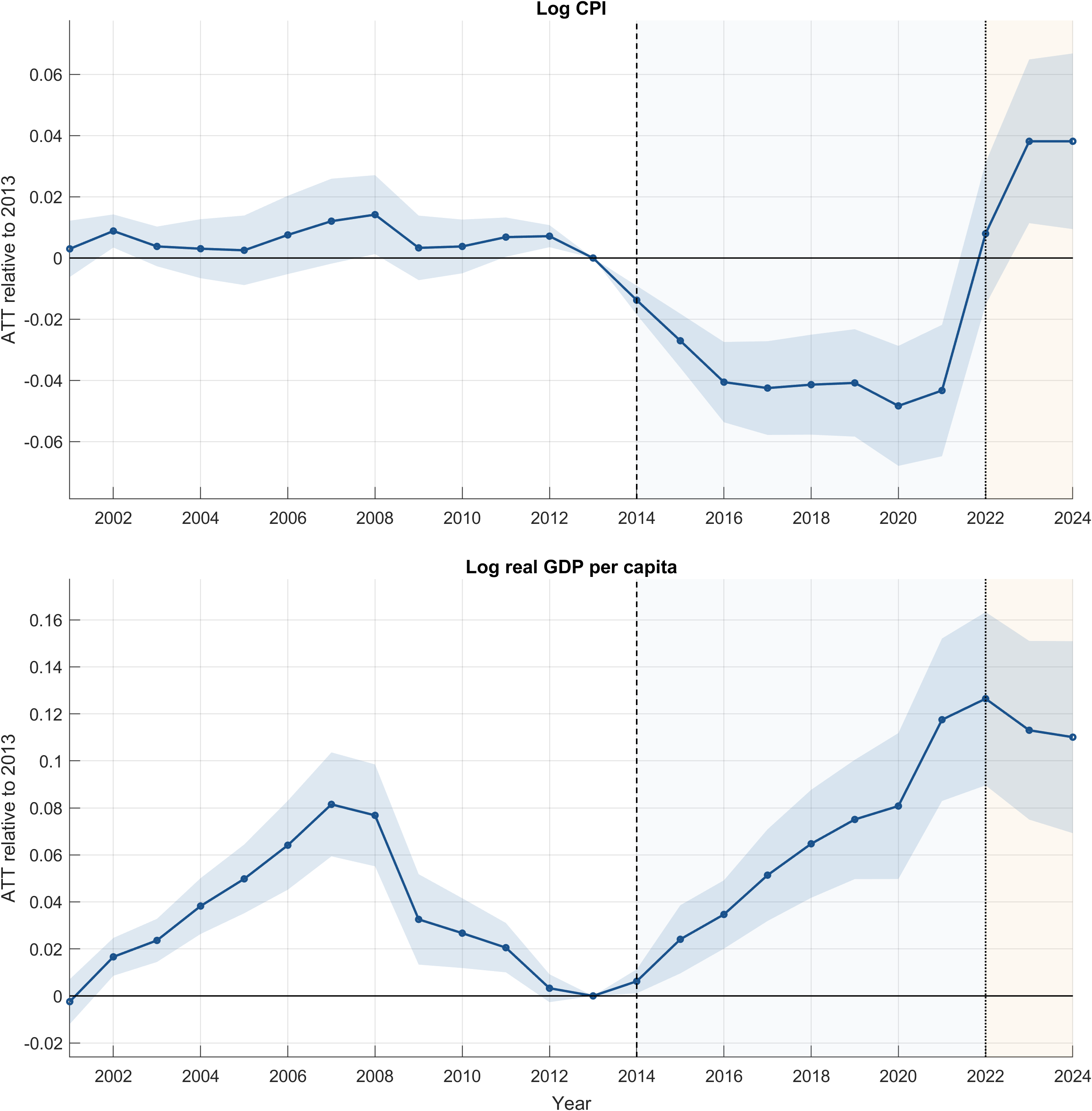}
		\label{fig:event_studies_sanctions}
		\begin{minipage}{0.96\textwidth}\footnotesize
			\textit{Notes:} Yearly averages of imputed gaps for the 42-country main sanctions target group, normalized to zero in 2013. Ribbons are pointwise 95\% intervals based on cross-country dispersion. Shading identifies Wave~1 (2014--2021) and Wave~2 (2022--2024); vertical lines mark 2014 and 2022.  Values before 2013 are normalized fitted gaps, not conventional treatment leads.
		\end{minipage}
	\end{figure}

	The first set of results examines the timing of the 42 sanctioning countries' deviations from their counterfactual paths. Figure~\ref{fig:event_studies_sanctions} plots yearly CCEDID gaps in log CPI and log real GDP per capita, set to zero in 2013 only so the two panels share a visual origin. That normalization is not a pre-trend test: values before 2013 are fitted gaps, not conventional leads. The event-study graphical evidence is central for interpretation because the wave averages in Table~\ref{tab:main_sanctions} are log-level gaps, not annual inflation or growth effects. A one-off movement can create a level gap that persists for several years even after annual rates return to normal. A positive Wave 2 average can therefore reflect a new post-2022 break, a gap inherited from Wave 1, or both; the dynamic paths and the rate-outcome checks distinguish these possibilities.

	The dynamic patterns in Figure~\ref{fig:event_studies_sanctions} place the main divergence in CPI rather than output. CPI gaps are negative or close to zero during most of Wave~1 and rise visibly after the 2022 escalation. GDP per capita remains above its imputed path in Wave~2, but the dynamic evidence does not show a large new post-2022 acceleration. The output estimates do not show a post-2022 fall in real GDP per capita. The positive Wave-2 level gap is mostly already there by 2019--2021; it should not be read as evidence that sanctions raised output.
	
	Do not read a 2022--2024 height in Figure~\ref{fig:event_studies_sanctions} as the Wave~2 ATT in Table~\ref{tab:main_sanctions}: the figure subtracts the 2013 gap, whereas the table averages unnormalized log-level gaps.

	\subsubsection{Main Wave Estimates and Incremental Wave 2 Effects}
	
    Table~\ref{tab:main_sanctions} reports the main CCEDID estimates for
sanctioning countries. The first two columns contain the average log-level
gaps for Wave~1 and Wave~2, while the third reports the incremental Wave~2
contrast $    \widehat{\mathrm{ATT}}_{\mathrm{Inc},2}$ defined in equation~\eqref{eq:incremental}. This incremental estimand is crucial because the Wave~2 ATT is a total level gap, not a marginal 2022 shock. Hence, the incremental contrast asks whether the 2022--2024 mean log-level gap rose or fell relative to each country's own 2019--2021 mean gap, not whether the Wave-2 ATT itself is large.

	\begin{table}[!htbp]
		\centering
		\caption{Main sanctions CCEDID estimates}
		\label{tab:main_sanctions}
		\small
		\setlength{\tabcolsep}{4pt}
		\begin{adjustbox}{max width=\textwidth}
			\begin{threeparttable}
				\begin{tabular}{lcccrr}
					\toprule
					Outcome & Wave 1 & Wave 2 & \makecell{Inc. W2 vs.\\2019--2021} & Treated & Donors \\
					\midrule
					Log CPI & \makecell{-0.043***\\(0.009)} & \makecell{0.022\\(0.015)} & \makecell{0.072***\\(0.007)} & 42 & 84 \\
					\addlinespace
					\makecell[l]{Log real GDP\\per capita} & \makecell{0.024**\\(0.011)} & \makecell{0.083***\\(0.019)} & \makecell{0.025***\\(0.008)} & 42 & 84 \\
					\bottomrule
				\end{tabular}
				\begin{tablenotes}[flushleft]
					\footnotesize
					\item \textit{Notes:} Wave 1 is 2014--2021 and Wave 2 is 2022--2024. The incremental contrast subtracts each country's 2019--2021 mean gap. The country-dispersion SE (in parentheses) is the sample SD of the complete country contrast divided by the square root of the treated-country count. Two-sided p-values use a Student t-distribution with $N-1$ degrees of freedom. *, **, and *** denote $p<0.10$, $p<0.05$, and $p<0.01$, respectively. Ukraine is excluded from the main ATT.
				\end{tablenotes}
			\end{threeparttable}
		\end{adjustbox}
	\end{table}
	
	The price-level results provide evidence of a macroeconomic boomerang effect, but its interpretation depends on when the gap emerged.    During the initial 2014--2021 period, the Log-CPI Wave~1 ATT is -0.043 (SE 0.009), meaning prices were approximately 4.2\% below the imputed untreated path. During the 2022--2024 escalation, the Wave~2 ATT swings to 0.022 (SE 0.015), placing prices roughly 2.2\% above that counterfactual path but this estimate is not statistically significant. However, the reversal becomes starker when evaluating the preferred incremental gap. By subtracting each country's late-Wave-1 baseline (2019--2021) from its Wave~2 mean (2022--2024), we find a highly statistically significant incremental post-2022 price effect of 0.072 log points (approximately 7.5\%; SE 0.007). The fact that this 7.5\% incremental shift is larger than the 2.2\% Wave~2 level gap is not a contradiction: because the sanctioning coalition entered 2022 from a negative CPI position, the incremental effect captures the sharp upward trajectory from that starting point, whereas the Wave~2 ATT simply measures the final standing relative to the zero-deviation baseline.  Taken together, the evidence therefore points to a sharp upward shift in price levels after the 2022 escalation, rather than to a large positive average gap over the whole post-2022 window.

	For real GDP per capita, the Wave~1 and Wave~2 level gaps are 0.024 and 0.083 log points (SEs 0.011 and 0.019), respectively. Much of the positive Wave~2 level gap was already present near the end of Wave~1. Once that inherited 2019--2021 position is removed, the incremental Wave-2 log-GDP-per-capita contrast is still positive and statistically significant, but much smaller than the Wave-2 ATT: 0.025 log points (about 2.5 percent; SE 0.008). 
	
	Table~\ref{tab:main_sanctions} reports model-imputed counterfactual gaps. Their causal interpretation depends on whether the pre-2014 relationship between each country's outcome, controls, and the donor factor would have remained informative without coalition participation. Section~\ref{sec:robustness} examines the empirical credibility of this assumption.

	\subsubsection{Regional Heterogeneity in Incremental Wave 2 Effects}
	
\begin{figure}[!htbp]
		\centering
		\caption{European heterogeneity in incremental Wave~2 effects}
		\includegraphics[height=0.76\textheight,keepaspectratio]{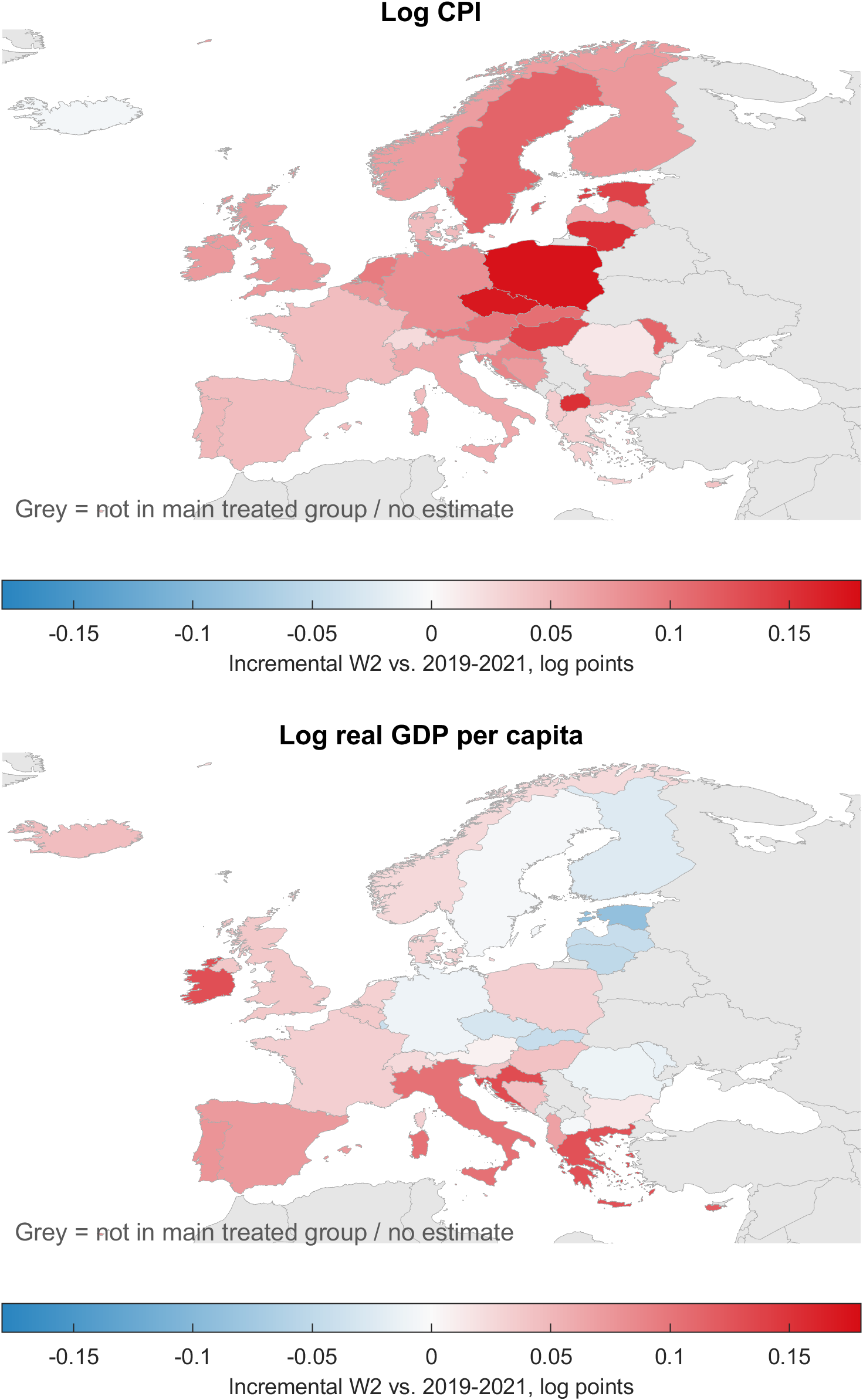}
		\label{fig:map_incremental_w2_europe}
		\begin{minipage}{0.96\textwidth}\footnotesize
			\textit{Notes:} Country-specific 2022--2024 gaps minus 2019--2021 gaps for the 42-country main sanctions target group. Grey denotes a country outside that group or without a main effect estimate. Ukraine has no main effect estimate and remains grey. The color scale extends to $\pm0.18$ log points so CPI estimates around 0.17--0.18 are not silently capped.
		\end{minipage}
	\end{figure}

	The average effects in Table ~\ref{tab:main_sanctions} summarize the mean response of the sanctioning coalition, but they do not show whether the post-2022 deviations are geographically concentrated or heterogeneous across treated countries. 
	To address this question, Figure~\ref{fig:map_incremental_w2_europe} maps country-specific incremental effects for the same 2019--2021 reference, focusing on Europe where the sanctions coalition and cross-country variation are concentrated.
	
	The CPI contrast is positive in 41 of the 42 target countries, so the coalition average is not driven by a few extreme observations. Magnitudes nevertheless vary considerably across countries revealing a distinct geographic gradient. The largest price-level deviations (the darkest red areas) are heavily concentrated in Central and Eastern Europe, whereas Western European nations experienced more moderate, albeit still positive, price increases. The map in Figure~\ref{fig:map_incremental_w2_europe} describes this geographic variation whereas the corresponding worldwide map is shown in Appendix Figure~\ref{fig:B1_world}.
	
	GDP-per-capita gaps are less uniform: 66.7\% are positive, compared with 97.6\% for CPI. A positive mean therefore summarizes a less uniform country distribution. The negative incremental GDP effects (depicted in blue) are distinctly clustered in Northern Europe, the Baltics, and parts of Central Europe (such as Germany), which are the regions historically most exposed to Russian energy supply chains and geopolitical proximity. Conversely, Southern Europe and certain outlier economies display strong positive GDP deviations. 
	
	\section{Validation and Robustness Checks \label{sec:robustness}}
	
	The validation exercises ask whether the main conclusions survive plausible
	changes in timing, outcome definition, reference period, counterfactual
	construction, comparability, and sample composition. Their
	purpose is to show which findings are stable and which depend on particular empirical 
	design choices.
	
	Unless stated otherwise, the sanctions checks use 42 target countries, 84
	donors, Wave~2 defined as 2022--2024, and the 2019--2021 reference. Ukraine is
	excluded throughout. Stars associated to statistical significance come from the country-dispersion test. Bootstrap intervals and placebo probabilities are separate inferential summaries.
	
	\subsection{Treatment Timing, Horizon, and Outcome Definition}

    This first group of checks investigates whether how the post-2022 effect is measured matters for the conclusions. It addresses three concerns: (i) some countries may have entered the sanctions coalition later than others, (ii) the result may be driven by the final year of the sample and (iii) a result expressed in price levels may not also appear in annual inflation.
    
	The baseline uses an ever-treated coalition: a country belongs to the
	sanctions group if it participates at any point in the sample. This definition
	captures coalition membership, but it allows late entrants to contribute
	Wave~2 observations from before their own sanctions indicator turns on.
	\footnote{Appendix Table~\ref{tab:A4_treatment} reports the country-specific entry
		years.}
	
	The active-year check presented instead averages only years in which each country's strict indicator is active. Table~\ref{tab:timing_outcome} shows that the CPI and GDP-per-capita gaps remain 0.072 and 0.025. The baseline findings are therefore not produced by counting pre-entry years as active-policy years. Excluding 2024 leaves estimates of 0.067 for log CPI and 0.029 for log real GDP per capita, so the patterns are already present in 2022--2023.  
    
    The rate outcomes qualify the level estimates. Although the incremental
log-GDP-per-capita gap is positive ($0.025$), annual GDP-per-capita growth
is 3.9 percentage points lower after 2022, while inflation is 2.5 percentage
points higher. The positive GDP-per-capita level gap therefore reflects an
accumulated position relative to the counterfactual, not faster post-2022
growth.\footnote{Appendix Table~\ref{tab:B7_secondary} provides two complementary timing and
		target-boundary checks: active-year timing for the direct-arming coalition and
		sensitivity to including Ukraine in the sanctions target set. Including
		Ukraine changes the preferred CPI contrast only from 0.072 to 0.071.}
	
	\begin{table}[!htbp]
		\centering
		\caption{Sanctions: robustness of the incremental Wave~2 gap}
		\label{tab:timing_outcome}
		\begin{threeparttable}
			\small
			\begin{tabular*}{\textwidth}
				{@{\extracolsep{\fill}} l l c r r @{}}
				\toprule
				Outcome
				& Robustness check
				& \makecell{Incremental Wave~2\\gap (SE)}
				& Treated
				& Donors \\
				\midrule
				\multicolumn{5}{l}{\textit{Price outcomes}} \\
				\addlinespace[2pt]
				Log CPI
				& Active sanctions years only
				& 0.072*** (0.007)
				& 42 & 84 \\
				Log CPI
				& Exclude 2024
				& 0.067*** (0.006)
				& 42 & 84 \\
				Inflation rate
				& Rate outcome
				& 0.025*** (0.002)
				& 42 & 84 \\
				\addlinespace[5pt]
				\multicolumn{5}{l}{\textit{Output outcomes}} \\
				\addlinespace[2pt]
				Log real GDP per capita
				& Active sanctions years only
				& 0.025*** (0.008)
				& 42 & 84 \\
				Log real GDP per capita
				& Exclude 2024
				& 0.029*** (0.007)
				& 42 & 84 \\
				GDP-per-capita growth
				& Rate outcome
				& -0.039*** (0.006)
				& 42 & 84 \\
				\bottomrule
			\end{tabular*}
			\begin{tablenotes}[flushleft]
				\footnotesize
				\item \textit{Notes:} The active-sanctions-years specification averages only years in 2022--2024 for which a country's strict-sanctions indicator is active and subtracts the 2019--2021 reference. The exclude-2024 specification compares 2022--2023 with 2019--2021. Rate outcomes compare 2022--2024 with 2019--2021and are expressed as decimals.
  Standard errors are country-dispersion standard
				errors; *, **, and *** denote $p<0.10$, $p<0.05$, and $p<0.01$.
			\end{tablenotes}
		\end{threeparttable}
	\end{table}

	\subsection{Reference Windows, Inference, and Counterfactual Construction}

	Our preferred incremental estimate subtracts the average 2019--2021 gap from
	the 2022--2024 gap. The three-year 2019--2021 reference is close to the escalation and
	does not rest on a single observation. However,  this is not the only possible benchmark. Table~\ref{tab:reference_inference} holds Wave~2 fixed
	and varies only the benchmark. This checks whether the result represents a
	post-2022 shift rather than an unusual starting point.
	
	For CPI, the estimate remains between 0.070 and 0.072 across all five
	references, and every country-history and region-block bootstrap interval stays above
	zero. In turn, the preferred size-matched placebo probability is 0.022. These measures
	answer complementary questions. The country-dispersion test evaluates whether
	the treated-country contrasts are centered away from zero, the placebo asks
	how often an equally large false coalition produces a contrast as extreme as
	the observed one\footnote{In the placebo exercise, we take countries that never imposed sanctions and pretend that some of them were treated. We then apply exactly the same method. This is repeated 500 times. Since these countries were not actually sanctioning countries, the estimated effects should usually be close to zero. The placebo distribution provides a direct check on whether the observed price effect is unusually large relative to false-treatment assignments.}, and the bootstraps re-estimate the complete design after
	resampling country histories or regions.
	
	GDP per capita is more reference-dependent. Its estimate ranges from
	$-0.001$ with 2021 alone to 0.053 with 2017--2019. The preferred bootstrap
	intervals are positive, but the placebo probability is 0.102. With the
	single-year reference, both intervals include zero and the placebo probability
	reaches 0.944, so the output result lacks the reference stability of CPI.\footnote{Appendix Table~\ref{tab:B6_windows} extends the reference-window and inference
		comparison to all four policy--outcome combinations, and Appendix
		Figure~\ref{fig:B3_reference_windows} displays the sanctions estimates across
		reference windows. The randomized reassignment audit and the corresponding
		placebo distributions are reported in Appendix Table~\ref{tab:B1_placebo} and
		Figure~\ref{fig:B2_placebos}, respectively.}

	\begin{table}[!htbp]
		\centering
		\caption{Reference-window and inference summary}
		\label{tab:reference_inference}
		\begin{adjustbox}{max width=\textwidth}
			\begin{threeparttable}
				\small
				\begin{tabular}{l c c c c c}
					\toprule
					Reference & ATT (SE) & Placebo $p$ & Country-history 95\% CI & Region-block 95\% CI & Draws \\
					\midrule
					\multicolumn{6}{l}{\textbf{Panel A. Log CPI}} \\
					2019--2021 & 0.072*** (0.007) & 0.022 & [0.054, 0.093] & [0.032, 0.082] & 500 \\
					2017--2019 & 0.070*** (0.008) & 0.106 & [0.045, 0.095] & [0.024, 0.086] & 500 \\
					2021 & 0.071*** (0.006) & 0.014 & [0.056, 0.090] & [0.035, 0.079] & 500 \\
					2017--2021 & 0.071*** (0.007) & 0.052 & [0.050, 0.095] & [0.029, 0.083] & 500 \\
					2018--2021 & 0.072*** (0.007) & 0.042 & [0.052, 0.093] & [0.030, 0.082] & 500 \\
					\addlinespace
					\multicolumn{6}{l}{\textbf{Panel B. Log real GDP per capita}} \\
					2019--2021 & 0.025*** (0.008) & 0.102 & [0.003, 0.043] & [0.011, 0.048] & 500 \\
					2017--2019 & 0.053*** (0.009) & 0.028 & [0.027, 0.080] & [0.034, 0.066] & 500 \\
					2021 & -0.001 (0.007) & 0.944 & [-0.020, 0.015] & [-0.017, 0.012] & 500 \\
					2017--2021 & 0.039*** (0.008) & 0.050 & [0.016, 0.060] & [0.021, 0.055] & 500 \\
					2018--2021 & 0.032*** (0.008) & 0.070 & [0.011, 0.050] & [0.017, 0.050] & 500 \\
					\bottomrule
				\end{tabular}
				\begin{tablenotes}[flushleft]
					\footnotesize
					\item \textit{Notes:} Wave~2 is 2022--2024 in every row. Placebo probabilities
					use 500 size-matched assignments with the plus-one correction. The two
					percentile intervals use 500 valid complete-design re-estimations. Bootstrap
					and placebo results do not determine the significance stars. Both outcomes
					use 42 target countries and 84 donors.
				\end{tablenotes}
			\end{threeparttable}
		\end{adjustbox}
	\end{table}
	
	The baseline uses all countries that never imposed sanctions. A possible concern is that this group includes economies that are very different from the sanctioning coalition. We therefore repeat the analysis after excluding groups that may be poor comparisons (i.e, major oil exporters, countries with unusual fuel-trade patterns, countries with unstable pre-2022 outcomes, and countries far below the income range of the sanctioning group). 

    Another concern is that the model must separate sanctions-related changes from events that affected many countries at the same time. In the baseline, shared comovements are summarized using the CSA of the outcome among the comparison countries.
 To test the robustness of the results to alternative factors specifications we consider (i) the baseline own-outcome average, (ii) the other-outcome average, using the average log GDP per capita in the price model and the average log price level in the GDP model, (iii) both outcome averages together, (iv) the own-outcome average plus average oil rents; and (v) the own-outcome average plus the average fuel-composition gap.

	Table~\ref{tab:robustness_summary} shows the results of varying these ingredients used to build
	the untreated counterfactual path.  Importantly, CPI remains positive across donor
	pools ([0.068, 0.106]) and all five factor proxies specifications considered ([0.072, 0.103]). 
    For GDP per capita, donor-pool changes barely move the estimate
([0.025, 0.026]), whereas alternative factor proxies produce a much wider range ([0.007, 0.099]). Part of the upper-end variation comes from richer but poorly conditioned factor matrices. The low $0.007$ estimate, however, uses the other-outcome proxy and is not explained by poor numerical conditioning. The GDP-per-capita result is therefore sensitive both to the substantive choice of proxy and, for some richer specifications, to numerical instability. \footnote{Appendix Table~\ref{tab:B3_counterfactual} reports every donor-pool and factor-proxy specification for sanctions and direct arming, together with standard errors, factor-matrix rank, and condition numbers.}

	\begin{table}[!htbp]
		\centering
		\caption{Placebo, donor-pool, and factor-proxy robustness}
		\label{tab:robustness_summary}
		\begin{adjustbox}{max width=\textwidth}
			\begin{threeparttable}
				\small
				\begin{tabular}{l c c c c c c}
					\toprule
					Outcome & Core ATT & Placebo $p$ & Donor-pool & Factor-proxy & Pool & Factor \\
					& (SE) & & range & range & specs. & specs. \\
					\midrule
					Log CPI & 0.072*** (0.007) & 0.022 & [0.068, 0.106] & [0.072, 0.103] & 3 & 5 \\
					Log real GDP per capita & 0.025*** (0.008) & 0.102 & [0.025, 0.026] & [0.007, 0.099] & 3 & 5 \\
					\bottomrule
				\end{tabular}
				\begin{tablenotes}[flushleft]
					\footnotesize
					\item \textit{Notes:} All entries use sanctions and the 2019--2021 reference.
					Ranges summarize separately estimated donor-pool and factor-proxy
					specifications. Placebo probabilities are empirical and do not determine the
					stars. The core sample contains 42 target countries and 84 donors.
				\end{tablenotes}
			\end{threeparttable}
		\end{adjustbox}
	\end{table}
	
	\subsection{Donor Comparability and Country Influence}
	The previous exercise removes some potentially unsuitable comparison countries. This section goes further and builds comparison groups that are deliberately closer to the sanctioning countries using two matching methods.
Sanctioning countries are generally richer and more institutionally developed than the average country in the full sample. They are also concentrated in Europe and North America. If very different countries provide a poor guide to what would have happened without sanctions, the estimated price gap could be a comparison-group artefact.

To address this concern we propose two designs. The high-income design restricts both groups at the median 2009--2013
	GDP-per-capita level whereas the Mahalanobis design selects the five nearest eligible
	donors for each treated country using pre-2014 characteristics and weights
	them by selection frequency.
	
\begin{figure}[!htbp]
		\centering
		\caption{Event-study paths under baseline, high-income, and Mahalanobis donor designs}
		\includegraphics[width=0.88\textwidth]{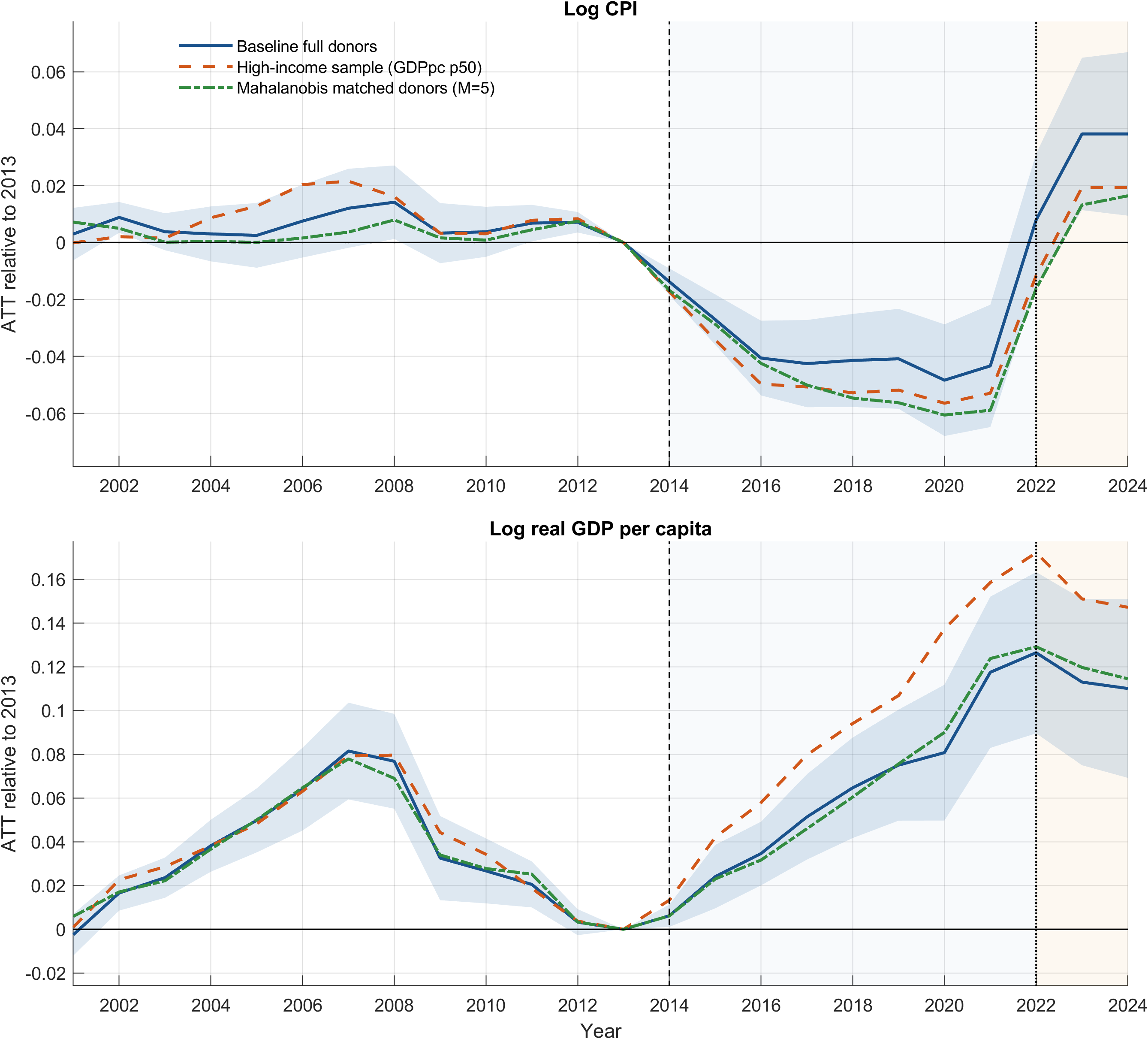}
		\label{fig:event_study_overlay_matched}
		\begin{minipage}{0.96\textwidth}\footnotesize
			\textit{Notes:} The baseline path includes its pointwise confidence band; the
			high-income and five-neighbor Mahalanobis paths are shown as lines to avoid
			visual clutter. Every path is normalized to 2013.
		\end{minipage}
	\end{figure}

	Both designs reduce, but do not eliminate, observable imbalance. The maximum
	absolute standardized difference falls from 1.498 to 1.175 and 1.052,
	respectively. Yet the estimates move little: the CPI contrast is 0.063 in
	both designs rather than 0.072, while the GDP-per-capita estimates are 0.023
	and 0.025 rather than 0.025. Figure~\ref{fig:event_study_overlay_matched}
	also preserves the post-2022 CPI break. Matching is not presented as
	randomized balance but it shows that the price result does not require the
	broadest set of dissimilar donors.\footnote{Appendix Table~\ref{tab:B4_matching} reports the complete balance vector, the corresponding direct-arming estimates, and the donor-selection frequencies used in the Mahalanobis design.}
	
	\begin{table}[!htbp]
		\centering
		\caption{Matched donor-pool comparability}
		\label{tab:matched}
		\begin{adjustbox}{max width=\textwidth}
			\begin{threeparttable}
				\small
				\begin{tabular}{l r r c c}
					\toprule
					Design & Treated & Unique donors & Inc.\ W2 ATT (SE) & Max.\ $|\mathrm{std.\ diff.}|$ \\
					\midrule
					\multicolumn{5}{l}{\textbf{Panel A. Log CPI}} \\
					Baseline full donors & 42 & 84 & 0.072*** (0.007) & 1.498 \\
					High-income sample (GDPpc p50) & 38 & 25 & 0.063*** (0.007) & 1.175 \\
					Mahalanobis matched donors (M=5) & 42 & 37 & 0.063*** (0.007) & 1.052 \\
					\addlinespace
					\multicolumn{5}{l}{\textbf{Panel B. Log real GDP per capita}} \\
					Baseline full donors & 42 & 84 & 0.025*** (0.008) & 1.498 \\
					High-income sample (GDPpc p50) & 38 & 25 & 0.023** (0.009) & 1.175 \\
					Mahalanobis matched donors (M=5) & 42 & 37 & 0.025*** (0.008) & 1.052 \\
					\bottomrule
				\end{tabular}
				\begin{tablenotes}[flushleft]
					\footnotesize
					\item \textit{Notes:} The high-income design restricts both groups at the
					median 2009--2013 log real GDP per capita. Mahalanobis matching uses five
					nearest eligible donors and frequency weights. Each row uses its displayed
					target count and the 2019--2021 reference.
				\end{tablenotes}
			\end{threeparttable}
		\end{adjustbox}
	\end{table}

	A stable average can still be misleading if it is generated by one treated
	country, one influential donor, or a small positive tail. Table~\ref{tab:influence}
	therefore examines all three possibilities. Treated-country leave-one-out
	recalculates the coalition average after removing each target in turn. Donor
	leave-one-out reconstructs the common factor after removing each donor. The
	median, trimmed mean, and positive share show how widely the country-level
	effects agree in sign and magnitude.
	
	For CPI, the median is 0.062, the 10\% trimmed mean is 0.071, and 97.6\% of
	contrasts are positive. Treated-country leave-one-out estimates remain between
	0.070 and 0.074, and the largest donor deletion changes the estimate by
	0.0049. GDP per capita is also insensitive to individual deletions, but only
	66.7\% of its country contrasts are positive. Thus, neither average is driven
	by one country, although cross-country support is much broader for CPI. \footnote{ Appendix Table~\ref{tab:B5_influence} reports the full treated- and donor-country influence audit, including all 42 CPI contrasts.}
	
	\begin{table}[!htbp]
		\centering
		\caption{Influence and distributional robustness}
		\label{tab:influence}
		\begin{threeparttable}
			\small
			\setlength{\tabcolsep}{5pt}
			\renewcommand{\arraystretch}{1.05}
			\begin{tabular}{l c c c c c c}
				\toprule
				\textit{Outcome} 
				& \makecell{\textit{Full ATT}\\ \textit{(SE)}} 
				& \textit{Median} 
				& \makecell{\textit{10\%}\\ \textit{trim}} 
				& \makecell{\textit{Positive}\\ \textit{share}} 
				& \makecell{\textit{Treated LOO}\\ \textit{range}} 
				& \makecell{\textit{Max donor}\\ \textit{LOO $\Delta$}} \\
				\midrule
				\addlinespace
				Log CPI & 0.072*** & 0.062 & 0.071 & 0.976 & [0.070, 0.074] & 0.0049 \\
				& (0.007) & & & & & \\
				\addlinespace
				Log real GDP per capita & 0.025*** & 0.026 & 0.025 & 0.667 & [0.023, 0.028] & 0.0019 \\
				& (0.008) & & & & & \\
				\bottomrule
			\end{tabular}
			\begin{tablenotes}[flushleft]
				\footnotesize
				\item \textit{Notes:} All diagnostics use sanctions and the 2019--2021 reference. Treated LOO deletes one treated country before re-aggregation. Donor LOO rebuilds the donor factor after deleting one eligible donor. The sample contains 42 targets and 84 donors.
			\end{tablenotes}
		\end{threeparttable}
	\end{table}
	
	\subsection{Direct Arming as a Narrower Coalition Definition
		\label{sec:direct_arming}}
	 
One issue with the sanctions indicator is that it may group together countries with different levels of involvement. Direct-arming countries form a smaller and more strongly committed group. 
We re-estimate the same design for the 29-country direct-arming coalition
defined in Section~\ref{subsec:policy_variables}. Because 28 of these
countries also impose strict sanctions, this exercise changes coalition
membership but does not isolate the effect of military assistance. We
therefore interpret it as a boundary check on whether the main pattern also
appears in a narrower, high-commitment coalition.
     
	For the 29-country direct-arming coalition, the incremental Wave-2 log-CPI contrast is 0.107 (SE 0.027; placebo p = 0.002). The incremental log-GDP-per-capita contrast is 0.016 (SE 0.010; placebo p = 0.339) and is not statistically significant. Excluding 2024 leaves a CPI gap of 0.095, and the inflation-rate gap
	is 3.7 percentage points. The positive output-level gap coexists with a
	3.7-percentage-point decline in GDP-per-capita growth. The positive
GDP-per-capita level gap therefore does not imply faster growth.\footnote{Appendix Tables~\ref{tab:B1_placebo},
		\ref{tab:B6_windows}, and~\ref{tab:B7_secondary} report, respectively, the
		complete placebo audit, alternative reference windows and bootstrap intervals,
		and the active-year timing exercise.}
	
	\begin{table}[!htbp]
		\centering
		\caption{Direct arming: baseline estimates and additional checks}
		\label{tab:direct_arming}
		\begin{threeparttable}
			\small
			\begin{tabular*}{\textwidth}
				{@{\extracolsep{\fill}} l c c c c @{}}
				\toprule
				\multicolumn{5}{l}{\textbf{Panel A. Level outcomes and preferred incremental gap}} \\
				\addlinespace[3pt]
				Outcome
				& \makecell{Wave~1\\level (SE)}
				& \makecell{Wave~2\\level (SE)}
				& \makecell{Incremental Wave~2\\gap (SE)}
				& \makecell{Placebo\\probability} \\
				\midrule
				Log CPI
				& -0.033** (0.013)
				& 0.077* (0.043)
				& 0.107*** (0.027)
				& 0.002 \\
				Log real GDP per capita
				& -0.001 (0.007)
				& 0.039*** (0.014)
				& 0.016 (0.010)
				& 0.339 \\
				\midrule
				\multicolumn{5}{l}{\textbf{Panel B. Horizon and rate-outcome checks}} \\
				\addlinespace[3pt]
				Outcome
				& \multicolumn{2}{l}{Robustness check}
				& \makecell{Incremental Wave~2\\gap (SE)}
				& Donors \\
				\midrule
				Log CPI
				& \multicolumn{2}{l}{Exclude 2024}
				& 0.095*** (0.021)
				& 96 \\
				Inflation rate
				& \multicolumn{2}{l}{Rate outcome}
				& 0.037*** (0.012)
				& 96 \\
				\addlinespace[3pt]
				Log real GDP per capita
				& \multicolumn{2}{l}{Exclude 2024}
				& 0.019** (0.009)
				& 97 \\
				GDP-per-capita growth
				& \multicolumn{2}{l}{Rate outcome}
				& -0.037*** (0.007)
				& 97 \\
				\bottomrule
			\end{tabular*}
			\begin{tablenotes}[flushleft]
				\footnotesize
				\item \textit{Notes:} Panel A reports level gaps for 2014--2021 and 2022--2024
				and the incremental contrast relative to 2019--2021. Panel B reports the
				2022--2023 contrast and the 2022--2024 rate contrasts, each relative to
				2019--2021. All specifications use 29 targets; CPI uses 96 donors and GDP per
				capita uses 97. Placebo probabilities do not determine the stars. Direct
				arming is a coalition-boundary check, not an independently identified
				military-aid effect.
			\end{tablenotes}
		\end{threeparttable}
	\end{table}
	
	\subsection{Overall Assessment \label{subsec:validation_stocktake}}

    The checks establish a clear hierarchy. The CPI gap survives changes in
policy timing, horizon, reference period, donor pool, factor proxy,
comparability restrictions, resampling scheme, reassignment, and country
influence. It is positive in 41 of 42 sanctioning countries and remains
positive under the narrower direct-arming coalition definition examined in
Section~\ref{sec:direct_arming}.
	
	The GDP-per-capita level result is less robust. It is stable to timing, the
	final sample year, donor composition, matching, and individual-country
	deletions, but it varies with the reference window and factor specification.
	Its preferred bootstrap intervals are positive, but its placebo probability
	is 0.102 and country-level agreement is weaker. Accordingly, the positive preferred level gap should not be interpreted as a post-2022 growth gain: it is specification-sensitive and accompanied by a negative annual growth differential.
	
	Overall, the validation exercises make it less likely that the CPI result is an artefact
	of one comparison group, one factor proxy, or a few influential countries.
	However, they cannot rule out anticipatory adjustment, differential spillovers, or an
	unrelated post-2013 structural change. Causal interpretation therefore still
	depends on the maintained counterfactual assumption, with considerably
	stronger validation support for CPI than for GDP per capita.

	\section{Russia-Energy Exposure and Post-2022 Heterogeneity}
	\label{sec:russia_energy}

Figure~\ref{fig:map_incremental_w2_europe} in our heterogeneity analysis documents substantial
cross-country variation in the coalition's post-2022 CPI and
real-GDP-per-capita gaps. This section examines whether that variation is
associated with pre-war Russian-energy exposure and whether the
corresponding exposure gradients differ between sanctioning and
non-sanctioning countries.

	\subsection{Exposure Measures and Hypotheses}
	
	For country $i$ and pre-war window $w$, exposure is defined as the monetary value of fossil-energy imports from Russia relative to the size of the importing economy:
	\begin{equation}
		E^{\mathrm{energy}}_{iw}
		=
		\frac{
			\sum_{t\in w}
			\left(
			\mathrm{Coal}^{RUS}_{it}
			+
			\mathrm{Crude}^{RUS}_{it}
			+
			\mathrm{Refined}^{RUS}_{it}
			+
			\mathrm{Gas}^{RUS}_{it}
			\right)
		}
		{
			\sum_{t\in w} \mathrm{GDP}_{it}
		}.
		\label{eq:energy_exposure}
	\end{equation}
	
	Trade flows and nominal GDP are measured in current US dollars. We rely on a 2018--2021 measure to capture immediate pre-invasion vulnerabilities, complementing it with a 2010--2013 measure to check sensitivity to any post-Crimea supply adjustments. This variable strictly measures monetary exposure, not physical energy dependence, and is standardized within each regression sample to facilitate interpretation.\footnote{Appendix Table~\ref{tab:A8_energy_exposure_construction} documents construction, verified zeros, missing values, and sample restrictions.}
	 
      The exposure regressions use 118 countries for the 2018--2021 exposure
measure and 119 for the 2010--2013 measure. Both samples contain 39
sanctioning countries rather than the 42 in the baseline ATT. The reason is that  Comtrade does not reliably verify pipeline-gas imports for Austria, Germany,
and Poland. The estimates therefore describe exposure gradients in a
reliable-data subsample, not a decomposition of the full 42-country
ATT.\footnote{The strict exposure sample retains 120 of 129 countries for
2018--2021 and 121 for 2010--2013. Intersecting them with the 126-country
mechanism universe removes Belarus and Ukraine and yields the 118- and
119-country regression samples. Russia already fails the exposure coverage
rule. Turkiye is also excluded for the pipeline-gas measurement reason but is
not a strict sanctioner. Appendix Table A8 reports the full reconciliation.}

Our primary hypothesis is that greater pre-war Russian-energy exposure is
associated within the coalition with larger post-2022 CPI and inflation
differentials and lower annual GDP-per-capita growth. We expect the
GDP-per-capita-level gradient to be weaker. The interaction specification
additionally tests whether these exposure gradients differ between sanctioning
and non-sanctioning countries.

	\subsection{Exposure-Interaction Model}
	
	This test uses a different estimator from the two-wave imputation. For each outcome and exposure window we estimate a CCE interaction on the observed panel, with $Post_t=1$ in 2022--2024 (the omitted period is all earlier years, not the 2019--2021 reference used for the incremental ATT) and without the eight lagged controls:
	\begin{equation}
		Y_{it}
		=
		\mu_i
		+
		\lambda_i
		\overline{Y}_{C,t}
		+
		\beta_1
		\left(S_i\times Post_t\right)
		+
		\beta_2
		\left(E_{iw}\times Post_t\right)
		+
		\beta_3
		\left(S_i\times E_{iw}\times Post_t\right)
		+
		u_{it},
		\label{eq:cce_ddd_mechanism}
	\end{equation}
	where $S_i$ identifies the sanctions coalition, $Post_t=1$ in 2022--2024, and $\overline{Y}_{C,t}$ is the CSA of the outcome outside the policy group. Country intercepts ($\mu_i$) absorb time-invariant traits, while $\lambda_i$ allows each economy to react heterogeneously to common global shocks. Standard errors are clustered at the country level.

     Because $E_{iw}$ is standardized, $E_{iw}=0$ denotes average exposure in the regression sample. The coefficients in equation (16) have four distinct meanings: (i) $\beta_1$ is the post-2022 sanctioner--non-sanctioner differential at average exposure, (ii) $\beta_2$ is the post-2022 exposure slope among non-sanctioning countries ($S_i=0$), (iii) $\beta_3$ is the difference between the sanctioner and non-sanctioner exposure slopes, and (iv) $\beta_2+\beta_3$ is the total post-2022 exposure slope among sanctioning countries. Thus, $\beta_2+\beta_3$ answers the within-coalition question of how the outcome differential changes with exposure among sanctioners whereas $\beta_3$ tests whether that gradient is steeper or flatter than among non-sanctioners.

	Table~\ref{tab:russia_energy_mechanism} reports the results of this exposure-heterogeneity regression.
	
	\paragraph{Post-2022 differential at average exposure}.\\
At sample-mean exposure, the estimated sanctioner--non-sanctioner
differential, $\widehat\beta_1$, is positive and statistically significant
for log CPI, inflation, and log real GDP per capita, and negative and
statistically significant for annual GDP-per-capita growth in both exposure
windows (Table~\ref{tab:russia_energy_mechanism}, Panels A and B).

	\paragraph{CPI and inflation}.\\
Within the coalition, a one-standard-deviation increase in Russian-energy
exposure over 2018--2021 is associated with a 0.0213-log-point
(approximately 2.15\%) larger post-2022 CPI differential
(Panel A, $\widehat\beta_2+\widehat\beta_3$). The positive triple
interaction ($\widehat\beta_3=0.0195$, $p<0.05$) indicates that this
exposure gradient is significantly steeper among sanctioning countries
than among non-sanctioning countries.

	While the pre-Crimea measure yields a similar internal coalition penalty of 0.0173 log points (Panel B, $\widehat\beta_2+\widehat\beta_3$), its triple interaction is statistically insignificant (Panel B, $\widehat\beta_3 = 0.0089$). The evidence that this price penalty was a coalition-specific phenomenon is therefore strongest when measuring exposure immediately before the full-scale invasion. Inflation moves in the same direction: a one-standard-deviation rise in exposure is associated with 0.64 and 0.60 percentage points more post-2022 inflation among sanctioners under the near-prewar and pre-Crimea measures. Unlike CPI, the difference between sanctioner and non-sanctioner inflation slopes is not statistically significant.
	
	\paragraph{GDP per capita and growth}.\\
	The real economy presents contrasting results between absolute output levels and growth rates.  We find no robust evidence that greater historical energy exposure is associated with a lower post-2022 level of real GDP per capita within the coalition. The estimated coalition exposure slope is (i) marginally negative for the near-prewar measure (Panel A, $\widehat\beta_2+\widehat\beta_3=-0.0147$), (ii) smaller and statistically insignificant for the pre-Crimea measure (Panel B,
$\widehat\beta_2+\widehat\beta_3=-0.0072$), and (iii) sensitive to the
estimator, including sign reversals under two-way fixed effects.

However, the drag on economic growth rates is unmistakable. A one-standard-deviation higher pre-war exposure penalized annual GDP-per-capita growth by roughly 0.96 percentage points (Panel A, $\widehat\beta_2+\widehat\beta_3$). This growth penalty is highly robust across sample definitions and estimators, and the highly significant triple interactions ($-1.2875$ and $-1.3592$ in $\widehat\beta_3$ of Panels A and B, respectively) suggest it was more severe for sanctioners than for the rest of the world. 
	
	\begin{table}[!htbp]
		\centering
		\caption{Pre-war Russian-energy exposure and post-2022 heterogeneity}
		\label{tab:russia_energy_mechanism}
		\begin{threeparttable}
			\small
			\setlength{\tabcolsep}{4.5pt} 
			\renewcommand{\arraystretch}{1.05} 
			\begin{tabular}{l c c c c c}
				\toprule
				\textit{Outcome} 
				& \makecell{\textit{Sanctions} \\ \textit{$\times$ Post}\\($\widehat\beta_1$)}
				& \makecell{\textit{Exposure} \\ \textit{$\times$ Post}\\($\widehat\beta_2$)}
				& \makecell{\textit{Sanc. $\times$ Exp.} \\ \textit{$\times$ Post}\\($\widehat\beta_3$)}
				& \makecell{\textit{Sanctioner}\\ \textit{Exp. Slope}\\($\widehat\beta_2+\widehat\beta_3$)}
				& \makecell{\textit{N}\\ \textit{(All/Sanc.)}} \\
				\midrule
				\multicolumn{6}{l}{\textbf{Panel A. Near-prewar exposure, 2018--2021}}\\
				\addlinespace
				Log CPI & 0.0579\sym{***} & 0.0018 & 0.0195\sym{**} & 0.0213\sym{***} & 118 / 39 \\
				& (0.0069) & (0.0077) & (0.0097) & (0.0060) & \\
				\addlinespace
				Inflation rate & 2.3054\sym{***} & 0.0449 & 0.5919 & 0.6368\sym{***} & 118 / 39 \\
				& (0.2852) & (0.4688) & (0.5147) & (0.2126) & \\
				\addlinespace
				Log real GDP per capita & 0.0596\sym{***} & 0.0042 & -0.0189 & -0.0147\sym{*} & 118 / 39 \\
				& (0.0114) & (0.0089) & (0.0120) & (0.0081) & \\
				\addlinespace
				GDP-per-capita growth & -0.9502\sym{***} & 0.3295 & -1.2875\sym{***} & -0.9580\sym{***} & 118 / 39 \\
				& (0.2950) & (0.3116) & (0.3955) & (0.2435) & \\
				\addlinespace
				\midrule
				\multicolumn{6}{l}{\textbf{Panel B. Pre-Crimea exposure, 2010--2013}}\\
				\addlinespace
				Log CPI & 0.0557\sym{***} & 0.0084 & 0.0089 & 0.0173\sym{***} & 119 / 39 \\
				& (0.0069) & (0.0123) & (0.0127) & (0.0034) & \\
				\addlinespace
				Inflation rate & 2.2009\sym{***} & 0.0818 & 0.5140 & 0.5959\sym{***} & 119 / 39 \\
				& (0.2969) & (0.8464) & (0.8551) & (0.1217) & \\
				\addlinespace
				Log real GDP per capita & 0.0588\sym{***} & 0.0134 & -0.0205\sym{*} & -0.0072 & 119 / 39 \\
				& (0.0114) & (0.0104) & (0.0115) & (0.0048) & \\
				\addlinespace
				GDP-per-capita growth & -0.9190\sym{***} & 0.7220\sym{**} & -1.3592\sym{***} & -0.6372\sym{***} & 119 / 39 \\
				& (0.3114) & (0.3188) & (0.3496) & (0.1434) & \\
				\bottomrule
			\end{tabular}
			\begin{tablenotes}[flushleft]
				\footnotesize
				\item \textit{Notes:} Exposure is Russian fossil-energy imports divided by nominal GDP over the indicated window and standardized within the regression sample. All rows use strict sanctions, the strict Comtrade sample, and equation~\eqref{eq:cce_ddd_mechanism}; $Post=1$ in 2022--2024. The reported 118 and 119 country counts are the actual regression samples after intersection with the 126-country mechanism universe, not the exposure-construction counts. Country intercepts and country-specific loadings on the non-policy-country outcome mean are partialled out. Country-clustered standard errors are in parentheses. Log outcomes are in log points; inflation and GDP-per-capita growth are in annual percentage points. $\beta_1$ is the sanctioning-country differential at mean exposure, not the baseline incremental ATT. The standard error of $\widehat\beta_2+\widehat\beta_3$ uses the full covariance matrix. \sym{*} $p<0.10$, \sym{**} $p<0.05$, \sym{***} $p<0.01$.
			\end{tablenotes}
		\end{threeparttable}
	\end{table}
	
	\paragraph{Summary}.\\
    Taken together, the results suggest that exposed sanctioners experienced larger CPI and inflation increases and slower GDP-per-capita growth after 2022. These within-coalition patterns remain stable across the main robustness checks. The comparison with non-sanctioning countries is strongest for GDP-per-capita growth, more limited for CPI, and weak for inflation. The GDP-per-capita level result is not robust. The evidence is therefore consistent with an energy-cost channel, but it does not establish that exposure and dependence on Russian energy is the sole cause of the aggregate post-2022 effect. \footnote{Table~\ref{tab:B8_energy_exposure_robustness} reports the sensitivity of these findings to the
		exposure window, sample-coverage rule, estimator, policy definition, and
		exposure measure. These checks assess the stability of the exposure gradients.}

	\section{Conclusions and Policy Implications \label{sec:conclusion}}
	
	Economic sanctions are designed to impose costs on a target, but they can also
	affect prices and economic activity in the countries that impose them.  Using a two-wave CCEDID design, this paper estimates the sender-side
macroeconomic incidence associated with participation in sanctions against
Russia across the post-2014 regime and the post-2022 escalation.
	
	The clearest result is the divergence between prices and output. Relative to
	2019--2021, the sanctions coalition's post-2022 CPI gap increased by 0.072 log
	points, or approximately 7.5\%.  The country-level CPI estimate is positive in 41 of the 42 coalition members
and remains similar across alternative reference periods, donor pools,
factor specifications, matched samples, and inference procedures. The incremental gap in the level of real GDP per capita is small, positive in the preferred specification (0.025), and not robust. It should not be read as a post-2022 growth gain: the same design gives a lower annual GDP-per-capita growth rate after 2022 (about 3.9 percentage points in the rate check).

    The pre-war energy-exposure analysis provides additional evidence on why domestic incidence differed across countries. Within the exposure sample, a one-standard-deviation increase in Russian fossil-energy imports relative to GDP during 2018--2021 is associated with a larger post-2022 CPI differential and lower annual GDP-per-capita
growth among sanctioning countries. Thus, more-exposed sanctioners
experienced larger post-2022 CPI and inflation differentials and slower
annual growth. Similar patterns appear with the pre-Crimea exposure measure. These findings complement structural evidence on
	heterogeneous energy costs \citep{Ari2022,Bachmann2024,DiBella2024}, but they
	should be interpreted as exposure gradients rather than as formal mediation. The reason is that the regressions use a smaller sample, measure monetary rather than physical
energy dependence, and cannot fully capture pipeline-gas exposure in several European economies. Moreover, evidence that the gradient differs from that among non-sanctioners is strongest for GDP-per-capita growth, more limited for CPI, and weak for inflation.

	Taken together, results have three relevant implications for policy evaluation and preparedness.
	
	First, assessing sender costs requires tracking prices and inflation, not just output. A coalition can show no robust loss in real GDP per capita and still record a broad post-2022 price-level increase. Who bears that price increase is a separate question; this paper does not estimate household or firm incidence.   Second, when sanctions escalate within an existing policy regime, new
effects should be measured relative to the immediate pre-escalation
reference. Level gaps should also be reported alongside annual rates so that
accumulated positions are not mistaken for continuing inflation or growth
effects. Third, the capacity to impose sanctions relies heavily on pre-crisis infrastructure and supply diversification. Using observable pre-existing exposures (like bilateral energy imports) for ex-ante vulnerability screening can predict where price and growth pressures will be most severe. Recognizing this unequal domestic incidence is vital for guiding coordinated procurement, emergency supply plans, and exposure-sensitive burden sharing.

	Finally, our evidence of sender costs does not establish that sanctions failed or
	reduced welfare. A complete assessment would also require evidence on the
	costs imposed on Russia and on the strategic benefits of coalition
	participation, neither of which is estimated here. Within its more limited
	scope, the paper shows that aggregate output resilience can conceal a broad
	price burden and that the distribution of this burden is related to observable
	pre-existing vulnerabilities. These are the margins that policy evaluation
	should monitor when sanctions are introduced or intensified.

	\paragraph{Data and code availability.}
	Appendix A documents the data sources, variable definitions, sample construction,
	and treatment coding. Appendix B details the estimation and inference procedures.
	This arXiv submission contains the manuscript and figure files; it does not
	include a replication archive.

	\clearpage
	\phantomsection
	\addcontentsline{toc}{section}{References}
	\begingroup
	\small
	\setlength{\bibsep}{4pt}
	
	\endgroup
	
	\clearpage
	\phantomsection
	\addcontentsline{toc}{section}{Supplementary Appendix A: Data and Sample Documentation}
	\section*{Supplementary Appendix A\\Data and Sample Documentation}
	\setcounter{table}{0}
	\renewcommand{\thetable}{A\arabic{table}}
	\renewcommand{\theHtable}{A\arabic{table}}
	\setcounter{figure}{0}
	\renewcommand{\thefigure}{A\arabic{figure}}
	\renewcommand{\theHfigure}{A\arabic{figure}}
	\setcounter{equation}{0}
	\renewcommand{\theequation}{A\arabic{equation}}
	\renewcommand{\theHequation}{A\arabic{equation}}
	
	Appendix A documents the data in the order in which they are constructed.
	Section~S.A.1 defines the variables and sources used in the main macroeconomic
	panel. Section~S.A.2 explains how the sample is selected from the raw data,
	how the remaining gaps are completed after eligibility is fixed, and how
	those repairs are audited. Section~S.A.3 documents treatment timing, donor
	restrictions, and each country's final estimation role. Section~S.A.4
	describes the pretreatment composition of the resulting groups.
	Section~S.A.5 explains the separately constructed measures of exposure to
	Russian energy. This sequence matters: repaired observations do not determine
	sample eligibility, and policy-based donor exclusions are distinct from
	missing-data exclusions.
	
	\subsection*{S.A.1 Main-Panel Data, Variables, and Sources}
	
	The main panel combines macroeconomic series from the World Development
	Indicators, rule-of-law data from the Worldwide Governance Indicators, and
	author-coded policy indicators. Table~\ref{tab:A1_variables} reports the exact
	series, transformations, and sources used for the outcomes and controls.
	Treatment coding and the additional trade data used in the energy-exposure
	analysis are documented separately below.
	
	\begin{table}[!htbp]
		\centering
		\caption{Main-panel variable definitions, transformations, and sources}
		\label{tab:A1_variables}
		\begin{adjustbox}{max width=\textwidth}
			\begin{threeparttable}
				\small
				\begin{tabular}{p{4.5cm} p{4cm} p{5cm} p{4.5cm}}
					\toprule
					Variable & Source code & Source & Transformation \\
					\midrule
					CPI / Log CPI & FP.CPI.TOTL & World Development Indicators & Index / natural log \\
					Real GDP per capita / log & NY.GDP.PCAP.KD & World Development Indicators & Constant-price / log \\
					Inflation rate & FP.CPI.TOTL.ZG & World Development Indicators & Percent; div. by 100 \\
					Real GDPpc growth & \makecell[l]{NY.GDP.PCAP.\\KD.ZG} & World Development Indicators & Percent; div. by 100 \\
					Oil rents & NY.GDP.PETR.RT.ZS & World Development Indicators & Percent of GDP \\
					Trade openness & NE.TRD.GNFS.ZS & World Development Indicators & Percent of GDP \\
					Terms of trade & \makecell[l]{TT.PRI.MRCH.\\XD.WD} & World Development Indicators & Index \\
					Investment rate & NE.GDI.TOTL.ZS & World Development Indicators & Percent of GDP \\
					Government consumption & NE.CON.GOVT.ZS & World Development Indicators & Percent of GDP \\
					Rule of law & WGI rule-of-law & Worldwide Governance Indicators & Level \\
					Population growth & SP.POP.GROW & World Development Indicators & Annual percent \\
					Fuel-share balance & TM.VAL.FUEL.ZS.UN; TX.VAL.FUEL.ZS.UN & World Development Indicators & Fuel-import share minus fuel-export share, percentage points \\
					Treatment indicators & Canonical annual files & Documented policy sources & Binary indicators \\
					\bottomrule
				\end{tabular}
				\begin{tablenotes}[flushleft]
					\footnotesize
					\item \textit{Notes:} Controls are completed before being lagged and enter the model with a one-year lag. CPI is transformed to its natural logarithm only after completion and is interpreted as a within-country index; index levels are not compared across countries. The fuel-share balance is constructed from WDI fuel-import and fuel-export shares and is not a measure of bilateral dependence on Russia. Neither treatment coding nor estimated treatment effects enter any variable-completion rule.
				\end{tablenotes}
			\end{threeparttable}
		\end{adjustbox}
	\end{table}
	
	\FloatBarrier
	
	\subsection*{S.A.2 Sample Eligibility, Panel Completion, and Missing-Data Audit}
	
	\paragraph{Coverage rules and estimation windows.}
	The eligibility rules are applied separately by variable to the raw
	2000--2024 observations. Each of the four outcomes---CPI, inflation, real GDP
	per capita, and GDP-per-capita growth---must have no more than 35\% raw
	missingness. The controls are subject to a two-part rule: at most one of the
	eight controls may exceed 25\% missingness and no control may exceed 65\%.
	This allows an isolated gap in a secondary covariate without permitting a
	country's control history to be predominantly reconstructed.
	
	An additional rule protects the quality and reliability of the preferred incremental comparison. CPI and
	real GDP per capita may each be missing in no more than 25\% of the six years
	from 2019 through 2024. Because the data are annual, this permits at most one
	missing observation in either key outcome during that window. The critical-window screen is applied to 131 otherwise eligible countries and excludes
	Syria and Sudan, yielding the final $N=129$ panel. These exclusions follow
	mechanically from the published rule: Sudan has two raw missing CPI
	observations in 2019--2024, while Syria has five raw missing CPI observations
	and two raw missing real-GDP-per-capita observations.
	
	The thresholds are concentration-sensitive rather than based on a single
	overall missingness score. A country may therefore remain in the panel when
	moderate missingness is dispersed across controls, but it fails when gaps are
	concentrated in a principal outcome or in the period required for the preferred
	estimand. Eligibility is finalized before any completion operation. Neither
	treatment indicators nor estimated treatment effects enter the coverage rules
	or the completion procedures.
	
	The underlying World Bank series are collected from 1996 onward, but
	1996--1999 are used only as supporting observations for beginning-of-series
	completion. Missingness rates, eligibility rules, the balanced panel, and all
	reported estimates refer to 2000--2024. Since controls enter the empirical
	model with a one-year lag, the effective estimation sample is 2001--2024 and
	the untreated model is estimated over 2001--2013.
	
	\begin{table}[!htbp]
		\centering
		\caption{Final sample coverage and estimation windows}
		\label{tab:A2_sample}
		\label{tab:A1_sample}
		\begin{adjustbox}{max width=\textwidth}
			\begin{threeparttable}
				\small
				\begin{tabular}{ll}
					\toprule
					Design component & Final rule \\
					\midrule
					Final balanced panel & 129 countries / 3,225 country-years \\
					Observed years & 2000--2024 \\
					Effective estimation years & 2001--2024 \\
					Untreated-model estimation window & 2001--2013 \\
					Whole-period outcome coverage & Each of four outcomes has at most 35\% raw missingness \\
					Whole-period control coverage & At most one of eight controls exceeds 25\% raw missingness; none exceeds 65\% \\
					Critical outcome window & CPI and real GDP per capita each have at most 25\% raw missingness in 2019--2024 \\
					Order of operations & Eligibility is evaluated on raw values before any interpolation \\
					Wave 1 reporting window & 2014--2021 \\
					Wave 2 reporting window & 2022--2024 \\
					Preferred reference & 2019--2021 \\
					Alternative references & 2017--2019; 2021; 2017--2021; 2018--2021 \\
					Main sanctions treated / donors & 42 / 84 \\
					Direct arming treated / CPI donors / GDPpc donors & 29 / 96 / 97 \\
					\bottomrule
				\end{tabular}
				\begin{tablenotes}[flushleft]
					\footnotesize
					\item \textit{Notes:} The four outcomes are GDP-per-capita growth, inflation, CPI, and real GDP per capita; the eight controls are listed in Table~\ref{tab:A1_variables}. All thresholds are applied by variable to raw 2000--2024 data before completion and without using treatment status. The 1996--1999 observations provide support for beginning-of-series completion but do not enter the eligibility calculations. In the six-year critical window, the 25\% ceiling permits at most one raw missing observation in CPI and at most one in real GDP per capita. The critical-window screen excludes Syria and Sudan from the 131 otherwise eligible countries and produces the final $N=129$ panel.
				\end{tablenotes}
			\end{threeparttable}
		\end{adjustbox}
	\end{table}
	
	\paragraph{Deterministic completion.}
	After sample eligibility has been fixed, missing control observations are
	completed using a common deterministic hierarchy. An internal gap bounded by
	observed values on both sides is filled by linear interpolation. A trailing
	gap is filled using the last available observation. A leading gap of no more
	than three years is filled using the first available observation. For a longer
	leading gap, the reconstruction code fits a local linear trend to the first
	five valid observations and extrapolates it backwards. The extrapolation is
	bounded by the country series' observed range and dispersion; for controls
	that are intrinsically nonnegative---trade openness, investment, and
	government consumption---the lower bound is additionally restricted to zero.
	The use of 1996--1999 as supporting data reduces the number of analytical-period
	observations requiring an edge extrapolation.
	
	Outcome completion is more limited. In the final sample, real GDP per capita
	and GDP-per-capita growth are fully observed in the raw data. CPI and inflation
	contain 25 and 32 raw missing cells, respectively. Fifty-five of these 57 cells
	are leading observations between 2000 and 2007. They are recorded by the audit
	as bounded local-trend backcasts based only on the country's own observed
	series. The other two cells are the 2024 CPI and inflation observations for
	Bosnia and Herzegovina; both equal the corresponding 2023 observation. Thus,
	the outcome-repair layer does not interpolate internal gaps, does not use
	cross-country information, and does not use treatment assignments or estimated
	effects.
	
	Across the eight controls, the raw data contain 1,471 missing cells out of
	25,800 possible control cells, or 5.70\%. These gaps are resolved in the final
	balanced panel using the documented hierarchy.  
	
	\paragraph{Missing-data audit.}
	Table~\ref{tab:A3_missing_variables} reports raw missingness by variable and
	audits the limited repairs made to the outcome series.
	
	\begin{table}[!htbp]
		\centering
		\caption{Variable-level raw-data coverage and audited outcome repairs}
		\label{tab:A3_missing_variables}
		\label{tab:A6_missing_variables}
		\begin{adjustbox}{max width=\textwidth}
			\begin{threeparttable}
				\small
				\begin{tabular}{l r r r r}
					\toprule
					\multicolumn{5}{l}{\textbf{Panel A. Raw missingness by variable, final $N=129$}} \\
					\addlinespace
					\textit{Variable} 
					& \makecell[r]{\textit{Raw missing}\\\textit{cells}} 
					& \makecell[r]{\textit{Possible}\\\textit{cells}} 
					& \makecell[r]{\textit{Raw missing}\\\textit{share}} 
					& \makecell[r]{\textit{Outcome-audit}\\\textit{repaired cells}} \\
					\midrule
					Real GDPpc growth & 0 & 3,225 & 0.00\% & 0 \\
					Inflation & 32 & 3,225 & 0.99\% & 32 \\
					Real GDP per capita & 0 & 3,225 & 0.00\% & 0 \\
					CPI & 25 & 3,225 & 0.78\% & 25 \\
					Oil rents & 403 & 3,225 & 12.50\% & -- \\
					Trade openness & 10 & 3,225 & 0.31\% & -- \\
					Terms of trade & 645 & 3,225 & 20.00\% & -- \\
					Investment rate & 46 & 3,225 & 1.43\% & -- \\
					Government consumption & 34 & 3,225 & 1.05\% & -- \\
					Rule of law & 129 & 3,225 & 4.00\% & -- \\
					Population growth & 1 & 3,225 & 0.03\% & -- \\
					Fuel-share balance & 203 & 3,225 & 6.29\% & -- \\
					\midrule
					\multicolumn{5}{l}{\textbf{Panel B. Audited outcome-repair methods}} \\
					\addlinespace
					\textit{Method} 
					& \multicolumn{1}{l}{\textit{Variable}} 
					& \makecell[r]{\textit{Repaired}\\\textit{cells}} 
					& \makecell[r]{\textit{Share of}\\\textit{all repairs}} 
					& \makecell[r]{\textit{Share within}\\\textit{variable}} \\
					\midrule
					Bounded trend backcast & \multicolumn{1}{l}{Inflation} & 31 & 54.39\% & 96.88\% \\
					2023 value carried forward & \multicolumn{1}{l}{Inflation} & 1 & 1.75\% & 3.12\% \\
					Bounded trend backcast & \multicolumn{1}{l}{CPI} & 24 & 42.11\% & 96.00\% \\
					2023 value carried forward & \multicolumn{1}{l}{CPI} & 1 & 1.75\% & 4.00\% \\
					\bottomrule
				\end{tabular}
				\begin{tablenotes}[flushleft]
					\footnotesize
					\item \textit{Notes:} Raw missingness is measured before any completion operation across 129 countries and 25 observed years. The four outcomes contain 57 raw missing cells out of 12,900 possible outcome cells (0.44\%): 32 for inflation and 25 for CPI. Fifty-five are leading observations between 2000 and 2007; the remaining two are the 2024 inflation and CPI observations for Bosnia and Herzegovina, both of which equal their respective 2023 values. No internal outcome gap is interpolated. Dashes for controls indicate that the final column summarizes the outcome-repair audit and is not applicable to control variables; they do not indicate that raw control gaps were left unresolved. The eight controls contain 1,471 raw missing cells out of 25,800 possible cells (5.70\%), all resolved in the final rectangular panel using the deterministic hierarchy described in the text. Because the archived control-method tag applies at the country-variable-series level, it is not converted into unsupported cell-level method shares.
				\end{tablenotes}
			\end{threeparttable}
		\end{adjustbox}
	\end{table}
	
	\noindent Country-level raw-missingness statistics are provided in the replication files. Every included country satisfies the ex ante coverage rules in Table~\ref{tab:A2_sample}; Table~\ref{tab:A3_missing_variables} reports the more informative variable-level totals and repair counts.
	
	\FloatBarrier
	
	\subsection*{S.A.3 Treatment Coding, Donor Eligibility, and Country Roles}
	
	\paragraph{Treatment definitions and timing.}
	Our sanctions indicator, \(S_{it}\), records participation in the sanctions coalition at the country--year level. It is zero until the first calendar year in which country \(i\) is legally covered by an official sanctions regime against Russia or formally aligns with one, and one from that year onward. EU members enter in 2014, when the EU adopted its first restrictive measures in response to Russia's annexation of Crimea. Other countries enter only when a dated national measure or official alignment record documents participation. We give priority to legal and government records, followed by official EU alignment declarations and European Commission documents. Secondary material is used only as a last resort and is not sufficient for contested cases.\footnote{Core sources include Council Regulation (EU) No.~833/2014 and the Council of the EU and European Commission sanctions records; dated Council declarations concerning non-EU alignment; Switzerland's ordinance of 2 April 2014; New Zealand's Russia Sanctions Act 2022; and the March 2022 government announcements by Singapore and South Korea.  These are cross-cehcked with the Ukraine Support Tracker developed by \cite{Trebesch2023}, The Global Sanctions Data Base \cite{YalcinEtAl2025} and Wikipedia.} The sanction series does not automatically assign a country's sanctions status to its territories or dependencies. Importantly, the indicator measures coalition participation, not the number, scope, enforcement, or intensity of the sanctions adopted. Table~\ref{tab:A4_treatment} reports the entry years used in the analysis.
	
	The direct-arming indicator follows the same country--year structure. It
	switches on in the first year for which publicly verifiable evidence documents
	a qualifying delivery or commitment under the definition in
	Section~\ref{sec:data}. It identifies participation in the narrower
	direct-arming coalition rather than the amount or intensity of military aid. 
	
	\begin{table}[!htbp]
		\centering
		\caption{Strict treatment coding and entry years}
		\label{tab:A4_treatment}
		\begin{adjustbox}{max width=\textwidth,max totalheight=.88\textheight,keepaspectratio}
			\begin{threeparttable}
				\scriptsize
				\begin{tabular}{l c c c c}
					\toprule
					Country & Code & Sanctions start & Direct-arming start & Arming active before 2022 \\
					\midrule
					Albania & ALB & 2014 & N/A & N/A \\
					Australia & AUS & 2014 & 2022 & No \\
					Austria & AUT & 2014 & N/A & N/A \\
					Belgium & BEL & 2014 & 2022 & No \\
					Bosnia and Herzegovina & BIH & 2022 & N/A & N/A \\
					Bulgaria & BGR & 2014 & 2022 & No \\
					Canada & CAN & 2014 & 2022 & No \\
					Croatia & HRV & 2014 & 2022 & No \\
					Cyprus & CYP & 2014 & N/A & N/A \\
					Czechia & CZE & 2014 & 2018 & Yes \\
					Denmark & DNK & 2014 & 2022 & No \\
					Estonia & EST & 2014 & 2022 & No \\
					Finland & FIN & 2014 & 2022 & No \\
					France & FRA & 2014 & 2022 & No \\
					Germany & DEU & 2014 & 2022 & No \\
					Greece & GRC & 2014 & 2022 & No \\
					Hungary & HUN & 2014 & N/A & N/A \\
					Iceland & ISL & 2014 & N/A & N/A \\
					Ireland & IRL & 2014 & N/A & N/A \\
					Italy & ITA & 2014 & 2022 & No \\
					Japan & JPN & 2014 & N/A & N/A \\
					Korea, Rep. & KOR & 2022 & N/A & N/A \\
					Latvia & LVA & 2014 & 2022 & No \\
					Lithuania & LTU & 2014 & 2022 & No \\
					Luxembourg & LUX & 2014 & 2022 & No \\
					Malta & MLT & 2014 & N/A & N/A \\
					Moldova & MDA & 2023 & N/A & N/A \\
					Netherlands & NLD & 2014 & 2022 & No \\
					New Zealand & NZL & 2022 & N/A & N/A \\
					North Macedonia & MKD & 2014 & 2022 & No \\
					Norway & NOR & 2014 & 2022 & No \\
					Poland & POL & 2014 & 2018 & Yes \\
					Portugal & PRT & 2014 & 2022 & No \\
					Romania & ROU & 2014 & 2022 & No \\
					Singapore & SGP & 2022 & N/A & N/A \\
					Slovak Republic & SVK & 2014 & 2022 & No \\
					Slovenia & SVN & 2014 & 2022 & No \\
					Spain & ESP & 2014 & 2022 & No \\
					Sweden & SWE & 2014 & 2022 & No \\
					Switzerland & CHE & 2014 & N/A & N/A \\
					Turkiye & TUR & N/A & 2019 & Yes \\
					Ukraine & UKR & 2014 & N/A & N/A \\
					United Kingdom & GBR & 2014 & 2022 & No \\
					United States & USA & 2014 & 2018 & Yes \\
					\bottomrule
				\end{tabular}
				\begin{tablenotes}[flushleft]
					\footnotesize
					\item \textit{Notes:} Direct-arming timing is reported as N/A for countries outside the strict direct-arming coalition. Lithuania begins strict direct arming in 2022; exactly four of the 29 direct-arming countries are active before 2022. For every positive country-year code, the replication ledger records the issuing authority, document title, publication date, archived source, policy type, entry year, and coding decision. Strict-sanctions codes require an official legal instrument or a formal alignment notice. Direct-arming codes require publicly verifiable evidence of a qualifying delivery or commitment under the definition in Section~\ref{sec:data}.
				\end{tablenotes}
			\end{threeparttable}
		\end{adjustbox}
	\end{table}
	
	\paragraph{Donor eligibility.}
	After treatment status is fixed, donor eligibility is determined separately
	for each policy definition and outcome. Table~\ref{tab:A5_donors} reports the
	candidate pools and the additional geopolitical and outcome-specific
	restrictions.
	
	\begin{table}[!htbp]
		\centering
		\caption{Donor eligibility by policy and outcome}
		\label{tab:A5_donors}
		\begin{adjustbox}{max width=\textwidth}
			\begin{threeparttable}
				\small
				\begin{tabular}{llrrp{5.5cm}}
					\toprule
					Policy & Outcome & Candidates & Eligible & Excluded candidates \\
					\midrule
					Sanctions & Log CPI & 86 & 84 & BLR, RUS \\
					Sanctions & Log real GDP per capita & 86 & 84 & BLR, RUS \\
					Direct arming & Log CPI & 100 & 96 & BIH, BLR, RUS, UKR \\
					Direct arming & Log real GDP per capita & 100 & 97 & BLR, RUS, UKR \\
					\bottomrule
				\end{tabular}
				\begin{tablenotes}[flushleft]
					\footnotesize
					\item \textit{Notes:} Candidates are countries that are never treated under the policy definition in the row. Russia, Belarus, and Ukraine are barred from every donor pool for geopolitical reasons. Ukraine is already outside the sanctions candidate set because it retains its canonical sanctions coding. Bosnia and Herzegovina remains in the $N=129$ panel, but is excluded from the direct-arming CPI donor factor because its 2024 CPI observation is repaired. This outcome-specific restriction prevents a repaired observation in the preferred 2019--2024 comparison window from entering the common donor factor; it is not a sample exclusion.
				\end{tablenotes}
			\end{threeparttable}
		\end{adjustbox}
	\end{table}
	
	\FloatBarrier
	
	\paragraph{Country-level estimation roles.}
	Table~\ref{tab:A6_countries} provides the country-level audit of these
	decisions. It distinguishes sanctions coding from inclusion in the main
	sanctions target group and reports treatment and donor status under both
	coalition definitions.

 \begingroup
\scriptsize
\setlength{\tabcolsep}{2.8pt} 
\renewcommand{\arraystretch}{0.92} 

\begin{longtable}{@{}
    >{\raggedright\arraybackslash}p{2.9cm}
    >{\centering\arraybackslash}p{1cm}
    *{7}{>{\centering\arraybackslash}p{\dimexpr(\textwidth-5.25cm-18\tabcolsep)/7\relax}}
    >{\centering\arraybackslash}p{1.35cm}@{}
}
    \caption{Countries and estimation roles in the final $N=129$ sample}\label{tab:A6_countries}\\
    \toprule
    Country & Code & \makecell{Sanc.\\coded} & \makecell{Sanc.\\target} & \makecell{Direct\\arming} & \makecell{Sanc.\\CPI\\donor} & \makecell{Sanc.\\GDPpc\\donor} & \makecell{Arming\\CPI\\donor} & \makecell{Arming\\GDPpc\\donor} & \makecell{Raw\\missing\\share (\%)} \\
    \midrule
    \endfirsthead

    \toprule
    Country & Code & \makecell{Sanc.\\coded} & \makecell{Sanc.\\target} & \makecell{Direct\\arming} & \makecell{Sanc.\\CPI\\donor} & \makecell{Sanc.\\GDPpc\\donor} & \makecell{Arming\\CPI\\donor} & \makecell{Arming\\GDPpc\\donor} & \makecell{Raw\\missing\\share (\%)} \\
    \midrule
    \endhead

    \midrule
    \multicolumn{10}{r}{\textit{Continued on next page}}\\
    \endfoot

    \bottomrule
    \multicolumn{10}{@{}p{\textwidth}@{}}{\vspace{3pt}\scriptsize \textit{Notes:} Russia, Belarus, and Ukraine are never donors. Ukraine is sanctions-coded but is not a main sanctions target. A country can be a treated unit under one policy definition and a donor under neither definition. The final column reports the share of originally missing cells across the four outcomes and eight controls over 2000--2024, before repair or interpolation. Each percentage uses 300 potential cells per country (12 variables $\times$ 25 years).}\\
    \endlastfoot

Albania & ALB & Yes & Yes & No & No & No & Yes & Yes & 3.00\% \\
Algeria & DZA & No & No & No & Yes & Yes & Yes & Yes & 5.33\% \\
Angola & AGO & No & No & No & Yes & Yes & Yes & Yes & 7.33\% \\
Armenia & ARM & No & No & No & Yes & Yes & Yes & Yes & 3.00\% \\
Australia & AUS & Yes & Yes & Yes & No & No & No & No & 3.00\% \\
Austria & AUT & Yes & Yes & No & No & No & Yes & Yes & 3.00\% \\
Azerbaijan & AZE & No & No & No & Yes & Yes & Yes & Yes & 3.00\% \\
Bahamas, The & BHS & No & No & No & Yes & Yes & Yes & Yes & 3.33\% \\
Bahrain & BHR & No & No & No & Yes & Yes & Yes & Yes & 3.00\% \\
Bangladesh & BGD & No & No & No & Yes & Yes & Yes & Yes & 5.33\% \\
Belarus & BLR & No & No & No & No & No & No & No & 4.33\% \\
Belgium & BEL & Yes & Yes & Yes & No & No & No & No & 3.00\% \\
Belize & BLZ & No & No & No & Yes & Yes & Yes & Yes & 3.67\% \\
Benin & BEN & No & No & No & Yes & Yes & Yes & Yes & 4.67\% \\
Bhutan & BTN & No & No & No & Yes & Yes & Yes & Yes & 6.33\% \\
Bolivia & BOL & No & No & No & Yes & Yes & Yes & Yes & 3.00\% \\
Bosnia and Herzegovina & BIH & Yes & Yes & No & No & No & No & Yes & 8.33\% \\
Botswana & BWA & No & No & No & Yes & Yes & Yes & Yes & 3.33\% \\
Brazil & BRA & No & No & No & Yes & Yes & Yes & Yes & 3.00\% \\
Brunei Darussalam & BRN & No & No & No & Yes & Yes & Yes & Yes & 3.67\% \\
Bulgaria & BGR & Yes & Yes & Yes & No & No & No & No & 3.00\% \\
Burkina Faso & BFA & No & No & No & Yes & Yes & Yes & Yes & 3.33\% \\
Cambodia & KHM & No & No & No & Yes & Yes & Yes & Yes & 4.67\% \\
Cameroon & CMR & No & No & No & Yes & Yes & Yes & Yes & 3.33\% \\
Canada & CAN & Yes & Yes & Yes & No & No & No & No & 3.00\% \\
Central African Republic & CAF & No & No & No & Yes & Yes & Yes & Yes & 5.00\% \\
Chile & CHL & No & No & No & Yes & Yes & Yes & Yes & 3.00\% \\
China & CHN & No & No & No & Yes & Yes & Yes & Yes & 3.00\% \\
Colombia & COL & No & No & No & Yes & Yes & Yes & Yes & 3.00\% \\
Comoros & COM & No & No & No & Yes & Yes & Yes & Yes & 6.67\% \\
Congo, Rep. & COG & No & No & No & Yes & Yes & Yes & Yes & 5.67\% \\
Costa Rica & CRI & No & No & No & Yes & Yes & Yes & Yes & 3.00\% \\
Cote d'Ivoire & CIV & No & No & No & Yes & Yes & Yes & Yes & 8.00\% \\
Croatia & HRV & Yes & Yes & Yes & No & No & No & No & 3.00\% \\
Cyprus & CYP & Yes & Yes & No & No & No & Yes & Yes & 3.00\% \\
Czechia & CZE & Yes & Yes & Yes & No & No & No & No & 3.00\% \\
Denmark & DNK & Yes & Yes & Yes & No & No & No & No & 3.00\% \\
Dominican Republic & DOM & No & No & No & Yes & Yes & Yes & Yes & 3.33\% \\
Ecuador & ECU & No & No & No & Yes & Yes & Yes & Yes & 3.00\% \\
Egypt, Arab Rep. & EGY & No & No & No & Yes & Yes & Yes & Yes & 3.00\% \\
El Salvador & SLV & No & No & No & Yes & Yes & Yes & Yes & 3.00\% \\
Estonia & EST & Yes & Yes & Yes & No & No & No & No & 3.00\% \\
Fiji & FJI & No & No & No & Yes & Yes & Yes & Yes & 4.00\% \\
Finland & FIN & Yes & Yes & Yes & No & No & No & No & 3.00\% \\
France & FRA & Yes & Yes & Yes & No & No & No & No & 3.00\% \\
Gabon & GAB & No & No & No & Yes & Yes & Yes & Yes & 3.33\% \\
Gambia, The & GMB & No & No & No & Yes & Yes & Yes & Yes & 4.33\% \\
Georgia & GEO & No & No & No & Yes & Yes & Yes & Yes & 3.00\% \\
Germany & DEU & Yes & Yes & Yes & No & No & No & No & 3.00\% \\
Ghana & GHA & No & No & No & Yes & Yes & Yes & Yes & 4.00\% \\
Greece & GRC & Yes & Yes & Yes & No & No & No & No & 3.00\% \\
Guatemala & GTM & No & No & No & Yes & Yes & Yes & Yes & 3.00\% \\
Guinea & GIN & No & No & No & Yes & Yes & Yes & Yes & 10.33\% \\
Honduras & HND & No & No & No & Yes & Yes & Yes & Yes & 4.00\% \\
Hong Kong SAR, China & HKG & No & No & No & Yes & Yes & Yes & Yes & 3.00\% \\
Hungary & HUN & Yes & Yes & No & No & No & Yes & Yes & 3.00\% \\
Iceland & ISL & Yes & Yes & No & No & No & Yes & Yes & 3.00\% \\
India & IND & No & No & No & Yes & Yes & Yes & Yes & 3.00\% \\
Indonesia & IDN & No & No & No & Yes & Yes & Yes & Yes & 3.00\% \\
Iran, Islamic Rep. & IRN & No & No & No & Yes & Yes & Yes & Yes & 5.00\% \\
Ireland & IRL & Yes & Yes & No & No & No & Yes & Yes & 3.00\% \\
Israel & ISR & No & No & No & Yes & Yes & Yes & Yes & 4.33\% \\
Italy & ITA & Yes & Yes & Yes & No & No & No & No & 3.00\% \\
Japan & JPN & Yes & Yes & No & No & No & Yes & Yes & 3.00\% \\
Kazakhstan & KAZ & No & No & No & Yes & Yes & Yes & Yes & 3.00\% \\
Kenya & KEN & No & No & No & Yes & Yes & Yes & Yes & 4.00\% \\
Kiribati & KIR & No & No & No & Yes & Yes & Yes & Yes & 15.33\% \\
Korea, Rep. & KOR & Yes & Yes & No & No & No & Yes & Yes & 3.00\% \\
Kuwait & KWT & No & No & No & Yes & Yes & Yes & Yes & 7.67\% \\
Kyrgyz Republic & KGZ & No & No & No & Yes & Yes & Yes & Yes & 3.00\% \\
Latvia & LVA & Yes & Yes & Yes & No & No & No & No & 3.00\% \\
Lithuania & LTU & Yes & Yes & Yes & No & No & No & No & 3.00\% \\
Luxembourg & LUX & Yes & Yes & Yes & No & No & No & No & 3.00\% \\
Macao SAR, China & MAC & No & No & No & Yes & Yes & Yes & Yes & 4.00\% \\
Madagascar & MDG & No & No & No & Yes & Yes & Yes & Yes & 3.00\% \\
Malaysia & MYS & No & No & No & Yes & Yes & Yes & Yes & 3.00\% \\
Mali & MLI & No & No & No & Yes & Yes & Yes & Yes & 4.67\% \\
Malta & MLT & Yes & Yes & No & No & No & Yes & Yes & 3.00\% \\
Mauritania & MRT & No & No & No & Yes & Yes & Yes & Yes & 6.67\% \\
Mauritius & MUS & No & No & No & Yes & Yes & Yes & Yes & 3.00\% \\
Mexico & MEX & No & No & No & Yes & Yes & Yes & Yes & 3.00\% \\
Moldova & MDA & Yes & Yes & No & No & No & Yes & Yes & 3.00\% \\
Mongolia & MNG & No & No & No & Yes & Yes & Yes & Yes & 5.33\% \\
Morocco & MAR & No & No & No & Yes & Yes & Yes & Yes & 3.00\% \\
Namibia & NAM & No & No & No & Yes & Yes & Yes & Yes & 4.67\% \\
Nepal & NPL & No & No & No & Yes & Yes & Yes & Yes & 7.33\% \\
Netherlands & NLD & Yes & Yes & Yes & No & No & No & No & 3.00\% \\
New Zealand & NZL & Yes & Yes & No & No & No & Yes & Yes & 3.00\% \\
Nicaragua & NIC & No & No & No & Yes & Yes & Yes & Yes & 5.00\% \\
Niger & NER & No & No & No & Yes & Yes & Yes & Yes & 3.00\% \\
North Macedonia & MKD & Yes & Yes & Yes & No & No & No & No & 3.00\% \\
Norway & NOR & Yes & Yes & Yes & No & No & No & No & 3.00\% \\
Oman & OMN & No & No & No & Yes & Yes & Yes & Yes & 3.33\% \\
Pakistan & PAK & No & No & No & Yes & Yes & Yes & Yes & 3.00\% \\
Panama & PAN & No & No & No & Yes & Yes & Yes & Yes & 3.00\% \\
Paraguay & PRY & No & No & No & Yes & Yes & Yes & Yes & 3.00\% \\
Peru & PER & No & No & No & Yes & Yes & Yes & Yes & 3.00\% \\
Philippines & PHL & No & No & No & Yes & Yes & Yes & Yes & 3.00\% \\
Poland & POL & Yes & Yes & Yes & No & No & No & No & 3.00\% \\
Portugal & PRT & Yes & Yes & Yes & No & No & No & No & 3.00\% \\
Romania & ROU & Yes & Yes & Yes & No & No & No & No & 3.00\% \\
Russian Federation & RUS & No & No & No & No & No & No & No & 4.00\% \\
Rwanda & RWA & No & No & No & Yes & Yes & Yes & Yes & 4.00\% \\
Saudi Arabia & SAU & No & No & No & Yes & Yes & Yes & Yes & 3.00\% \\
Senegal & SEN & No & No & No & Yes & Yes & Yes & Yes & 3.00\% \\
Seychelles & SYC & No & No & No & Yes & Yes & Yes & Yes & 3.33\% \\
Singapore & SGP & Yes & Yes & No & No & No & Yes & Yes & 3.00\% \\
Slovak Republic & SVK & Yes & Yes & Yes & No & No & No & No & 3.00\% \\
Slovenia & SVN & Yes & Yes & Yes & No & No & No & No & 6.33\% \\
Solomon Islands & SLB & No & No & No & Yes & Yes & Yes & Yes & 8.00\% \\
South Africa & ZAF & No & No & No & Yes & Yes & Yes & Yes & 3.00\% \\
Spain & ESP & Yes & Yes & Yes & No & No & No & No & 3.00\% \\
Sri Lanka & LKA & No & No & No & Yes & Yes & Yes & Yes & 8.33\% \\
Sweden & SWE & Yes & Yes & Yes & No & No & No & No & 3.00\% \\
Switzerland & CHE & Yes & Yes & No & No & No & Yes & Yes & 3.00\% \\
Tanzania & TZA & No & No & No & Yes & Yes & Yes & Yes & 3.00\% \\
Thailand & THA & No & No & No & Yes & Yes & Yes & Yes & 3.00\% \\
Togo & TGO & No & No & No & Yes & Yes & Yes & Yes & 4.33\% \\
Tonga & TON & No & No & No & Yes & Yes & Yes & Yes & 6.33\% \\
Tunisia & TUN & No & No & No & Yes & Yes & Yes & Yes & 3.00\% \\
Turkiye & TUR & No & No & Yes & Yes & Yes & No & No & 3.00\% \\
Uganda & UGA & No & No & No & Yes & Yes & Yes & Yes & 3.00\% \\
Ukraine & UKR & Yes & No & No & No & No & No & No & 3.00\% \\
United Arab Emirates & ARE & No & No & No & Yes & Yes & Yes & Yes & 15.00\% \\
United Kingdom & GBR & Yes & Yes & Yes & No & No & No & No & 3.00\% \\
United States & USA & Yes & Yes & Yes & No & No & No & No & 3.00\% \\
Uruguay & URY & No & No & No & Yes & Yes & Yes & Yes & 3.00\% \\
Viet Nam & VNM & No & No & No & Yes & Yes & Yes & Yes & 3.33\% \\
West Bank and Gaza & PSE & No & No & No & Yes & Yes & Yes & Yes & 3.33\% \\
\end{longtable}
\endgroup
	
	\FloatBarrier
	
	\subsection*{S.A.4 Pretreatment Composition of the Estimation Groups}
	
	Table~\ref{tab:A7_characteristics} describes the pretreatment composition of
	the resulting target and donor groups. These comparisons are descriptive:
	they are not used to select donors or to claim balance in raw country
	characteristics.
	
	\begin{table}[!htbp]
		\centering
		\caption{Pre-treatment characteristics by estimation group, 2009--2013}
		\label{tab:A7_characteristics}
		\label{tab:A8_characteristics}
		\begin{adjustbox}{max width=\textwidth}
			\begin{threeparttable}
				\small
				\begin{tabular}{l c c c c}
					\toprule
					\textit{Variable} 
					& \makecell{\textit{Main sanctions}\\\textit{targets}} 
					& \makecell{\textit{Sanctions}\\\textit{CPI donors}} 
					& \makecell{\textit{Direct-arming}\\\textit{targets}} 
					& \makecell{\textit{Direct-arming}\\\textit{CPI donors}} \\
					\midrule
					Log CPI & 4.633 (0.011) & 4.662 (0.041) & 4.635 (0.011) & 4.658 (0.040) \\
					Log real GDP per capita & 10.061 (0.893) & 8.184 (1.209) & 10.084 (0.787) & 8.430 (1.354) \\
					Inflation (\%) & 2.159 (1.220) & 5.171 (3.656) & 2.282 (1.253) & 4.765 (3.607) \\
					Real GDPpc growth (\%) & 0.104 (1.870) & 2.307 (2.475) & -0.140 (1.816) & 2.087 (2.481) \\
					Oil rents & 0.383 (1.023) & 5.459 (11.235) & 0.448 (1.171) & 4.808 (10.645) \\
					Trade openness & 108.069 (65.478) & 82.342 (49.703) & 97.250 (54.036) & 89.079 (57.679) \\
					Terms of trade & 100.645 (6.922) & 105.895 (18.932) & 100.937 (7.451) & 105.169 (17.919) \\
					Investment rate & 21.760 (3.540) & 24.471 (7.731) & 21.440 (3.050) & 24.223 (7.431) \\
					Government consumption & 19.577 (3.814) & 14.828 (6.880) & 20.254 (3.234) & 15.180 (6.684) \\
					Rule of law & 1.029 (0.712) & -0.361 (0.629) & 1.000 (0.690) & -0.165 (0.821) \\
					Population growth & 0.270 (0.891) & 1.861 (1.263) & 0.257 (0.858) & 1.688 (1.297) \\
					Fuel-share balance & 6.336 (13.333) & -3.635 (34.432) & 3.716 (14.049) & -1.601 (32.810) \\
					\bottomrule
				\end{tabular}
				\begin{tablenotes}[flushleft]
					\footnotesize
					\item \textit{Notes:} Entries are unweighted means of country-level 2009--2013 averages, with cross-country standard deviations in parentheses. Groups overlap across policy definitions. This descriptive table carries no significance stars.
				\end{tablenotes}
			\end{threeparttable}
		\end{adjustbox}
	\end{table}
	
	\FloatBarrier
	
	\subsection*{S.A.5 Construction and Coverage of Exposure to Russian Energy}
	
	The exposure analysis uses additional trade data and therefore has its own
	coverage and completion rules. It is documented separately from the
 129-country macroeconomic panel so that Comtrade coverage restrictions are not
	confused with the eligibility rules used for the baseline estimates.
	
	\paragraph{Source series and construction universe.}
	We construct the exposure measures from UN Comtrade for the same 129 countries used in the macroeconomic panel. The two four-year windows, 2010--2013 and 2018--2021, produce 1,032 potential country-year observations. Russia, Belarus, and Ukraine do not enter the 126-country mechanism universe. Nominal GDP is taken from the World Development Indicators and is positive in all 1,032 cells.
	
	We query three trade blocks separately:(i)  imports from Russia of coal, crude oil, and refined petroleum; (ii) imports from Russia of liquefied and gaseous natural gas; (iii) and total imports of the same gas products from all suppliers. Keeping these requests separate prevents observed oil trade from being mistaken for evidence of gas coverage.
	
	\paragraph{Distinguishing zero trade from missing data.}
	A positive value reported by the importing country is retained. If a country is missing from a multi-country request, we repeat the request for that country alone. We classify an empty result as a verified zero only when the individual request succeeds and Comtrade confirms that the country reported HS trade data in that year. Otherwise, the observation remains missing. Thus, an absent API row is not automatically treated as zero trade.
	
	\paragraph{Mirror flows and construction of the gas denominator.}
	For gas, a positive exporter-mirror value---that is, trade reported by Russia rather than by the importing country---may recover a missing bilateral flow. Importer-reported positives always have priority. The available Russian exporter data contain LNG observations but no positive pipeline-gas observations. They therefore cannot provide complete coverage for Austria, Germany, Poland, or Turkiye. We exclude the affected country-windows instead of assigning a false zero.
	
	The gas-share denominator normally uses the importer's reported total gas imports. A targeted sum of exporter reports is used only when that total is missing, non-positive despite a positive Russian flow, or smaller than a mirror-recovered Russian numerator. We retain a country-year only when the Russian numerator is between zero and the world total.
	
	\paragraph{Within-window completion.}
	Missing trade values are never borrowed from another country and no
	interpolation crosses the 2013--2018 gap. Completion is applied separately to
	Russian gas, world gas, and Russian non-gas energy within each four-year
	window. An internal gap of no more than two consecutive years is interpolated
	linearly in $\log(1+x)$. A one-year gap at the beginning or end of a window can
	be extrapolated only when at least three anchors are available. The
	extrapolated log growth rate is the median adjacent growth rate in the
	country-window, bounded by the 5th and 95th percentiles of observed Comtrade
	growth rates. Values are constrained to be non-negative. Russian gas is never
	imputed for Austria, Germany, Poland, or Turkiye.
	
	The completion log contains 37 primitive component imputations: 13 for
	Russian gas, 13 for world gas, and 11 for Russian non-gas energy. Ratios are
	not themselves interpolated. They are recomputed from the completed monetary
	components so that numerator and denominator can evolve separately.
	
	\paragraph{Exposure measures, standardization, and final coverage.}
	The two reported exposures are ratios of sums:
	\begin{align}
		\mathrm{GasShare}_{iw}
		&=
		\frac{\sum_{t\in w}\mathrm{RussianGas}_{it}}
		{\sum_{t\in w}\mathrm{WorldGas}_{it}},
		\label{eq:A_gas_share}\\
		\mathrm{EnergyGDP}_{iw}
		&=
		\frac{\sum_{t\in w}
			\left(\mathrm{RussianNonGas}_{it}+\mathrm{RussianGas}_{it}\right)}
		{\sum_{t\in w}\mathrm{GDP}_{it}}.
		\label{eq:A_energy_gdp}
	\end{align}
	The strict sample requires at least three usable years out of four and no more
	than one imputed year for each component needed by the ratio. The extended
	sample requires at least two usable years and allows no more than two imputed
	years per component. Both rules exclude the four priority pipeline countries.
	Within each window, the completed ratio is standardized among the countries
	that satisfy the applicable sample rule and belong to the main mechanism
	universe.
	
	It is useful to distinguish the \emph{construction sample} from the
	\emph{regression sample}. Construction counts are calculated on the full
	129-country exposure grid; regression counts additionally intersect those
	country-windows with the 126-country mechanism universe. Russia already fails
	the exposure-coverage rule, while Belarus and Ukraine are removed by mechanism
	eligibility, so the regression counts are two below the corresponding
	construction counts. Under the strict rule, the energy-imports/GDP regressions
	contain 119 countries for 2010--2013 and 118 for 2018--2021, with 39
	sanctioners in each. Relative to the 42-country baseline sanctions group,
	Austria, Germany, and Poland are unavailable because Comtrade cannot certify
	their pipeline-gas component. Turkiye is screened for the same measurement
	reason but is not a strict sanctioner.
	
	The extended rule does not change every sample. For energy imports/GDP in
	2018--2021, Strict and Extended retain the same 120 construction countries and
	118 regression countries. For 2010--2013, Extended adds Kenya, increasing the
	construction sample from 121 to 122 and the regression sample from 119 to 120.
	For the gas share, Extended likewise adds Kenya in 2010--2013 and Bangladesh
	in 2018--2021. Other exclusions follow mechanically from insufficient usable
	years or unresolved trade pairs.
	\begin{table}[!htbp]
		\centering
		\caption{Construction and coverage of the Comtrade-only exposure measures}
		\label{tab:A8_energy_exposure_construction}
		\label{tab:A9_energy_exposure_construction}
		\begin{adjustbox}{max width=\textwidth}
			\begin{threeparttable}
				\small
				\begin{tabular}{l l l r r r}
					\toprule
					\multicolumn{6}{l}{\textbf{Panel A. Annual cell audit, 2010--2013 and 2018--2021}} \\
					\addlinespace
					\multicolumn{2}{l}{\textit{Series}} & \textit{Positive} & \makecell[r]{\textit{Verified}\\\textit{zero}} & \textit{Missing} & \makecell[r]{\textit{Imputed}\\\textit{cells}} \\
					\midrule
					\multicolumn{2}{l}{Russian gas, screened} & 229 & 733 & 70 & 13 \\
					\multicolumn{2}{l}{World gas, screened} & 809 & 185 & 38 & 13 \\
					\multicolumn{2}{l}{Russian non-gas fossil energy} & 769 & 224 & 39 & 11 \\
					\multicolumn{2}{l}{Russian total fossil energy, screened} & 738 & 223 & 71 & -- \\
					\midrule
					\multicolumn{6}{l}{\textbf{Panel B. Window-level construction and regression sample accounting}} \\
					\addlinespace
					\makecell[l]{\textit{Exposure}\\\textit{window}} & \textit{Measure} & \textit{Rule} & \makecell[r]{\textit{Construction}\\\textit{countries}} & \makecell[r]{\textit{Regression}\\\textit{countries}} & \makecell[r]{\textit{Strict}\\\textit{sanctioners}} \\
					\midrule
					2010--2013 & Russian gas share & Strict & 121 & 119 & 39 \\
					           &                   & Extended & 122 & 120 & 39 \\
					2010--2013 & Russian energy imports / GDP & Strict & 121 & 119 & 39 \\
					           &                              & Extended & 122 & 120 & 39 \\
					2018--2021 & Russian gas share & Strict & 120 & 118 & 39 \\
					           &                   & Extended & 121 & 119 & 39 \\
					2018--2021 & Russian energy imports / GDP & Strict & 120 & 118 & 39 \\
					           &                              & Extended & 120 & 118 & 39 \\
					\bottomrule
				\end{tabular}
				\begin{tablenotes}[flushleft]
					\footnotesize
					\item \textit{Notes:} Panel A covers 1,032 potential country-year cells. Positive, zero, and missing are mutually exclusive statuses in the screened pre-completion series. The imputation column is a separate audit count and can end as either a positive or a zero in the completed series; the total-energy series is constructed from its gas and non-gas components, so no separate primitive imputation count is reported. In Panel B, construction counts refer to the full 129-country exposure files, whereas regression counts intersect each construction sample with the 126-country mechanism universe. Russia already fails the exposure-coverage rule; Belarus and Ukraine account for the two-country difference between construction and regression counts. The strict rule requires at least three usable years and at most one imputed year per required component; the extended rule requires at least two usable years and permits at most two. Austria, Germany, Poland, and Turkiye are excluded under both rules because of unresolved pipeline-gas undercoverage. Extended adds Kenya in both 2010--2013 measures and Bangladesh only for the 2018--2021 gas-share measure; it adds no country for 2018--2021 energy imports/GDP.
				\end{tablenotes}
			\end{threeparttable}
		\end{adjustbox}
	\end{table}
	
	\clearpage
	\phantomsection
	\addcontentsline{toc}{section}{Supplementary Appendix B: CCEDID Implementation and Robustness Checks}
	\section*{Supplementary Appendix B\\CCEDID Implementation and Robustness Checks}
	\label{sec:appendix_B}
	\setcounter{table}{0}
	\renewcommand{\thetable}{B\arabic{table}}
	\renewcommand{\theHtable}{B\arabic{table}}
	\setcounter{figure}{0}
	\renewcommand{\thefigure}{B\arabic{figure}}
	\renewcommand{\theHfigure}{B\arabic{figure}}
	\setcounter{equation}{0}
	\renewcommand{\theequation}{B\arabic{equation}}
	\renewcommand{\theHequation}{B\arabic{equation}}
	\small
	
	This appendix states the exact computational form of the estimator and documents the additional diagnostic checks. These checks assess sensitivity to alternative modeling and resampling choices; they do not verify the identifying assumption. The corresponding intuition is provided in Section~\ref{sec:econometric_strategy}.
	
\begin{figure}[!htbp]
		\centering
		\includegraphics[width=\textwidth]{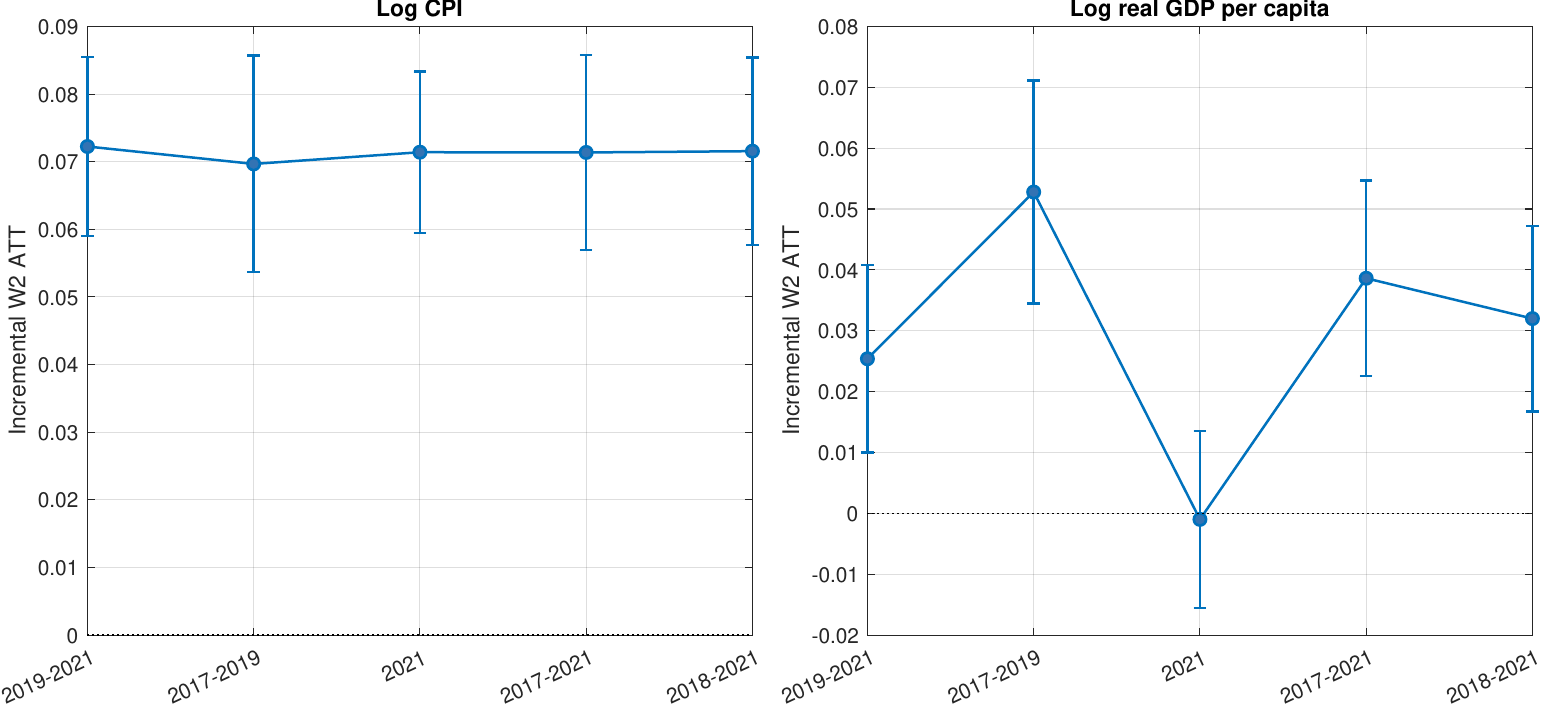}
		\caption{Sanctions estimates across alternative Wave~2 reference windows}
		\label{fig:B3_reference_windows}
		\begin{minipage}{0.96\textwidth}\footnotesize
			\textit{Notes:} Wave~2 is fixed at 2022--2024; only the subtracted reference interval changes. Error bars use the country-dispersion standard error based on one complete contrast per target country.
		\end{minipage}
	\end{figure}

	The estimator is run separately for each policy--outcome combination; as in the main methods, the notation below describes one generic run. Tables~\ref{tab:B1_placebo} and~\ref{tab:B3_counterfactual}--\ref{tab:B7_secondary} retain the numerical results needed in the paper, while Figures~\ref{fig:B1_world}--\ref{fig:B3_reference_windows} display the worldwide distribution, placebo distributions, and reference-window estimates.

	\subsection*{S.B.1 Exact Computational Form of the CCEDID Estimator}
	
	This section gives the exact computational form of the estimator summarized in the main text. All quantities are constructed separately for each policy--outcome pair.
	
	Let $\mathcal T^*$ denote the final target set and let $\mathcal C_Y$ denote the eligible donor set for outcome $Y$. Let $\mathcal S_Y$ be the set of target and donor countries included in the pooled pre-treatment regression. The effective pre-treatment period is
	\[
	\mathcal P=\{2001,\ldots,2013\},
	\qquad
	T_0=|\mathcal P|=13.
	\]
	For country $i$, let $Y_i$ be the $T_0\times1$ vector that stacks $Y_{it}$ over $t\in\mathcal P$, and let $X_i$ be the $T_0\times K$ matrix whose row for year $t$ is $X_{i,t-1}'$. In the preferred specification, $K=8$.
	
	The common-factor proxy contains a constant and the contemporaneous mean of the outcome among eligible donors:
	\begin{equation}
		\overline Y_{t,\mathcal C_Y}
		=
		\frac{1}{|\mathcal C_Y|}
		\sum_{j\in\mathcal C_Y}Y_{jt},
		\qquad
		\widehat f_t
		=
		\begin{pmatrix}
			1\\[2pt]
			\overline Y_{t,\mathcal C_Y}
		\end{pmatrix}.
		\label{eq:B_factor_proxy}
	\end{equation}
	Stacking the proxy over the pre-treatment period gives the $T_0\times2$ matrix
	\begin{equation}
		\widehat F
		=
		\begin{bmatrix}
			\widehat f_{2001}'\\[2pt]
			\vdots\\[2pt]
			\widehat f_{2013}'
		\end{bmatrix}.
		\label{eq:B_factor_matrix}
	\end{equation}
	Within a given policy--outcome estimation, the same matrix $\widehat F$ is used for every country.
	
	Let $(\cdot)^+$ denote the Moore--Penrose inverse. The residual-maker for the donor proxy is
	\begin{equation}
		M_{\widehat F}
		=
		I_{T_0}-\widehat F\widehat F^+,
		\qquad
		\widetilde Y_i=M_{\widehat F}Y_i,
		\qquad
		\widetilde X_i=M_{\widehat F}X_i.
		\label{eq:B_factor_residualization}
	\end{equation}
	Thus, within each country, both the outcome and the lagged controls are residualized with respect to the constant and the donor mean. When $\widehat F$ has full column rank,
	\[
	\widehat F^+=(\widehat F'\widehat F)^{-1}\widehat F',
	\]
	so equation~\eqref{eq:B_factor_residualization} reduces to the ordinary-inverse expression reported in the main text.
	
	The common control coefficient is estimated from the pooled factor-adjusted pre-treatment observations:
	\begin{equation}
		\widehat\beta
		=
		\left(
		\sum_{i\in\mathcal S_Y}
		\widetilde X_i'\widetilde X_i
		\right)^+
		\left(
		\sum_{i\in\mathcal S_Y}
		\widetilde X_i'\widetilde Y_i
		\right).
		\label{eq:B_beta}
	\end{equation}
	The coefficient vector $\beta$ is common within a policy--outcome estimation, whereas exposure to the donor proxy is country-specific. In the implementation, any control column with negligible residual pre-treatment variance is omitted when equation~\eqref{eq:B_beta} is estimated and is assigned a coefficient of zero. This prevents a numerically unidentified column from destabilizing the fit.
	
	For each country for which a counterfactual is required, and in particular for each $i\in\mathcal T^*$, the pre-treatment data are used to estimate a proxy-to-covariate coefficient matrix and an outcome-loading vector:
	\begin{align}
		\widehat\Lambda_i
		&=\widehat F^+X_i,
		\label{eq:B_lambda}\\
		\widehat\gamma_i
		&=\widehat F^+
		\left(
		Y_i-X_i\widehat\beta
		\right).
		\label{eq:B_gamma}
	\end{align}
	Here, $\widehat\Lambda_i$ is a $2\times K$ matrix and $\widehat\gamma_i$ is a $2\times1$ vector. Under full column rank of $\widehat F$, these expressions are equivalent to
	\[
	\widehat\Lambda_i=(\widehat F'\widehat F)^{-1}\widehat F'X_i,
	\qquad
	\widehat\gamma_i=(\widehat F'\widehat F)^{-1}\widehat F'
	\left(Y_i-X_i\widehat\beta\right).
	\]
	
	For each post-2013 year, the untreated covariate path is obtained by applying the target country's pre-treatment coefficients to the contemporaneous donor proxy:
	\begin{equation}
		\widehat X_{i,t-1}(\infty)
		=
		\widehat\Lambda_i'\widehat f_t,
		\qquad
		t\geq2014.
		\label{eq:B_xhat}
	\end{equation}
	The corresponding untreated outcome is
	\begin{equation}
		\widehat Y_{it}(\infty)
		=
		\widehat X_{i,t-1}(\infty)'\widehat\beta
		+
		\widehat\gamma_i'\widehat f_t,
		\qquad
		t\geq2014.
		\label{eq:B_yhat}
	\end{equation}
	Observed post-2013 controls for the target country are therefore not inserted into the counterfactual. Instead, their untreated components are predicted from the donor proxy and the target country's pre-treatment relationship with that proxy.
	
	The estimated country-year gap is
	\begin{equation}
		\widehat g_{it}
		=
		Y_{it}-\widehat Y_{it}(\infty).
		\label{eq:B_gap}
	\end{equation}
	For any reporting window $\mathcal W$, the estimator first forms one time-averaged contrast for each target country:
	\begin{equation}
		\overline g_{i,\mathcal W}
		=
		\frac{1}{|\mathcal W|}
		\sum_{t\in\mathcal W}\widehat g_{it}.
		\label{eq:B_country_window_gap}
	\end{equation}
	It then assigns equal weight to the country-specific contrasts:
	\begin{equation}
		\widehat{\mathrm{ATT}}_{\mathcal W}
		=
		\frac{1}{N_T}
		\sum_{i\in\mathcal T^*}
		\overline g_{i,\mathcal W},
		\qquad
		N_T=|\mathcal T^*|.
		\label{eq:B_wave_att}
	\end{equation}
	Thus, countries receive equal weight regardless of population or economic size.
	
	For the incremental Wave~2 estimand, let
	\[
	\mathcal R=\{2019,2020,2021\}
	\]
	denote the preferred late-Wave-1 reference window. The country-specific incremental contrast is
	\begin{equation}
		d_{i,\mathrm{Inc},2}
		=
		\overline g_{i,\mathcal W_2}
		-
		\overline g_{i,\mathcal R},
		\label{eq:B_incremental_country}
	\end{equation}
	and its equally weighted coalition average is
	\begin{equation}
		\widehat{\mathrm{ATT}}_{\mathrm{Inc},2}
		=
		\frac{1}{N_T}
		\sum_{i\in\mathcal T^*}
		d_{i,\mathrm{Inc},2}
		=
		\frac{1}{N_T}
		\sum_{i\in\mathcal T^*}
		\left(
		\overline g_{i,\mathcal W_2}
		-
		\overline g_{i,\mathcal R}
		\right).
		\label{eq:B_incremental}
	\end{equation}
	The implementation requires every included target country to have complete observations for the relevant reporting and reference windows; it does not average different numbers of years across countries. For an estimate $\widehat\theta$ expressed in log points, the exact percentage transformation is $100[\exp(\widehat\theta)-1]$.
	
	\subsection*{S.B.2 Identification Conditions}
	
	Let $\mathbf D_i^{obs}$ denote country $i$'s observed policy history and let $\mathbf D_{-i}^{obs}$ denote the observed policy histories of all other countries. Write $Y_{it}(\mathbf d_i;\mathbf d_{-i})$ for the potential outcome under those policy histories, and use $\infty$ to denote non-participation by country $i$. The causal object associated with the country-level gap is
	\begin{equation}
		\tau_{it}^{part}
		=
		Y_{it}(\mathbf D_i^{obs};\mathbf D_{-i}^{obs})
		-
		Y_{it}(\infty;\mathbf D_{-i}^{obs}).
		\label{eq:B_participation_estimand}
	\end{equation}
	It is the effect of country $i$'s participation under the global policy environment that was actually observed. Averaging equation~\eqref{eq:B_participation_estimand} over the target countries and reporting windows produces the corresponding coalition estimands. This object is distinct from the general-equilibrium effect of moving the entire world from the observed sanctions regime to a regime in which no country sanctions Russia.
	
	\paragraph{Consistency and treatment versions.}
	Consistency requires the observed outcome to equal the potential outcome under the policy history actually followed:
	\begin{equation}
		Y_{it}^{obs}
		=
		Y_{it}(\mathbf D_i^{obs};\mathbf D_{-i}^{obs}).
		\label{eq:B_consistency}
	\end{equation}
	The strict-sanctions indicator identifies participation in the coalition; it does not imply that every country adopted an identical legal package. The ATT therefore averages the effects of the country-specific policy versions actually implemented by coalition members. Consistency supports this interpretation as long as the coding records the relevant participation history. It would not justify interpreting the estimate as the effect of one uniform legal instrument applied identically across countries.
	
	In this application, consistency is plausible at the level of coalition participation because positive sanctions codes are tied to official legal coverage or a formal alignment notice, and every entry year is documented in the treatment ledger summarized in Appendix Table~\ref{tab:A4_treatment}. The core policy version is also relatively well defined for the many European countries covered by common EU measures. The condition is less credible if treatment is understood as a common dose: enforcement, exemptions, exposure to Russian countermeasures, and domestic mitigation differed across countries. For that reason, the estimates are interpreted as an average over the policy versions actually adopted, not as the effect of a standardized sanctions package.
	
	\paragraph{No anticipation.}
	For the pre-2014 observations to identify the untreated relationship, future participation must not materially affect outcomes or controls during the estimation period $\mathcal P$:
	\begin{equation}
		Y_{it}(\mathbf D_i^{obs};\mathbf D_{-i}^{obs})
		=
		Y_{it}(\infty;\mathbf D_{-i}^{obs}),
		\qquad
		X_{it}(\mathbf D_i^{obs};\mathbf D_{-i}^{obs})
		=
		X_{it}(\infty;\mathbf D_{-i}^{obs}),
		\qquad t\in\mathcal P.
		\label{eq:B_no_anticipation}
	\end{equation}
	The incremental Wave~2 contrast requires a related but narrower condition. Because 2019--2021 is already part of the first sanctions regime, it need not be untreated. Instead, it must not contain a material response to the subsequent 2022 escalation that determines the estimated change in gaps. The alternative reference-window results assess sensitivity to possible anticipatory adjustment in 2021 but cannot rule it out directly.
	
	The timing makes this condition reasonably plausible for Wave~1. The estimation sample ends in 2013, before the sanctions adopted after the annexation of Crimea, and most of the 2001--2013 period is far removed from that policy change. The main concern is the end of 2013, when the political crisis in Ukraine had already begun and some economic actors may have adjusted before formal sanctions were imposed. Anticipation is more plausible before Wave~2 because Russia's military build-up was visible during 2021. Reassuringly, the CPI estimate changes little when the reference excludes 2021 or is restricted to 2021 alone (Appendix Table~\ref{tab:B6_windows}); this suggests that the result is not generated by one particular choice of reference period, although it does not prove that all anticipatory responses were absent.
	
	\paragraph{Validity of the donor factor.}
	Let
	\begin{equation}
		q_t=
		\begin{pmatrix}
			1\\
			\overline Y_{t,\mathcal C}
		\end{pmatrix},
		\qquad
		\overline Y_{t,\mathcal C}
		=
		\frac{1}{|\mathcal C^Y|}
		\sum_{j\in\mathcal C^Y}Y_{jt}^{obs},
		\label{eq:B_donor_factor_validity}
	\end{equation}
	denote the preferred donor-factor proxy. If untreated outcomes depend on latent common shocks $F_t$, validity requires $q_t$ to span the component of those shocks that is relevant for the target countries. Formally, after reparameterization, the relevant common component must admit a representation of the form $\lambda_i'F_t=q_t'\alpha_i$, up to an idiosyncratic error with conditional mean zero. The condition does not require donors to be insulated from the war, global energy prices, pandemic recovery, or monetary tightening. It requires their average outcome to trace the common movements that would also have affected coalition members without their own participation.
	
	The donor countries must remain untreated under the policy definition, and policy-related spillovers must not make their mean a systematically different object from the common counterfactual component. A large donor pool reduces sampling noise in $\overline Y_{t,\mathcal C}$, while the alternative donor pools, factor definitions, and donor leave-one-out exercises assess whether the estimates depend on a particular construction. These checks support factor relevance and stability, but they cannot establish that every relevant post-2013 shock is spanned by the preferred one-factor proxy.
	
	Several features make factor validity plausible for CPI. The preferred sanctions factor averages 84 countries, excludes the sanctioning coalition and the three direct war participants, and draws on economies from several regions. It should therefore capture broad movements such as pandemic recovery, commodity-price inflation, and global monetary tightening without mechanically importing the coalition's realized outcome into its counterfactual. The CPI contrast also remains positive across alternative donor pools, matched donors, richer factor proxies, and donor leave-one-out exercises (Appendix Tables~\ref{tab:B3_counterfactual}--\ref{tab:B5_influence}). The main remaining concern is that a worldwide mean may not span a specifically European energy or monetary shock. This concern is substantive because most target countries are advanced European economies, whereas many donors are not, and because the GDP-per-capita estimate is more sensitive to the choice of factor proxy. The evidence for factor validity is therefore stronger for CPI than for GDP per capita.
	
	\paragraph{Spillovers and the scope of the estimand.}
	Equation~\eqref{eq:B_participation_estimand} does not require complete absence of interference across countries. Policies adopted by the rest of the coalition, and their worldwide consequences, are held fixed in both potential outcomes. Russian countermeasures directed at country $i$, energy-market adjustments induced by its participation, and domestic mitigation policies may therefore form part of the total participation effect rather than separate confounders.
	
	Identification does require the donor factor used for country $i$'s counterfactual to be invariant, or approximately invariant, to switching that country's own participation status:
	\begin{equation}
		q_t(\mathbf D_i^{obs};\mathbf D_{-i}^{obs})
		\simeq
		q_t(\infty;\mathbf D_{-i}^{obs}).
		\label{eq:B_spillover_condition}
	\end{equation}
	This condition can fail if one country's participation materially changes donor outcomes through commodity prices, trade diversion, or other general-equilibrium channels. It can also fail when donors experience policy spillovers that do not represent the common path relevant for target countries. The design therefore identifies a participation effect conditional on the observed global regime; it does not identify the worldwide effect of eliminating the sanctions coalition itself.
	
	Approximate invariance is plausible for most individual coalition members because the target country's own outcome never enters its donor factor and that factor averages a large set of economies. Switching the participation status of one small or medium-sized country would be unlikely to move that worldwide average appreciably. The donor leave-one-out results further show that no single donor determines the estimated CPI contrast. The argument is weaker for large economies and for measures coordinated at the EU level: their participation may affect world energy prices, trade routes, or expectations, and an individual country's policy may not be meaningfully separable from the coalition's joint action. These possibilities are why the conditional estimand in equation~\eqref{eq:B_participation_estimand} is narrower than a total coalition effect.
	
	\paragraph{Stability of the untreated relationship.}
	The main counterfactual condition is that the relationships learned over $\mathcal P$ continue to describe the target country's untreated outcome and untreated controls after 2013:
	\begin{align}
		Y_{it}(\infty;\mathbf D_{-i}^{obs})
		&=
		X_{i,t-1}(\infty;\mathbf D_{-i}^{obs})'\beta
		+q_t'\alpha_i+\varepsilon_{it},
		\label{eq:B_stable_outcome}\\
		X_{i,t-1}(\infty;\mathbf D_{-i}^{obs})
		&=
		q_t'\Lambda_i+v_{i,t-1}.
		\label{eq:B_stable_controls}
	\end{align}
	The coefficients $\beta$, $\alpha_i$, and $\Lambda_i$ must remain informative after 2013, and the post-2013 disturbances $\varepsilon_{it}$ and $v_{i,t-1}$ must not contain a systematic coalition-specific component unrelated to participation. This allows the size and direction of global shocks to change sharply and allows countries to respond differently through $\alpha_i$ and $\Lambda_i$. It rules out an unmodeled structural break that changes the target country's relationship with the common component for reasons independent of its policy participation.
	
	The incremental contrast is less sensitive to a persistent country-specific counterfactual bias because the 2019--2021 gap is subtracted from the 2022--2024 gap. A time-invariant prediction error therefore cancels from the contrast. A prediction error that emerges or changes in 2022 does not. The alternative donor pools, factors, matched samples, reference windows, and influence checks provide evidence on the stability of the estimated CPI contrast. They cannot directly test the unobserved post-2013 counterfactual, and the greater sensitivity of the GDP-per-capita estimates indicates that the same causal weight should not automatically be assigned to every outcome.
	
	Stability is plausible here for three reasons. First, the 2001--2013 estimation period contains substantial international variation, including commodity-price movements, the global financial crisis, and the euro-area crisis, rather than only tranquil years. Second, country-specific factor loadings allow the same global movement to have different effects across countries, while the lagged controls capture observable differences in their macroeconomic structure. Third, the preferred CPI contrast is stable across donor definitions, factor proxies, matched samples, reference windows, and influence checks. None of this eliminates the central threat: the post-2022 break in European energy relations, together with unusually large fiscal and monetary responses, may have changed the coalition's relationship with global inflation for reasons that would have arisen even without an individual country's participation. Such a break is observationally difficult to separate from the participation effect. The robustness of the CPI result makes the stability assumption plausible, but not verifiable; the less stable GDP-per-capita results warrant a more cautious causal interpretation.

	\subsection*{S.B.3 Inferential Procedures}
	
	The paper reports one baseline test, two bootstrap intervals, and one placebo diagnostic. They are kept separate because each uses a different source of variation. The notation below describes the procedure for one policy, outcome, and reference interval $r$; the complete procedure is repeated for every reported design.
	
	\paragraph{Country-dispersion test.}
	Let $\mathcal R_r$ denote reference window $r$. The starting observations for
	inference are the complete target-country contrasts
	$d_i(r)=\overline g_{i,\mathcal W_2}-\overline g_{i,\mathcal R_r}$, not the
	underlying country-year observations. Their equally weighted coalition average
	is $\widehat\tau(r)=N_T^{-1}\sum_{i\in\mathcal T^*}d_i(r)$. Let
	\begin{equation}
		s_d^2(r)
		=
		\frac{1}{N_T-1}
		\sum_{i\in\mathcal T^*}
		\left[d_i(r)-\widehat\tau(r)\right]^2.
		\label{eq:B_country_variance}
	\end{equation}
	The reported standard error, test statistic, and two-sided probability are
	\begin{equation}
		\widehat{SE}_{CD}\{\widehat\tau(r)\}
		=\frac{s_d(r)}{\sqrt{N_T}},
		\qquad
		t_{CD}(r)=
		\frac{\widehat\tau(r)}{\widehat{SE}_{CD}\{\widehat\tau(r)\}},
		\qquad
		p_{CD}(r)=2\left[1-F_{t_{N_T-1}}\left(|t_{CD}(r)|\right)\right].
		\label{eq:B_country_test}
	\end{equation}
	Here $F_{t_{N_T-1}}$ is the Student $t$ cumulative distribution function with $N_T-1$ degrees of freedom. These probabilities, and only these probabilities, determine the stars in the tables. Collapsing each country's full path into one contrast avoids treating its annual observations as independent. The calculation nevertheless treats the target-country contrasts as independent across countries and conditions on the estimated donor factor and counterfactual paths. It therefore does not propagate all first-stage estimation uncertainty.
	
	\paragraph{Country-history bootstrap.}
	Let $N_C=|\mathcal C^Y|$ and $B=500$. In bootstrap draw $b$, the implementation samples $N_T$ target-country identifiers and $N_C$ donor-country identifiers independently and with replacement:
	\begin{equation}
		I^T_{\ell b}\overset{iid}{\sim}\operatorname{Unif}(\mathcal T^*),
		\quad \ell=1,\ldots,N_T,
		\qquad
		I^C_{\ell b}\overset{iid}{\sim}\operatorname{Unif}(\mathcal C^Y),
		\quad \ell=1,\ldots,N_C.
		\label{eq:B_country_bootstrap_draw}
	\end{equation}
	A selected identifier brings the country's entire 2000--2024 history into the draw. The code gives every selected copy a new synthetic identifier, so a country selected more than once enters the estimation with the same multiplicity. The draw-specific donor mean is therefore
	\begin{equation}
		\overline Y^{CH,b}_{t,\mathcal C}
		=
		\frac{1}{N_C}
		\sum_{\ell=1}^{N_C}Y_{I^C_{\ell b},t}.
		\label{eq:B_country_bootstrap_factor}
	\end{equation}
	Using this mean, the code reconstructs the factor and re-estimates $\widehat\beta$, the country-specific control and outcome loadings, all untreated control paths, all untreated outcome paths, and all target-country contrasts. The estimate stored for the draw is
	\begin{equation}
		\widehat\tau_b^{CH}(r)
		=
		\frac{1}{N_T}
		\sum_{\ell=1}^{N_T}
		d_{I^T_{\ell b}}^{CH,b}(r).
		\label{eq:B_country_bootstrap_att}
	\end{equation}
	Thus the procedure varies both sides of the comparison: which target histories contribute to the ATT and which donor histories construct the counterfactual. The same re-estimation produces all five reference-window contrasts in each draw. The engine requires every selected history to contain all 25 annual observations, every selected target copy to reach the final aggregation, and every required reporting window to be complete. After exactly 500 draws, the code verifies that every stored estimate is finite. A failed country-history draw is not silently discarded or replaced; it stops the run.
	
	\paragraph{World Bank region-block bootstrap.}
	Let $R_i\in\mathcal R=\{1,\ldots,G\}$ denote country $i$'s World Bank region, and define $\mathcal T_g=\{i\in\mathcal T^*:R_i=g\}$ and $\mathcal C_g^Y=\{j\in\mathcal C^Y:R_j=g\}$. A country without a usable region code is assigned to a common residual block, labelled \texttt{OTHER}. In draw $b$, the code samples $G$ region identifiers with replacement:
	\begin{equation}
		J^R_{\ell b}\overset{iid}{\sim}\operatorname{Unif}(\mathcal R),
		\quad \ell=1,\ldots,G,
		\qquad
		a_{gb}=\sum_{\ell=1}^{G}\mathbf 1\{J^R_{\ell b}=g\}.
		\label{eq:B_region_bootstrap_draw}
	\end{equation}
	Every selected region contributes the complete histories of all its target and donor countries. Repeated regions, like repeated countries in the country-history bootstrap, enter through distinct synthetic identifiers and therefore receive multiplicity $a_{gb}$. The resulting target and donor counts are
	\begin{equation}
		N_{T,b}^{RB}=\sum_{g=1}^{G}a_{gb}|\mathcal T_g|,
		\qquad
		N_{C,b}^{RB}=\sum_{g=1}^{G}a_{gb}|\mathcal C_g^Y|,
		\label{eq:B_region_bootstrap_counts}
	\end{equation}
	and the draw-specific donor mean is
	\begin{equation}
		\overline Y^{RB,b}_{t,\mathcal C}
		=
		\frac{
			\sum_{g=1}^{G}a_{gb}
			\sum_{j\in\mathcal C_g^Y}Y_{jt}
		}{N_{C,b}^{RB}}.
		\label{eq:B_region_bootstrap_factor}
	\end{equation}
	The factor and every subsequent estimation step are reconstructed from this block-resampled panel. The stored estimate is
	\begin{equation}
		\widehat\tau_b^{RB}(r)
		=
		\frac{
			\sum_{g=1}^{G}a_{gb}
			\sum_{i\in\mathcal T_g}d_i^{RB,b}(r)
		}{N_{T,b}^{RB}}.
		\label{eq:B_region_bootstrap_att}
	\end{equation}
	The code rejects a proposed region draw only when it contains no target histories or no donor histories. It then samples another set of regions and continues until 500 non-empty region draws have been estimated. Once a proposed draw reaches the estimator, the same completeness and finite-output assertions used above apply; a failure at that stage stops the run rather than being silently replaced. A hard attempt limit prevents indefinite resampling: 20$B$ proposals for the preserved sanctions--CPI run and 30$B$ for each of the other three policy--outcome runs. This procedure preserves dependence among countries in the same selected region. Because the number of World Bank regions is small, the resulting interval is reported as a sensitivity interval rather than as exact many-cluster inference.
	
	For $k\in\{CH,RB\}$, the reported interval is the unstudentized percentile interval
	\begin{equation}
		CI_{0.95}^{k}(r)
		=
		\left[
		Q_{0.025}\left(\{\widehat\tau_b^{k}(r)\}_{b=1}^{B}\right),
		Q_{0.975}\left(\{\widehat\tau_b^{k}(r)\}_{b=1}^{B}\right)
		\right],
		\qquad B=500,
		\label{eq:B_bootstrap_interval}
	\end{equation}
	where $Q_p$ denotes the empirical $p$th quantile. The estimates are neither recentered nor divided by a bootstrap standard error before the quantiles are taken. The code also stores the standard deviation of the 500 estimates and a two-sided sign probability,
	\begin{equation}
		\begin{aligned}
			\widehat{SE}_{B}^{k}(r)
			&=sd\!\left(\{\widehat\tau_b^k(r)\}_{b=1}^{B}\right),\\
			\widehat\pi_{-,B}^{k}(r)
			&=\frac{1+\sum_{b=1}^{B}\mathbf 1\{\widehat\tau_b^k(r)\leq0\}}{B+1},\\
			\widehat\pi_{+,B}^{k}(r)
			&=\frac{1+\sum_{b=1}^{B}\mathbf 1\{\widehat\tau_b^k(r)\geq0\}}{B+1},\\
			p_{B}^{k}(r)
			&=\min\!\left\{1,2\min\!\left[\widehat\pi_{-,B}^{k}(r),
			\widehat\pi_{+,B}^{k}(r)\right]\right\}.
		\end{aligned}
		\label{eq:B_bootstrap_sign_test}
	\end{equation}
	The paper reports the percentile intervals, not these sign probabilities, and neither bootstrap quantity determines the significance stars.\footnote{The archived implementation uses MATLAB's \texttt{twister} generator. The bootstrap seeds for sanctions--CPI, sanctions--GDP per capita, direct-arming--CPI, and direct-arming--GDP per capita are 24680, 26680, 27680, and 28680, respectively. Within each policy--outcome run, the region draws continue from the generator state reached after the country-history draws.}
	
	\paragraph{Size-matched placebo assignments.}
	The placebo procedure changes estimation roles rather than resampling the actual coalition. In draw $b$, it selects uniformly and without replacement a pseudo-target set and assigns all remaining eligible countries to the donor role:
	\begin{equation}
		\mathcal T_b^P
		\sim
		\operatorname{Unif}\!\left(
		\left\{\mathcal S\subset\mathcal C^Y:|\mathcal S|=N_T\right\}
		\right),
		\qquad
		\mathcal C_b^P=\mathcal C^Y\setminus\mathcal T_b^P.
		\label{eq:B_placebo_assignment}
	\end{equation}
	The code reconstructs the donor factor from $\mathcal C_b^P$, re-estimates the complete pre-treatment model, predicts the pseudo-target counterfactuals, and forms
	\begin{equation}
		\widehat\tau_b^P(r)
		=
		\frac{1}{N_T}
		\sum_{i\in\mathcal T_b^P}d_i^{P,b}(r).
		\label{eq:B_placebo_att}
	\end{equation}
	For sanctions, this divides the 84 eligible donors into 42 pseudo-targets and 42 remaining donors. For direct arming, it selects 29 pseudo-targets and leaves 67 CPI or 68 GDP-per-capita donors. One assignment is used to compute all five reference-window contrasts before the next pseudo-coalition is drawn. The routine starts from the final $N=129$ eligibility rules, initializes one \texttt{twister} stream at seed 12345 before processing the four policy--outcome designs, records the complete pseudo-target list for every draw, and requires all 500 estimates to be finite; it neither imports earlier assignments nor replaces failed draws. The reported two-sided finite-draw probability applies the plus-one correction in equation~\eqref{eq:placebo_p}. The code also stores the corresponding one-sided positive-tail probability,
	\begin{equation}
		\widehat p_{P,+}(r)
		=
		\frac{1+\sum_{b=1}^{B}\mathbf 1\{\widehat\tau_b^P(r)\geq\widehat\tau(r)\}}{B+1},
		\qquad B=500,
		\label{eq:B_placebo_positive_p}
	\end{equation}
	although the paper reports only the two-sided probability. This diagnostic measures how often an equally sized group of eligible non-participants produces an estimate at least as large in absolute value as the actual estimate. Because coalition membership was not randomized, it is not a randomization $p$-value or a confidence interval.
	
	The four quantities therefore answer different questions. The country-dispersion test summarizes variation among the actual target-country contrasts. The country-history bootstrap also varies the composition of target and donor histories. The region-block bootstrap asks how the result behaves when broad geographic groups move together. The placebo asks whether the observed coalition is unusual relative to equally sized groups formed from eligible non-participants. None of them tests the identifying conditions stated above.

	\subsection*{S.B.4 Worldwide Distribution of the Preferred Country Contrasts}
	
\begin{figure}[!htbp]
		\centering
		\includegraphics[width=0.78\textwidth]{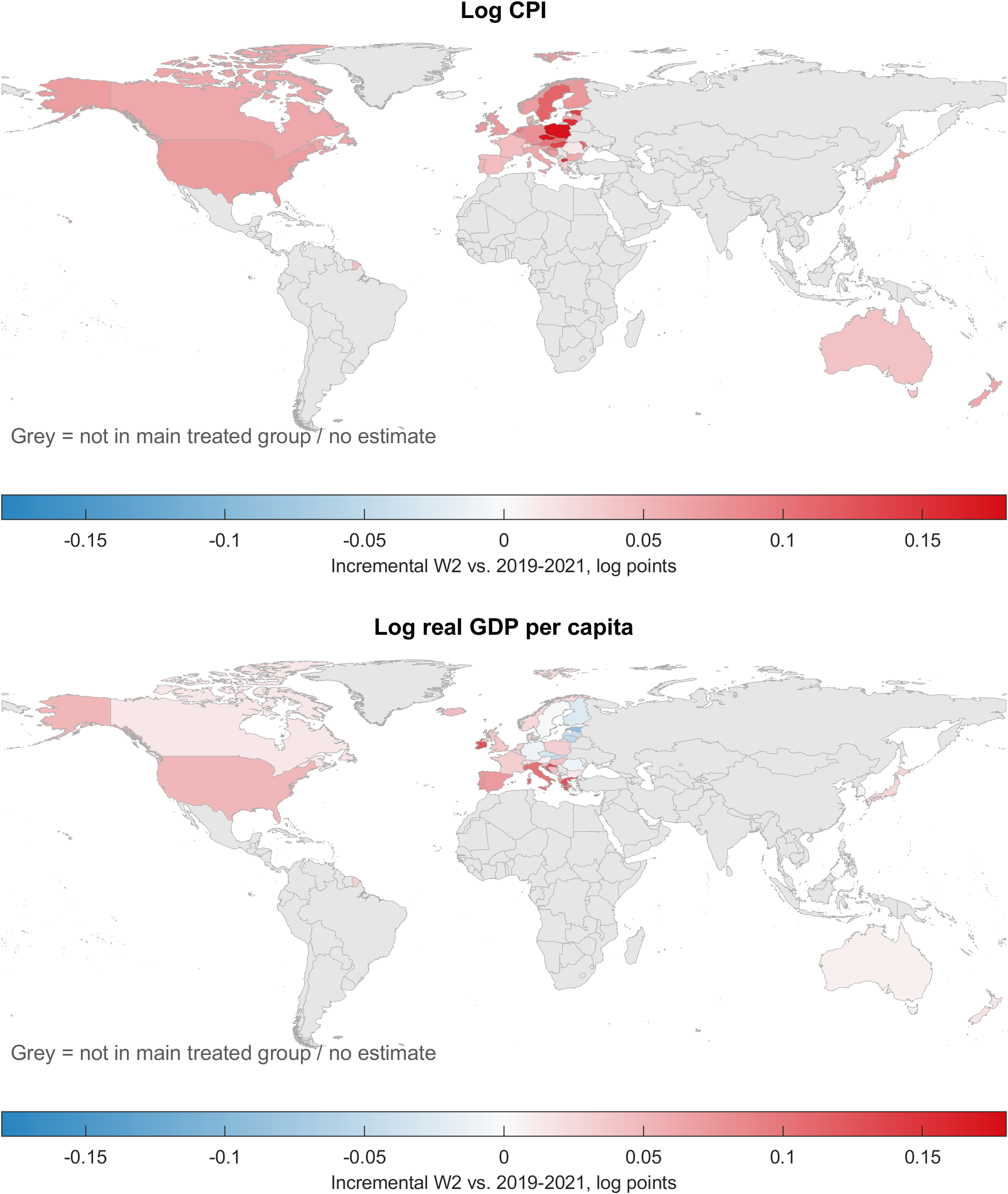}
		\caption{Worldwide heterogeneity in incremental Wave~2 effects}
		\label{fig:B1_world}
		\begin{minipage}{0.96\textwidth}\footnotesize
			\textit{Notes:} Country-specific 2022--2024 gaps minus 2019--2021 gaps for the 42-country main sanctions target group. Grey denotes a country outside that target group or without a main effect estimate. Ukraine is deliberately grey.
		\end{minipage}
	\end{figure}

	Figure~\ref{fig:B1_world} maps the preferred country-specific incremental
	contrasts $d_{i,\mathrm{Inc},2}$ defined in
	equation~\eqref{eq:B_incremental_country} for the main sanctions targets.
	It is a descriptive display of heterogeneity, not a model of spatial spillovers. Grey countries either lie outside the 42-country target set or have no main effect estimate. Ukraine is deliberately grey because it is not included in the sender-country aggregation.
	
	\subsection*{S.B.5 Size-Matched Placebo Assignments }
	
	To address the concern that the estimated macroeconomic gaps might be a statistical artifact of the imputation model rather than a distinct feature of the sanctioning coalition, we generate false coalitions of an identical size and ask whether they commonly generate an incremental estimate as large as the observed one. 
	
	Let $\widehat\tau$ denote the actual incremental estimate and $\widehat\tau_b^{P}$ the estimate from placebo assignment $b=1,\ldots,B$. The reported two-sided finite-draw probability is
	\begin{equation}
		\widehat p_P
		=
		\frac{1+\sum_{b=1}^{B}
			\mathbf 1\!\left\{|\widehat\tau_b^P|\geq|\widehat\tau|\right\}}
		{B+1},
		\qquad B=500.
		\label{eq:placebo_p}
	\end{equation}
	The plus-one correction prevents a simulated probability of exactly zero and includes the observed assignment in the finite randomization comparison.
	
	For sanctions, each draw selects 42 pseudo-targets from the 84 eligible donors, leaving 42 pseudo-donors. For direct arming, it selects 29 pseudo-targets and leaves 67 CPI or 68 GDPpc pseudo-donors. Every draw rebuilds the donor factor and re-estimates the same incremental contrast. Ukraine is absent from every assignment.

	\begin{table}[!htbp]
		\centering
		\caption{Complete size-matched placebo audit}
		\label{tab:B1_placebo}
		\begin{adjustbox}{max width=\textwidth}
			\begin{threeparttable}
				\small
				\begin{tabular}{lcccc}
					\toprule
					Reference window & Actual ATT (SE) & Placebo $p$ & Pseudo-treated & Pseudo-donors \\
					\midrule
					\multicolumn{5}{l}{\textit{Panel A. Sanctions: Log CPI}} \\
					2019--2021 & 0.072*** (0.007) & 0.022 & 42 & 42 \\
					2017--2019 & 0.070*** (0.008) & 0.106 & 42 & 42 \\
					2021 & 0.071*** (0.006) & 0.014 & 42 & 42 \\
					2017--2021 & 0.071*** (0.007) & 0.052 & 42 & 42 \\
					2018--2021 & 0.072*** (0.007) & 0.042 & 42 & 42 \\
					\addlinespace
					\multicolumn{5}{l}{\textit{Panel B. Sanctions: Log real GDP per capita}} \\
					2019--2021 & 0.025*** (0.008) & 0.102 & 42 & 42 \\
					2017--2019 & 0.053*** (0.009) & 0.028 & 42 & 42 \\
					2021 & -0.001 (0.007) & 0.944 & 42 & 42 \\
					2017--2021 & 0.039*** (0.008) & 0.050 & 42 & 42 \\
					2018--2021 & 0.032*** (0.008) & 0.070 & 42 & 42 \\
					\addlinespace
					\multicolumn{5}{l}{\textit{Panel C. Direct arming: Log CPI}} \\
					2019--2021 & 0.107*** (0.027) & 0.002 & 29 & 67 \\
					2017--2019 & 0.114*** (0.033) & 0.002 & 29 & 67 \\
					2021 & 0.101*** (0.024) & 0.002 & 29 & 67 \\
					2017--2021 & 0.111*** (0.030) & 0.002 & 29 & 67 \\
					2018--2021 & 0.109*** (0.028) & 0.002 & 29 & 67 \\
					\addlinespace
					\multicolumn{5}{l}{\textit{Panel D. Direct arming: Log real GDP per capita}} \\
					2019--2021 & 0.016 (0.010) & 0.339 & 29 & 68 \\
					2017--2019 & 0.035*** (0.009) & 0.186 & 29 & 68 \\
					2021 & -0.003 (0.009) & 0.820 & 29 & 68 \\
					2017--2021 & 0.025** (0.010) & 0.210 & 29 & 68 \\
					2018--2021 & 0.021** (0.010) & 0.250 & 29 & 68 \\
					\bottomrule
				\end{tabular}
				\begin{tablenotes}[flushleft]
					\footnotesize
					\item \textit{Notes:} Each empirical $p$-value uses 500 reproducible size-matched assignments and the finite-draw plus-one correction. The country-dispersion SE is the sample SD of the corresponding complete country contrast divided by the square root of the actual treated count. Stars use only a two-sided Student $t$-test with treated count minus one degrees of freedom: *, **, and *** denote $p<0.10$, $p<0.05$, and $p<0.01$. Placebo probabilities do not determine stars. Wave~2 is 2022--2024; Ukraine is absent from the main sanctions targets and every placebo assignment. Actual treated counts are 42 for sanctions and 29 for direct arming. Core donor counts are 84/84 and 96/97 for CPI/GDPpc, respectively.
				\end{tablenotes}
			\end{threeparttable}
		\end{adjustbox}
	\end{table}

	Table~\ref{tab:B1_placebo} shows that the reassignment evidence differs
	substantially across outcomes. For the preferred sanctions Log-CPI contrast,
	the two-sided empirical placebo probability is 0.022. Thus, after applying the
	finite-draw correction, only about 2\% of the size-matched pseudo-coalitions
	produce an absolute estimate at least as large as the observed 0.072 contrast.
	The corresponding probability for direct-arming CPI is 0.002, the smallest
	value attainable with 500 draws under the plus-one correction.
	
	The GDP-per-capita evidence is weaker and more dependent on the reference
	window. Under the preferred 2019--2021 reference, the placebo probabilities
	are 0.102 for sanctions and 0.339 for direct arming. Across the five reference
	windows, the sanctions-GDPpc probabilities range from 0.028 to 0.944, while
	the direct-arming GDPpc probabilities range from 0.186 to 0.820. The placebo
	exercise therefore provides much stronger reassignment evidence for the
	post-2022 Log-CPI contrasts than for the GDP-per-capita contrasts.
	
	These probabilities should be interpreted as diagnostic empirical tail
	probabilities rather than as the source of the significance stars. Each
	pseudo-coalition is constructed from countries that are eligible donors in the
	corresponding baseline design, and the donor factor is reconstructed after the
	pseudo-targets have been removed. Consequently, the exercise asks whether
	similarly sized false coalitions drawn from the untreated pool commonly produce
	contrasts as extreme as the observed coalition; it is not a confidence interval
	for the observed estimate and does not replace the country-dispersion test.
	
\begin{figure}[!htbp]
		\centering
		\includegraphics[width=\textwidth]{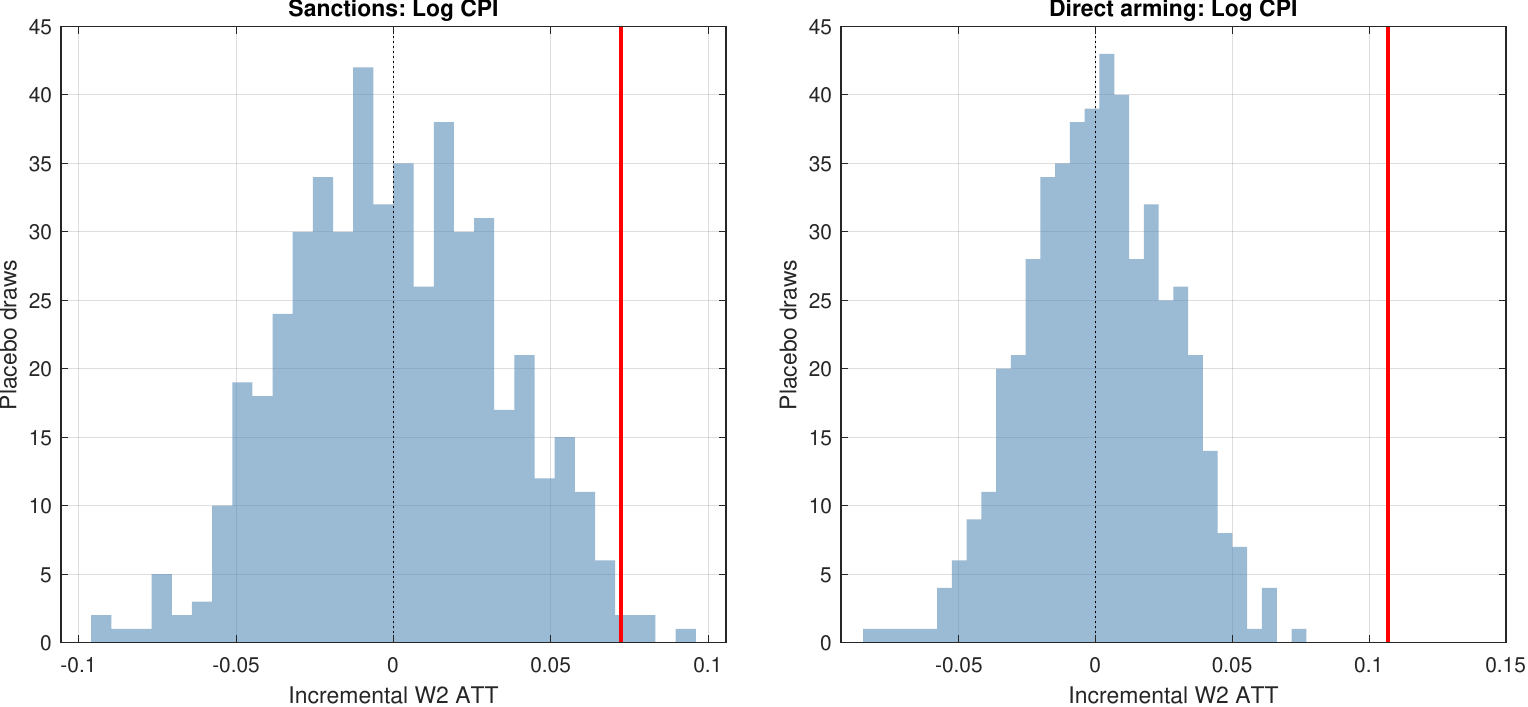}
		\caption{Size-matched placebo distributions for the preferred incremental estimator}
		\label{fig:B2_placebos}
		\begin{minipage}{0.96\textwidth}\footnotesize
			\textit{Notes:} The red line is the observed preferred incremental ATT; the dotted line marks zero. Each distribution contains 500 size-matched pseudo-treatment assignments.
		\end{minipage}
	\end{figure}

	Full placebo assignments, including the 20 largest absolute estimates, are provided in the replication files. Table~\ref{tab:B1_placebo} and Figure~\ref{fig:B2_placebos} contain the inferential information needed here.
	
	\subsection*{S.B.6 Donor Pools and Common-Factor Proxies }
	
	To verify that the estimated macroeconomic gaps are not driven by the idiosyncratic inclusion of a specific donor country or an overly narrow definition of the global factor proxy, Table~\ref{tab:B3_counterfactual} disentangles these two conceptual dimensions. Panel A changes the countries that contribute to the own-outcome mean while retaining the baseline factor definition. Panel B keeps the corrected donor pool intact but expands how the international component is analytically represented (e.g., substituting the other outcome mean, including both, or augmenting with oil rents or fuel-share balances). 
	
	The table also reports the rank and condition number of the pre-2014 factor matrix. Rank records the number of linearly independent columns. The condition number measures near-collinearity: a large value indicates that small changes in the factor inputs may produce larger changes in fitted coefficients. These are numerical diagnostics, not measures of pre-treatment fit.
	
	For sanctions Log CPI, the incremental estimate ranges from 0.068 to 0.106
	across the three donor-pool definitions and from 0.072 to 0.103 across the five
	factor proxies. Every displayed estimate remains positive and statistically
	different from zero under the country-dispersion test. The corresponding
	sanctions-GDPpc estimate is almost unchanged across donor pools, ranging from
	0.025 to 0.026, but varies much more across factor specifications, from 0.007
	to 0.099. The estimate based only on the other outcome's donor mean is small
	and not statistically different from zero. The sensitivity of GDPpc therefore
	comes primarily from how the common component is represented rather than from
	the membership of the donor pool.
	
	The direct-arming results display a similar outcome asymmetry. Its Log-CPI
	estimate remains positive across the donor-pool and factor variants, whereas
	the GDPpc estimates range from values close to zero to substantially larger
	positive values under the richer factor specifications. These results reinforce
	the interpretation of direct arming as a narrower coalition comparison: its
	CPI pattern is comparatively stable, while its GDPpc result is not robust to
	the representation of the common component.
	
	The rank and condition number help distinguish substantive sensitivity from
	numerical instability. Rank records the number of linearly independent factor
	columns. The condition number is a scale-dependent diagnostic of numerical
	conditioning: large values warn that the fitted coefficients may be sensitive
	to small changes in highly correlated or differently scaled factor proxies.
	In particular, estimates from the most poorly conditioned specifications should
	not automatically be treated as improvements over the parsimonious
	own-outcome factor. The preferred specification is retained because it provides
	a direct outcome-specific proxy, preserves full pre-treatment rank, and avoids
	introducing several unstable factor columns into a 13-year pre-treatment
	period.
	
	The high-income donor restriction in Panel A should not be confused with the
	high-GDPpc matched design in Table~\ref{tab:B4_matching}. The former changes
	the donor pool while retaining the full target group, whereas the latter
	restricts both target and donor countries according to pre-treatment GDP per
	capita.
	\begin{table}[!htbp]
		\centering
		\caption{Donor-pool and factor-proxy robustness}
		\label{tab:B3_counterfactual}
		\begin{adjustbox}{max width=\textwidth,max totalheight=.88\textheight,keepaspectratio}
			\begin{threeparttable}
				\scriptsize
				\begin{tabular}{lcccc}
					\toprule
					Variant & Donors & Pre-rank & Condition & ATT (SE) \\
					\midrule
					\multicolumn{5}{l}{\textbf{Panel A. Donor-pool variants}} \\
					\textit{Sanctions: Log CPI} & & & & \\
					Corrected full pool & 84 & -- & -- & 0.072*** (0.007) \\
					No repaired main-outcome cells & 80 & -- & -- & 0.068*** (0.007) \\
					High-income donors & 15 & -- & -- & 0.106*** (0.007) \\
					\addlinespace
					\textit{Sanctions: Log real GDP per capita} & & & & \\
					Corrected full pool & 84 & -- & -- & 0.025*** (0.008) \\
					No repaired main-outcome cells & 84 & -- & -- & 0.025*** (0.008) \\
					High-income donors & 15 & -- & -- & 0.026*** (0.008) \\
					\addlinespace
					\textit{Direct arming: Log CPI} & & & & \\
					Corrected full pool & 96 & -- & -- & 0.107*** (0.027) \\
					No repaired main-outcome cells & 92 & -- & -- & 0.103*** (0.027) \\
					High-income donors & 26 & -- & -- & 0.111*** (0.027) \\
					\addlinespace
					\textit{Direct arming: Log real GDP per capita} & & & & \\
					Corrected full pool & 97 & -- & -- & 0.016 (0.010) \\
					No repaired main-outcome cells & 97 & -- & -- & 0.016 (0.010) \\
					High-income donors & 26 & -- & -- & 0.000 (0.012) \\
					\midrule
					\multicolumn{5}{l}{\textbf{Panel B. Factor-proxy variants}} \\
					\textit{Sanctions: Log CPI} & & & & \\
					Own-outcome donor mean & 84 & 2 & 91.47 & 0.072*** (0.007) \\
					Other-outcome donor mean & 84 & 2 & 637.31 & 0.103*** (0.007) \\
					Both outcome means & 84 & 3 & 4409.35 & 0.076*** (0.007) \\
					Own outcome + oil rents & 84 & 3 & 147.99 & 0.072*** (0.007) \\
					Own outcome + fuel-share balance & 84 & 3 & 650.05 & 0.075*** (0.007) \\
					\addlinespace
					\textit{Sanctions: Log real GDP per capita} & & & & \\
					Own-outcome donor mean & 84 & 2 & 637.31 & 0.025*** (0.008) \\
					Other-outcome donor mean & 84 & 2 & 91.47 & 0.007 (0.010) \\
					Both outcome means & 84 & 3 & 4409.35 & 0.062*** (0.009) \\
					Own outcome + oil rents & 84 & 3 & 799.54 & 0.025*** (0.008) \\
					Own outcome + fuel-share balance & 84 & 3 & 3374.14 & 0.099*** (0.012) \\
					\addlinespace
					\textit{Direct arming: Log CPI} & & & & \\
					Own-outcome donor mean & 96 & 2 & 99.15 & 0.107*** (0.027) \\
					Other-outcome donor mean & 96 & 2 & 704.84 & 0.133*** (0.029) \\
					Both outcome means & 96 & 3 & 4303.36 & 0.111*** (0.028) \\
					Own outcome + oil rents & 96 & 3 & 148.25 & 0.107*** (0.027) \\
					Own outcome + fuel-share balance & 96 & 3 & 579.91 & 0.114*** (0.031) \\
					\addlinespace
					\textit{Direct arming: Log real GDP per capita} & & & & \\
					Own-outcome donor mean & 97 & 2 & 700.03 & 0.016 (0.010) \\
					Other-outcome donor mean & 97 & 2 & 99.60 & 0.002 (0.012) \\
					Both outcome means & 97 & 3 & 4280.63 & 0.051*** (0.011) \\
					Own outcome + oil rents & 97 & 3 & 838.20 & 0.016 (0.010) \\
					Own outcome + fuel-share balance & 97 & 3 & 3335.39 & 0.091*** (0.015) \\
					\bottomrule
				\end{tabular}
				\begin{tablenotes}[flushleft]
					\scriptsize
					\item \textit{Notes:} All rows use the 2019--2021 reference. Panel A changes donor eligibility while retaining the own-outcome factor. Panel B changes the factor proxy with the corrected pool. Pre-rank is the rank of the pre-2014 factor matrix; Condition is its condition number. The country-dispersion SE is the sample SD of the complete country contrast divided by the square root of the treated count; two-sided Student $t$-tests use treated count minus one degrees of freedom. *, **, and *** denote $p<0.10$, $p<0.05$, and $p<0.01$. The WDI fuel-share balance is a control proxy, not a measure of bilateral Russia exposure. Treated counts are 42 sanctions countries and 29 direct-arming countries.
				\end{tablenotes}
			\end{threeparttable}
		\end{adjustbox}
	\end{table}
	
	\subsection*{S.B.7 Observable Comparability and Matched Donors }
	
	To ensure that structural baseline differences between the wealthy target coalition and the global donor pool are not subtly driving the imputation results, we apply two restrictive matching designs to the 2009--2013 country averages. The high-GDPpc design deliberately discards any country (target or donor) below the median pre-treatment log GDP per capita. The Mahalanobis design enforces local structural similarity by forcing each target to select its five most observably similar donors with replacement, weighting the factor accordingly. 
	
	For the Mahalanobis design, we standardize the matching features, and for target $i$ and eligible donor $j$, we compute:
	\begin{equation}
		m_{ij}
		=
		\left[
		(Z_i-Z_j)'
		\left(\widehat S_Z+10^{-6}I\right)^{+}
		(Z_i-Z_j)
		\right]^{1/2},
		\label{eq:mahalanobis}
	\end{equation}
	where $Z$ contains the standardized features and $\widehat S_Z$ is their covariance matrix in the combined target--donor sample. For raw matching feature $Q_k$, balance is summarized by the standardized difference:
	\begin{equation}
		SD_k
		=
		\frac{\overline Q_{T,k}-\overline Q_{C,k}}
		{s_{Q_k,\mathcal I}},
		\label{eq:std_diff}
	\end{equation}
	where $s_{Q_k,\mathcal I}$ is the pre-matching standard deviation across the full combined estimation sample. 
	
	Table~\ref{tab:B4_matching} examines whether the principal estimates depend on
	using the broad worldwide donor pool despite substantial observable differences
	between target and donor countries. The two restricted designs address this
	concern in different ways. The high-GDPpc design removes countries below the
	median pre-treatment GDP-per-capita threshold from both sides of the comparison.
	The Mahalanobis design retains the full target group but constructs a
	frequency-weighted donor factor from the five nearest eligible donors selected
	for each target.
	
	The balance results show improvement along several dimensions, but they do not
	imply randomized or uniformly close comparability. Under Mahalanobis matching,
	for example, the absolute standardized difference in terms of trade falls from
	0.326 to 0.060, while those for oil rents, population growth, and the fuel-share
	balance also decline substantially. The trade-openness difference falls more
	moderately, from 0.455 to 0.312. Important differences nevertheless remain:
	the maximum absolute standardized difference is 1.052 under Mahalanobis
	matching, compared with 1.498 in the baseline design. The restricted designs
	should therefore be interpreted as improvements in observable comparability,
	not as elimination of all pre-treatment imbalance.
	
	Despite these changes, the sanctions Log-CPI contrast remains positive and
	statistically different from zero. It is 0.063 under both the high-GDPpc and
	Mahalanobis designs, compared with 0.072 in the baseline. The corresponding
	GDPpc estimates are 0.023 and 0.025, compared with 0.025 in the baseline.
	For direct arming, the CPI estimate declines from 0.107 to 0.105 in the
	high-GDPpc sample and to 0.076 under Mahalanobis matching, while the GDPpc
	estimates remain small and statistically indistinguishable from zero.
	
	The matching exercise therefore shows that the positive CPI result does not
	require the broadest possible donor pool. It does not by itself establish
	causal identification, because substantial observable imbalance remains and
	matching cannot address unobserved differences between the geopolitical
	coalition and untreated countries.
	\begin{table}[!htbp]
		\centering
		\caption{Matched-donor design, balance, and selection details}
		\label{tab:B4_matching}
		\begin{adjustbox}{max width=\textwidth,max totalheight=.88\textheight,keepaspectratio}
			\begin{threeparttable}
				\scriptsize
				\begin{tabular}{l c c c}
					\toprule
					\multicolumn{4}{l}{\textbf{Panel A. Design sizes and preferred estimates}} \\
					\addlinespace
					\textit{Design} & \textit{Treated} & \textit{Donors} & \textit{ATT (SE)} \\
					\midrule
					\multicolumn{4}{l}{\textit{Sanctions: Log CPI}} \\
					Baseline full donors & 42 & 84 & 0.072*** (0.007) \\
					High-income sample (GDPpc p50) & 38 & 25 & 0.063*** (0.007) \\
					Mahalanobis matched donors (M=5) & 42 & 37 & 0.063*** (0.007) \\
					\addlinespace
					\multicolumn{4}{l}{\textit{Sanctions: Log real GDP per capita}} \\
					Baseline full donors & 42 & 84 & 0.025*** (0.008) \\
					High-income sample (GDPpc p50) & 38 & 25 & 0.023** (0.009) \\
					Mahalanobis matched donors (M=5) & 42 & 37 & 0.025*** (0.008) \\
					\addlinespace
					\multicolumn{4}{l}{\textit{Direct arming: Log CPI}} \\
					Baseline full donors & 29 & 96 & 0.107*** (0.027) \\
					High-income sample (GDPpc p50) & 28 & 35 & 0.105*** (0.028) \\
					Mahalanobis matched donors (M=5) & 29 & 35 & 0.076*** (0.025) \\
					\addlinespace
					\multicolumn{4}{l}{\textit{Direct arming: Log real GDP per capita}} \\
					Baseline full donors & 29 & 97 & 0.016 (0.010) \\
					High-income sample (GDPpc p50) & 28 & 35 & 0.007 (0.011) \\
					Mahalanobis matched donors (M=5) & 29 & 35 & 0.009 (0.011) \\
					\midrule
					\multicolumn{4}{l}{\textbf{Panel B. Sanctions-design standardized differences}} \\
					\addlinespace
					\textit{Macroeconomic Feature} & \textit{Baseline} & \textit{High-GDPpc} & \textit{Mahalanobis} \\
					\midrule
					Log GDPpc & 1.320 & 0.441 & 0.971 \\
					Log CPI & -0.799 & -0.356 & -0.380 \\
					Inflation & -0.908 & -0.526 & -0.493 \\
					GDPpc growth & -0.863 & -0.864 & -0.718 \\
					Oil rents & -0.535 & -1.052 & -0.131 \\
					Trade openness & 0.455 & 0.052 & 0.312 \\
					Terms of trade & -0.326 & -0.488 & 0.060 \\
					Investment rate & -0.402 & -0.489 & -0.283 \\
					Government consumption & 0.739 & 0.719 & 0.565 \\
					Rule of law & 1.498 & 1.015 & 1.052 \\
					Population growth & -1.159 & -1.175 & -0.497 \\
					Fuel-share balance & 0.338 & 0.721 & -0.106 \\
					\midrule
					\multicolumn{4}{l}{\textbf{Panel C. Most frequently selected sanctions/Log-CPI Mahalanobis donors}} \\
					\addlinespace
					\textit{Donor Country} & \textit{Frequency} & \textit{Donor Country} & \textit{Frequency} \\
					\midrule
					Tunisia (TUN) & 30 & Bahamas, The (BHS) & 7 \\
					Tonga (TON) & 24 & El Salvador (SLV) & 5 \\
					South Africa (ZAF) & 24 & Indonesia (IDN) & 5 \\
					Thailand (THA) & 20 & Georgia (GEO) & 5 \\
					Malaysia (MYS) & 14 & Brazil (BRA) & 4 \\
					Mauritius (MUS) & 9 & Uruguay (URY) & 4 \\
					Costa Rica (CRI) & 9 & Namibia (NAM) & 4 \\
					Chile (CHL) & 8 & & \\
					\bottomrule
				\end{tabular}
				\begin{tablenotes}[flushleft]
					\scriptsize
					\item \textit{Notes:} Matching uses only 2009--2013 features. In Panel A, the country-dispersion SE (in parentheses) is the sample SD of the complete country contrast divided by the square root of the treated count shown in the row; two-sided Student $t$-tests use that count minus one degrees of freedom. *, **, and *** denote $p<0.10$, $p<0.05$, and $p<0.01$. Panel B gives standardized treated--donor differences for the baseline, high-GDPpc, and Mahalanobis designs. Panel C reports donor-selection frequency; repeated selection generates frequency weights, it does not create additional countries. The main sanctions designs exclude Ukraine.
				\end{tablenotes}
			\end{threeparttable}
		\end{adjustbox}
	\end{table}
	
	\subsection*{S.B.8 Influence, Distribution, and Country Heterogeneity }
	
	To confirm that the coalition averages are not being mechanically skewed by a single extreme country on either the target or donor side, we implement two strict leave-one-out (LOO) sensitivity tests. 
	
	Target-country leave-one-out holds the estimated country contrasts fixed and removes one treated country before re-aggregating:
	\begin{equation}
		\widehat\tau_{(-i)}
		=\frac{1}{N_T-1}
		\sum_{k\in\mathcal T^*,\,k\neq i}d_k.
		\label{eq:treated_loo}
	\end{equation}
	Donor leave-one-out is computationally more demanding: for each donor $j$, it entirely removes that donor, reconstructs the common factor, re-estimates the pre-treatment model and every target counterfactual, and recalculates the aggregate. It thereby measures the donor's structural influence on the complete estimated design rather than merely deleting one final observation.
	
	The leave-one-out diagnostics in Table~\ref{tab:B5_influence}  distinguish robustness of the coalition average
	from agreement across individual countries. For Log CPI, both dimensions are
	strong. The mean incremental contrast is 0.072, the median is 0.062, and 41 of
	the 42 target-country contrasts are positive. Removing any single target leaves
	the coalition estimate between 0.070 and 0.074. The largest donor influence
	comes from excluding Turkey, which raises the estimate by 0.0049 log points.
	No individual target or donor therefore accounts for the positive coalition
	average.
	
	The GDPpc average is similarly insensitive to deletion of a particular
	country, but its underlying distribution is less uniform. The mean is 0.025,
	the median is 0.026, and 66.7\% of the country contrasts are positive. The
	treated-country leave-one-out range is [0.023, 0.028], and the largest donor
	leave-one-out change is only 0.0019. These results establish that the average
	GDPpc estimate is not mechanically generated by one influential country, but
	they do not give it the broad sign agreement observed for CPI.
	
	Panel D reports the complete distribution of sanctions Log-CPI country
	contrasts. These entries are descriptive country-specific differences between
	the 2022--2024 and 2019--2021 counterfactual gaps; they are the components
	averaged to obtain the coalition ATT. They should not be interpreted as 42
	separately identified country-level treatment effects. Their main contribution
	is to show that the positive coalition estimate is broad-based rather than
	being generated by a small upper tail.
	
	\begin{table}[!htbp]
		\centering
		\caption{Influence diagnostics and country-specific effects}
		\label{tab:B5_influence}
		\begin{adjustbox}{max width=\textwidth}
			\begin{threeparttable}
				\footnotesize
				\begin{tabular}{l c l c l c}
					\toprule
					\multicolumn{6}{l}{\textbf{Panel A. Distribution and treated-country leave-one-out}} \\
					\addlinespace
					\textit{Outcome} & \textit{Mean} & \textit{Median} & \textit{Positive Share} & \multicolumn{2}{c}{\textit{Treated LOO Range}} \\
					\midrule
					Log CPI & 0.072 & 0.062 & 97.6\% & \multicolumn{2}{c}{[0.070, 0.074]} \\
					Log real GDP per capita & 0.025 & 0.026 & 66.7\% & \multicolumn{2}{c}{[0.023, 0.028]} \\
					\midrule
					\multicolumn{3}{l}{\textbf{Panel B. Top 10 donor LOO changes: Log CPI}} & \multicolumn{3}{l}{\textbf{Panel C. Top 10 donor LOO changes: GDPpc}} \\
					\addlinespace
					\textit{Excluded Donor} & \textit{New ATT} & \textit{$\Delta$ ATT} & \textit{Excluded Donor} & \textit{New ATT} & \textit{$\Delta$ ATT} \\
					\midrule
					TUR & 0.0771 & +0.0049 & PSE & 0.0235 & -0.0019 \\
					GIN & 0.0692 & -0.0031 & LKA & 0.0237 & -0.0017 \\
					IRN & 0.0753 & +0.0031 & BHS & 0.0269 & +0.0015 \\
					LKA & 0.0738 & +0.0015 & AZE & 0.0240 & -0.0014 \\
					VNM & 0.0709 & -0.0013 & SYC & 0.0241 & -0.0013 \\
					GHA & 0.0735 & +0.0013 & MAC & 0.0241 & -0.0013 \\
					CRI & 0.0711 & -0.0012 & SLB & 0.0241 & -0.0013 \\
					SYC & 0.0711 & -0.0011 & AGO & 0.0242 & -0.0012 \\
					PAK & 0.0732 & +0.0010 & BLZ & 0.0265 & +0.0011 \\
					KEN & 0.0713 & -0.0009 & FJI & 0.0264 & +0.0010 \\
					\midrule
					\multicolumn{6}{l}{\textbf{Panel D. All 42 country-specific sanctions Log-CPI effects}} \\
					\addlinespace
					\textit{Country} & \textit{ATT} & \textit{Country} & \textit{ATT} & \textit{Country} & \textit{ATT} \\
					\midrule
					POL & 0.1754 & FIN & 0.0745 & DNK & 0.0482 \\
					CZE & 0.1721 & IRL & 0.0738 & ESP & 0.0471 \\
					LTU & 0.1541 & GBR & 0.0734 & FRA & 0.0470 \\
					MKD & 0.1530 & BIH & 0.0725 & CYP & 0.0427 \\
					EST & 0.1393 & USA & 0.0704 & AUS & 0.0417 \\
					HUN & 0.1373 & NOR & 0.0697 & LUX & 0.0372 \\
					SWE & 0.1145 & ITA & 0.0619 & SGP & 0.0361 \\
					MDA & 0.1130 & CAN & 0.0618 & ALB & 0.0340 \\
					SVK & 0.1051 & NZL & 0.0616 & MLT & 0.0326 \\
					AUT & 0.0992 & BGR & 0.0604 & GRC & 0.0311 \\
					NLD & 0.0961 & LVA & 0.0594 & CHE & 0.0239 \\
					HRV & 0.0884 & JPN & 0.0571 & ROU & 0.0146 \\
					DEU & 0.0807 & SVN & 0.0524 & KOR & 0.0015 \\
					BEL & 0.0755 & PRT & 0.0500 & ISL & -0.0059 \\
					\bottomrule
				\end{tabular}
				\begin{tablenotes}[flushleft]
					\scriptsize
					\item \textit{Notes:} All diagnostics use sanctions and the 2019--2021 reference. Treated LOO deletes one treated contrast and re-aggregates. Donor LOO removes one donor, reconstructs the factor, and re-estimates the complete model; it is reported for both outcomes. For Panels B and C, each iteration leaves 83 donors remaining. Country-specific effects are 2022--2024 mean gaps minus 2019--2021 mean gaps. The GDPpc donor-LOO values were independently reproduced from the archived panel using the documented CCEDID engine and match the 0.0019 maximum change reported in Table~\ref{tab:influence}.
				\end{tablenotes}
			\end{threeparttable}
		\end{adjustbox}
	\end{table}
	
	\subsection*{S.B.9 Reference Windows and Inferential Summaries }
	
	We use alternative reference windows to examine whether the estimated post-2022 contrast depends on an unusual baseline period. We report the three inferential summaries separately because they answer different questions.
	
	The reference-window exercise changes only $\mathcal R_r$, the pre-2022
	interval whose average gap is subtracted from the 2022--2024 gap. It does not
	alter Wave~2, target status, donor eligibility, or the counterfactual model.
	It changes exclusively the pre-2022 gap subtracted from 2022--2024. Table~\ref{tab:B6_windows} places three inferential summaries side by side:
	\begin{itemize}
		\item \textit{Country Dispersion}: Uses cross-country variation in the complete target-country contrasts and determines the significance stars.
		\item \textit{Placebo Simulation}: Asks whether size-matched false coalitions drawn from the eligible donor pool commonly produce an estimate as extreme as the observed one.
		\item \textit{Percentile Bootstraps (Country-History and Region-Block)}: Re-estimate the complete counterfactual under two resampling schemes and provide separate sensitivity intervals.
	\end{itemize}

	The sanctions Log-CPI estimate is exceptionally stable in magnitude, ranging
	from 0.070 to 0.072 across the five reference windows. Every
	country-dispersion test rejects zero at the 1\% level, and every country-history
	and region-block interval lies above zero. The placebo evidence is less uniform:
	the probabilities range from 0.014 to 0.106 and equal 0.022 for the preferred
	2019--2021 reference. Thus, the magnitude and bootstrap evidence are stable,
	while the strength of the reassignment evidence depends somewhat on the
	reference window.
	
	Direct-arming CPI produces a similarly consistent pattern. Its estimates range
	from 0.101 to 0.114, every country-dispersion test rejects zero, every placebo
	probability equals 0.002, and all country-history and region-block intervals
	exclude zero. This comparison concerns a narrower high-commitment coalition and
	should not be interpreted as an independent military-aid effect holding
	sanctions constant.
	
	The GDPpc results are substantially more reference-dependent. For sanctions,
	the estimates range from -0.001 with the 2021 reference to 0.053 with
	2017--2019. Under the preferred reference, the country-dispersion test and both
	bootstrap intervals are positive, but the placebo probability is 0.102. For
	direct arming, the preferred estimate is 0.016: it has no
	country-dispersion star, its placebo probability is 0.339, its country-history
	interval includes zero, and its region-block interval lies above zero.
	Alternative references likewise produce different conclusions across the three
	inferential summaries.
	
	The table therefore supports a clear hierarchy. The post-2022 CPI contrast is
	stable in magnitude across references and receives broad inferential support,
	especially under the preferred reference. GDPpc is more sensitive both to the
	reference interval and to the inferential procedure. No single probability or
	interval should be substituted for the complete comparison.
	
	\begin{table}[!htbp]
		\centering
		\caption{Full reference-window and inference results}
		\label{tab:B6_windows}
		\begin{threeparttable}
			\small
			\setlength{\tabcolsep}{6pt}
			\renewcommand{\arraystretch}{1.1}
			\begin{tabular}{l c c c c}
				\toprule
				\textit{Reference} 
				& \makecell{\textit{ATT}\\ \textit{(country-disp. SE)}} 
				& \textit{Placebo $p$} 
				& \makecell{\textit{Country-history}\\ \textit{CI}} 
				& \makecell{\textit{Region-block}\\ \textit{CI}} \\
				\midrule
				\multicolumn{5}{l}{\textbf{Panel A. Sanctions: Log CPI}} \\
				\addlinespace
				2019--2021 & 0.072*** (0.007) & 0.022 & [0.054, 0.093] & [0.032, 0.082] \\
				2017--2019 & 0.070*** (0.008) & 0.106 & [0.045, 0.095] & [0.024, 0.086] \\
				2021 & 0.071*** (0.006) & 0.014 & [0.056, 0.090] & [0.035, 0.079] \\
				2017--2021 & 0.071*** (0.007) & 0.052 & [0.050, 0.095] & [0.029, 0.083] \\
				2018--2021 & 0.072*** (0.007) & 0.042 & [0.052, 0.093] & [0.030, 0.082] \\
				\addlinespace
				\midrule
				\multicolumn{5}{l}{\textbf{Panel B. Sanctions: Log real GDP per capita}} \\
				\addlinespace
				2019--2021 & 0.025*** (0.008) & 0.102 & [0.003, 0.043] & [0.011, 0.048] \\
				2017--2019 & 0.053*** (0.009) & 0.028 & [0.027, 0.080] & [0.034, 0.066] \\
				2021 & -0.001 (0.007) & 0.944 & [-0.020, 0.015] & [-0.017, 0.012] \\
				2017--2021 & 0.039*** (0.008) & 0.050 & [0.016, 0.060] & [0.021, 0.055] \\
				2018--2021 & 0.032*** (0.008) & 0.070 & [0.011, 0.050] & [0.017, 0.050] \\
				\addlinespace
				\midrule
				\multicolumn{5}{l}{\textbf{Panel C. Direct arming: Log CPI}} \\
				\addlinespace
				2019--2021 & 0.107*** (0.027) & 0.002 & [0.067, 0.170] & [0.044, 0.118] \\
				2017--2019 & 0.114*** (0.033) & 0.002 & [0.065, 0.195] & [0.044, 0.128] \\
				2021 & 0.101*** (0.024) & 0.002 & [0.066, 0.153] & [0.039, 0.111] \\
				2017--2021 & 0.111*** (0.030) & 0.002 & [0.067, 0.183] & [0.045, 0.123] \\
				2018--2021 & 0.109*** (0.028) & 0.002 & [0.066, 0.176] & [0.044, 0.120] \\
				\addlinespace
				\midrule
				\multicolumn{5}{l}{\textbf{Panel D. Direct arming: Log real GDP per capita}} \\
				\addlinespace
				2019--2021 & 0.016 (0.010) & 0.339 & [-0.005, 0.039] & [0.004, 0.031] \\
				2017--2019 & 0.035*** (0.009) & 0.186 & [0.010, 0.060] & [0.017, 0.040] \\
				2021 & -0.003 (0.009) & 0.820 & [-0.024, 0.017] & [-0.008, 0.013] \\
				2017--2021 & 0.025** (0.010) & 0.210 & [0.003, 0.048] & [0.012, 0.035] \\
				2018--2021 & 0.021** (0.010) & 0.250 & [-0.000, 0.043] & [0.008, 0.032] \\
				\bottomrule
			\end{tabular}
			\begin{tablenotes}[flushleft]
				\footnotesize
				\item \textit{Notes:} Wave 1 is 2014--2021 and Wave 2 is 2022--2024. The country-dispersion SE is the sample SD of one complete treated-country contrast divided by the square root of the treated-country count. Two-sided p-values use a Student t distribution with N treated minus one degrees of freedom. *, **, and *** denote p<0.10, p<0.05, and p<0.01, respectively, from that country-dispersion test. Wave 2 is 2022--2024 throughout. All four policy-outcome designs have 500 country-history and 500 valid region-block re-estimations. Placebo p-values and bootstrap intervals are separate inferential summaries and do not determine stars. Sanctions use 42 treated and 84 donors; direct arming uses 29 treated and 96/97 CPI/GDPpc donors. Ukraine is excluded from the main sanctions ATT and from every donor pool.
			\end{tablenotes}
		\end{threeparttable}
	\end{table}

	\subsection*{S.B.10 Active Timing and the Role of Ukraine }
	
	To verify the conceptual boundaries of the target group, Table~\ref{tab:B7_secondary} accounts for staggered entry timing and formally evaluates the mechanical influence of the conflict's epicenter.
	
	The ever-treated direct-arming estimand describes the experience of the high-commitment coalition over common geopolitical windows, even though entry dates differ. The active-timing diagnostic instead weights the analysis strictly by genuine participation, defining for country $i$ and window $\mathcal W$:
	\begin{equation}
		\overline g^{A}_{i,\mathcal W}
		=
		\frac{\sum_{t\in\mathcal W}A_{it}\widehat g_{it}}
		{\sum_{t\in\mathcal W}A_{it}},
		\label{eq:active_timing}
	\end{equation}
	whenever the denominator is positive. 
	
	Table~\ref{tab:B7_secondary} examines two boundaries of the baseline
	aggregation. Panel A restricts each direct-arming country's window average to
	years in which its strict annual indicator is active. Panel B examines the
	sensitivity of the sanctions estimates to including Ukraine in the target set,
	while continuing to exclude Ukraine, Russia, and Belarus from every donor pool.
	
	Only four direct-arming countries are active during Wave~1. Their active-year
	estimates should therefore be treated as descriptive and should not be used to
	draw general conclusions about the pre-2022 high-commitment coalition. During
	Wave~2, all 29 direct-arming countries are active. The resulting active-year
	level gaps are 0.077 for Log CPI and 0.039 for Log real GDP per capita, matching
	the Wave~2 level estimates in Table~\ref{tab:direct_arming}. The CPI estimate is
	statistically different from zero at the 10\% level under the
	country-dispersion test, while the GDPpc estimate is significant at the 1\%
	level. These are Wave~2 level gaps, not incremental contrasts relative to
	2019--2021. Because the full target group is active during this window, the
	exercise primarily confirms the interpretation of the common Wave~2
	aggregation rather than providing an independent estimate.
	
	Panel B shows why Ukraine is excluded from the preferred target group while
	also documenting that this decision is not responsible for the principal CPI
	result. Including Ukraine changes the preferred sanctions Log-CPI incremental
	contrast only from 0.072 to 0.071. The estimates remain between 0.069 and
	0.071 across the alternative references. For GDPpc, the preferred contrast
	declines from 0.025 to 0.020, while the 2021-reference estimate remains close
	to zero and statistically insignificant. The Wave~1 GDPpc level estimate also
	loses statistical significance when Ukraine is included.
	
	The Ukraine-inclusive exercise therefore shows that the preferred incremental
	CPI estimate is not driven by Ukraine's exclusion. It does not imply that
	Ukraine belongs in the substantive sender-country estimand: its macroeconomic
	outcomes reflect direct invasion, destruction, displacement, and territorial
	loss, making it fundamentally different from the countries whose sender-side
	incidence the main analysis is intended to summarize.

	\begin{table}[!htbp]
		\centering
		\caption{Secondary analyses: active direct-arming timing and Ukraine sensitivity}
		\label{tab:B7_secondary}
		\begin{adjustbox}{max width=\textwidth}
			\begin{threeparttable}
				\small
				\begin{tabular}{l l c c c}
					\toprule
					\multicolumn{5}{l}{\textbf{Panel A. Direct-arming active-year timing}} \\
					\addlinespace
					\textit{Outcome} & \textit{Window} & \textit{Active-Year ATT (SE)} & \multicolumn{2}{c}{\textit{Active N}} \\
					\midrule
					Log CPI & Wave 1 (2014--2021) & 0.082 (0.083) & \multicolumn{2}{c}{4} \\
					Log CPI & Wave 2 (2022--2024) & 0.077* (0.043) & \multicolumn{2}{c}{29} \\
					Log real GDP per capita & Wave 1 (2014--2021) & 0.071** (0.018) & \multicolumn{2}{c}{4} \\
					Log real GDP per capita & Wave 2 (2022--2024) & 0.039*** (0.014) & \multicolumn{2}{c}{29} \\
					\midrule
					\multicolumn{5}{l}{\textbf{Panel B. Sensitivity to including Ukraine as sanctions-treated}} \\
					\addlinespace
					\textit{Outcome} & \textit{Window / Reference} & \textit{Baseline ATT (SE)} & \textit{Inclusive ATT (SE)} & \textit{N (Base / Incl)} \\
					\midrule
					Log CPI & 2014--2021 level & -0.043*** (0.009) & -0.034** (0.013) & 42 / 43 \\
					Log CPI & 2022--2024 level & 0.022 (0.015) & 0.033* (0.018) & 42 / 43 \\
					Log CPI & 2019--2021 & 0.072*** (0.007) & 0.071*** (0.007) & 42 / 43 \\
					Log CPI & 2017--2019 & 0.070*** (0.008) & 0.069*** (0.008) & 42 / 43 \\
					Log CPI & 2021 & 0.071*** (0.006) & 0.070*** (0.006) & 42 / 43 \\
					Log CPI & 2017--2021 & 0.071*** (0.007) & 0.070*** (0.007) & 42 / 43 \\
					Log CPI & 2018--2021 & 0.072*** (0.007) & 0.070*** (0.007) & 42 / 43 \\
					\addlinespace
					Log real GDP per capita & 2014--2021 level & 0.024** (0.011) & 0.018 (0.012) & 42 / 43 \\
					Log real GDP per capita & 2022--2024 level & 0.083*** (0.019) & 0.071*** (0.022) & 42 / 43 \\
					Log real GDP per capita & 2019--2021 & 0.025*** (0.008) & 0.020** (0.009) & 42 / 43 \\
					Log real GDP per capita & 2017--2019 & 0.053*** (0.009) & 0.048*** (0.010) & 42 / 43 \\
					Log real GDP per capita & 2021 & -0.001 (0.007) & -0.006 (0.009) & 42 / 43 \\
					Log real GDP per capita & 2017--2021 & 0.039*** (0.008) & 0.033*** (0.010) & 42 / 43 \\
					Log real GDP per capita & 2018--2021 & 0.032*** (0.008) & 0.027*** (0.009) & 42 / 43 \\
					\bottomrule
				\end{tabular}
				\begin{tablenotes}[flushleft]
					\footnotesize
					\item \textit{Notes:} The country-dispersion SE (in parentheses) is the sample SD of the corresponding complete country contrast divided by the square root of its treated count; two-sided Student $t$-tests use that count minus one degrees of freedom. *, **, and *** denote $p<0.10$, $p<0.05$, and $p<0.01$. Only four direct-arming countries are active before 2022, so Wave~1 active-year estimates are descriptive; Lithuania enters in 2022. The Ukraine-inclusive column in Panel B has 43 sanctions targets and is confined strictly to this sensitivity check. Direct-arming CPI/GDPpc donor counts are 96/97; sanctions donors number 84. Russia, Belarus, and Ukraine remain excluded from every donor pool.
				\end{tablenotes}
			\end{threeparttable}
		\end{adjustbox}
	\end{table}

	\label{sec:appendix_energy_exposure_robustness}
	\subsection*{S.B.11 Robustness of the Russia-Energy Exposure Results }
	\addcontentsline{toc}{subsection}{Robustness of the Russia-Energy Exposure Results}
	
	Table~\ref{tab:B8_energy_exposure_robustness} examines two separate quantities: the difference in exposure slopes between policy and non-policy countries, $\widehat\beta_3$, and the exposure slope within the policy group, $\widehat\beta_2+\widehat\beta_3$. The within-policy CPI, inflation, and growth slopes retain their signs across exposure windows, sample rules, estimators, and policy definitions. Evidence on $\widehat\beta_3$ is less uniform for CPI and inflation, while the negative GDP-per-capita growth interaction is strongest when exposure is measured relative to GDP. The GDP-level result changes under TWFE, and the gas-share results are less precise. These checks support a stable pattern of within-coalition heterogeneity; they do not prove a structural mechanism or identify an independent military-aid effect.
	The Extended row uses the 2010--2013 pre-Crimea window and adds Kenya, increasing the regression sample from 119 to 120 countries. For energy imports/GDP in 2018--2021, Strict and Extended select the same 118-country regression sample, so no duplicate near-prewar Extended row is reported.
	
	\begin{landscape}
	\begin{table}[p]
		\centering
		\caption{Robustness and alternative definitions for Russia-energy exposure heterogeneity}
		\label{tab:B8_energy_exposure_robustness}
		\begin{adjustbox}{max width=\linewidth}
			\begin{threeparttable}
				\small
				\begin{tabular}{l c c c c c}
					\toprule
					\multicolumn{6}{l}{\textbf{Panel A. Price outcomes}} \\
					\addlinespace
					& \multicolumn{2}{c}{\textit{Log CPI}} & \multicolumn{2}{c}{\textit{Inflation (\%)}} & \\
					\cmidrule(lr){2-3} \cmidrule(lr){4-5}
					\textit{Specification} & $\widehat\beta_3$ (Diff.) & $\widehat\beta_2+\widehat\beta_3$ (Total) & $\widehat\beta_3$ (Diff.) & $\widehat\beta_2+\widehat\beta_3$ (Total) & \textit{N (All / Treated)} \\
					\midrule
					Baseline configuration & 0.0195\sym{**} (0.0097) & 0.0213\sym{***} (0.0060) & 0.5919 (0.5147) & 0.6368\sym{***} (0.2126) & 118 / 39 \\
					Pre-Crimea exposure (2010--2013) & 0.0089 (0.0127) & 0.0173\sym{***} (0.0034) & 0.5140 (0.8551) & 0.5959\sym{***} (0.1217) & 119 / 39 \\
					Extended pre-Crimea sample (2010--2013) & 0.0073 (0.0130) & 0.0173\sym{***} (0.0034) & 0.4655 (0.8543) & 0.5937\sym{***} (0.1213) & 120 / 39 \\
					Two-way fixed effects (TWFE) & 0.0559 (0.0400) & 0.0725\sym{***} (0.0197) & 1.0142\sym{***} (0.2871) & 1.0817\sym{***} (0.1849) & 118 / 39 \\
					Strict direct arming definition & 0.0120 (0.0104) & 0.0155\sym{**} (0.0067) & 0.4496 (0.4941) & 0.5185\sym{**} (0.2205) & 118 / 26 \\
					Alternative exposure: gas share & 0.0081 (0.0180) & 0.0162\sym{***} (0.0052) & 0.1084 (1.1670) & 0.3632\sym{*} (0.1917) & 118 / 39 \\
					\midrule
					\multicolumn{6}{l}{\textbf{Panel B. Output outcomes}} \\
					\addlinespace
					& \multicolumn{2}{c}{\textit{Log real GDP per capita}} & \multicolumn{2}{c}{\textit{GDPpc growth (\%)}} & \\
					\cmidrule(lr){2-3} \cmidrule(lr){4-5}
					\textit{Specification} & $\widehat\beta_3$ (Diff.) & $\widehat\beta_2+\widehat\beta_3$ (Total) & $\widehat\beta_3$ (Diff.) & $\widehat\beta_2+\widehat\beta_3$ (Total) & \textit{N (All / Treated)} \\
					\midrule
					Baseline configuration & -0.0189 (0.0120) & -0.0147\sym{*} (0.0081) & -1.2875\sym{***} (0.3955) & -0.9580\sym{***} (0.2435) & 118 / 39 \\
					Pre-Crimea exposure (2010--2013) & -0.0205\sym{*} (0.0115) & -0.0072 (0.0048) & -1.3592\sym{***} (0.3496) & -0.6372\sym{***} (0.1434) & 119 / 39 \\
					Extended pre-Crimea sample (2010--2013) & -0.0194\sym{*} (0.0113) & -0.0072 (0.0048) & -1.3423\sym{***} (0.3514) & -0.6349\sym{***} (0.1429) & 120 / 39 \\
					Two-way fixed effects (TWFE) & 0.0214 (0.0253) & 0.0611\sym{***} (0.0157) & -1.2403\sym{***} (0.3815) & -0.8801\sym{***} (0.2390) & 118 / 39 \\
					Strict direct arming definition & -0.0163 (0.0115) & -0.0134\sym{**} (0.0067) & -1.1524\sym{***} (0.3951) & -0.9239\sym{***} (0.2387) & 118 / 26 \\
					Alternative exposure: gas share & -0.0061 (0.0194) & -0.0130\sym{*} (0.0075) & -0.7079 (0.5985) & -0.6567\sym{**} (0.2564) & 118 / 39 \\
					\bottomrule
				\end{tabular}
				\begin{tablenotes}[flushleft]
					\footnotesize
					\item \textit{Notes:} Entries are coefficient estimates with country-clustered standard errors in parentheses. $\beta_3$ represents the policy-by-exposure-by-post-2022 interaction (the differential gradient); $\beta_2+\beta_3$ represents the total post-2022 exposure slope exclusively within the targeted policy group. The standard error for $\beta_2+\beta_3$ uses the full covariance matrix. Unless otherwise specified, the baseline configuration utilizes the CCE estimator (with country-specific intercepts and donor-factor loadings), the strict inclusion rule, the strict sanctions definition, and Russian fossil-energy imports/GDP measured over 2018--2021. The extended-sample row applies the Extended coverage rule to the 2010--2013 exposure window and adds Kenya relative to Strict. Strict and Extended energy-imports/GDP samples are identical for 2018--2021, so no separate near-prewar Extended row is shown. TWFE substitutes the donor-factor loadings for rigid country and year fixed effects. Inflation and GDP-per-capita growth are measured in annual percentage points; logged outcomes are in log points. The gas-share alternative remains subject to the pipeline-reporting limitations documented in Appendix Table~\ref{tab:A9_energy_exposure_construction}. \sym{*} $p<0.10$, \sym{**} $p<0.05$, and \sym{***} $p<0.01$.
				\end{tablenotes}
			\end{threeparttable}
		\end{adjustbox}
	\end{table}
	\end{landscape}

\end{document}